\documentclass[12pt,final,epsfig]{styles/ucscthesis}

\usepackage{styles/epsf}
\usepackage{styles/apjuc}
\usepackage{graphicx}
\usepackage{amsmath} 
\usepackage{amsxtra} 
\usepackage{amssymb} 
\usepackage{amsfonts} 
\usepackage{amstext} 
\usepackage{amsgen} 
\usepackage{amsbsy} 
\usepackage{amsopn} 
\usepackage{amscd} 

\begin{document}


\title{A Search for TeV Gamma-Ray Burst Emission with the Milagro Observatory}

\author{Miguel F. Morales}

\degreeyear{2002}
\degreemonth{September}
\degree{Doctor of Philosophy}
\field{Physics}

\chair{Professor David Dorfan}
\committeememberone{Professor David A. Williams}
\committeemembertwo{Professor Joel Primack}
\numberofmembers{3}

\campus{Santa Cruz}

\maketitle
\copyrightpage


\begin{frontmatter}
\tableofcontents
\listoffigures
\listoftables

\begin{abstract}
The Milagro telescope monitors the northern sky for 100 GeV -- 100 TeV transient emission through continuous very high energy wide-field observations.  The large effective area and low energy threshold of Milagro allow it to detect very high energy gamma-ray burst emission with much higher sensitivity than previous instruments, and a fluence sensitivity at TeV energies comparable to dedicated gamma-ray burst satellites at keV-MeV energies.  Observation of gamma-ray burst emission at TeV energies could place important constraints on  gamma-ray burst progenitor and emission models.  This study details the development of a weighted analysis technique; the implementation of this technique to perform a real time search for TeV transients of 40 seconds to 3 hours duration in the Milagro data; and the results from more than one year of observation.  Between May $2^\text{nd}$, 2001, and May $22^\text{nd}$, 2002, no TeV transients of 40 seconds to 3 hours duration were observed.  Upper limits on both observed and emitted high energy gamma-ray burst emission are presented.
\end{abstract}

\begin{dedication}
\null\vfil
{\large
\begin{center}
To Ginger\\\vspace{4pt}
For all her love and making this possible
\end{center}}
\vfil\null
\end{dedication}

\begin{acknowledgements}

There are many, many people without whose support and encouragement this thesis would not have been possible.

First I would like to thank my advisor David Williams for wading through all my half-formed, crackpot ideas and having the patience to manage a graduate student who is politely described as ``independently minded."  I would also like to thank the rest of the UCSC Milagro group for all their help. Don for his quiet guidance, long statistics discussions, and letting me see how the ``real" physics world works.  Linda for letting me break her computers again and again, and Michael for fixing the detector every summer.  Ty for more statistics discussions --- Feldman and Cousins is cool --- and Wystan who has been a compatriot through six years of Milagro.

A special thanks goes out to Jay Norris for showing me joys of working with actual observations and his patience as the BATSE analysis fell further and further behind --- I still plan to do some of it.  Being introduced to the Goddard Space Flight Center and the people in the GRB community has been an enormous help, as has the cash to allow me to concentrate on research.

The entire Milagro collaboration has built a neat instrument and a supportive environment for graduate students.  I'd particularly like to thank Brenda and Gus for their suggestions and constructive critiques and Jordan for promoting my movies.  I'd also like to thank Scott and the DeLay family for taking me fishing and inviting me into their home while I was out in New Mexico babysitting the detector.

David Dorfan and Joel Primack not only served on my committee, but have been very helpful in guiding me through some of the interesting physics ideas over the years.  I will miss the passionate discussions with David on everything from soccer to particle physics.

Lowell has been a fabulous office mate and a source of much entertainment.  I will miss the long discussions figuring out exactly how things work and having someone to bounce ideas off of.  Someday we should write that paper on Cherenkov radiation.

I'd also like to thank all of the institutions which I've hoodwinked into giving me money over the years:  the Harvey N. Collison Scholarships, the Mellon Foundation, the Cota Robles Fellowships, the GAANN Fellowships, and NASA.  The financial support has been crucial in allowing me to obtain a great education.

On a more personal note, I'd like to thank my family for all their support over the years --- look what happens when you stress the importance of education!  Thanks Mom and Dad for teaching me how to do math and giving me the inheritance of a Swarthmore education.  Your love of science fostered my interest in studying astrophysics.

And finally I'd like to thank Ginger for all her love and enabling me to pursue my dream.  This could not have been done without you, and I thank you with all my heart.

\hspace{4in}-- Miguel

\end{acknowledgements}

\end{frontmatter}


\chapter{Gamma-Ray Bursts}

\section{Introduction}

Since their discovery in 1967, gamma-ray bursts have remained one of the most enigmatic astrophysical phenomena.  For thirty years after their discovery even basic questions such as whether they were local or cosmological in origin were open to debate.  Over the past five years the knowledge of gamma-ray bursts (GRBs) has been revolutionized with the discovery of transient x-ray, optical and radio counterparts and the advent of the ``afterglow era" of GRB science.  Afterglow observations have settled the old debate on the distance scale of GRBs, helped determine GRB energetics, and allowed tentative early studies of the host galaxies and environments.  However, we are still in the early stages of GRB science, with key questions about the progenitors and emission mechanisms still to be determined, and almost no observational constraints for a second class of GRBs.  The field of GRB science is evolving so rapidly any review is destined to be immediately out of date.  This chapter attempts to give a broad overview of the current understanding of GRBs and put the science motivations for this thesis into context.

\section{BATSE Observations and the Beginning of the Afterglow Era}

No review of gamma-ray bursts is complete without discussing the pioneering results from the burst and transient search experiment (BATSE) on the Compton gamma-ray observatory (CGRO).  CGRO was launched in April 1991 and BATSE observed nearly four thousand GRBs over the next nine years.  BATSE combined extraordinarily high gamma-ray sensitivity and full sky coverage with good energy resolution and localization abilities.  BATSE's sensitivity to GRBs will only be surpassed with the launch of SWIFT and the next generation of GRB satellite experiments.

\begin{figure}
\begin{center}
\includegraphics[width=5.75in]{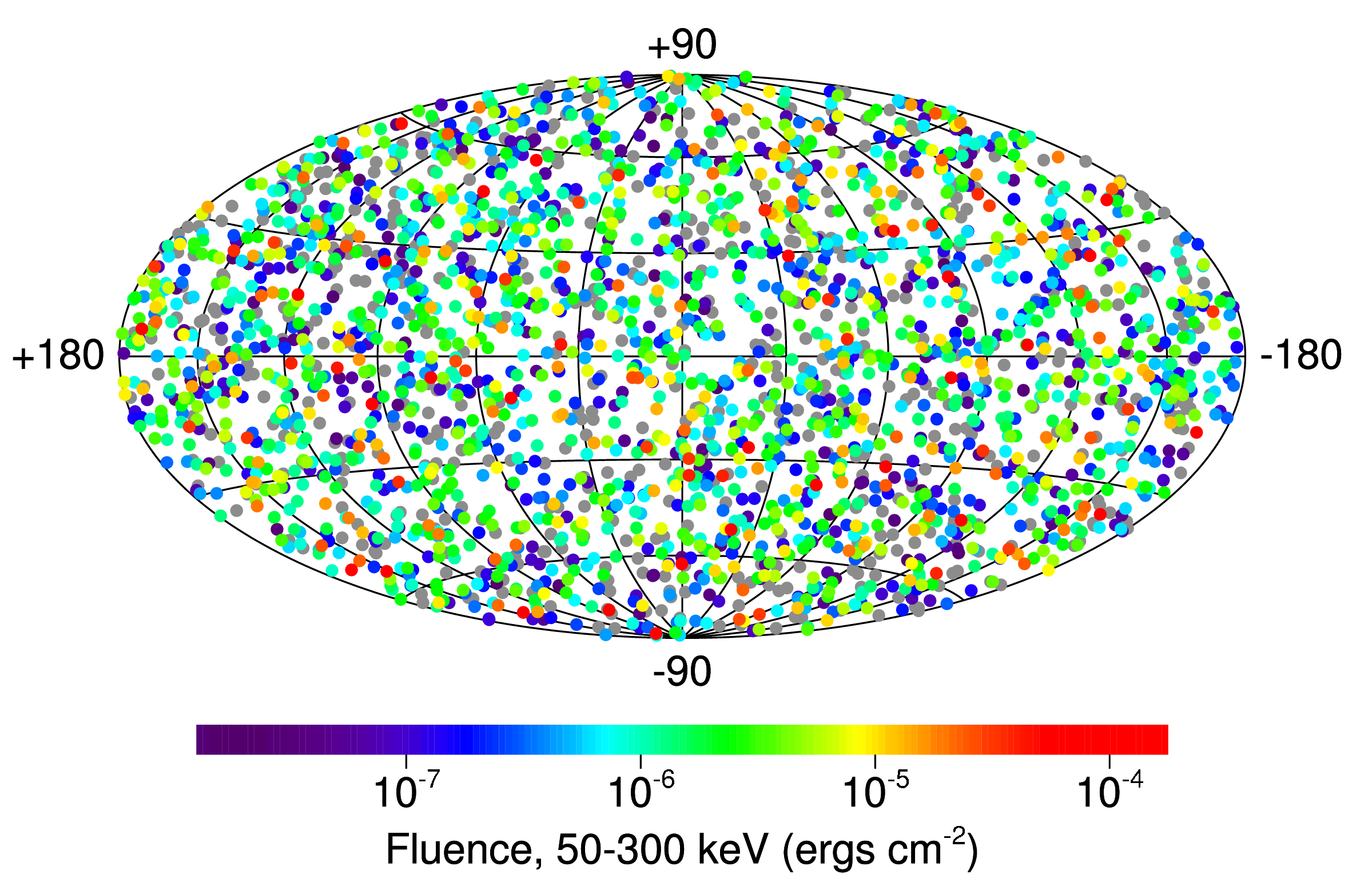}
\caption[Map of the GRB locations observed by BATSE]{The locations of 2704 BATSE GRBs in galactic coordinates.  The isotropic distribution of events is clearly evident with no strengthening towards the galactic plane.  In addition, there is no evidence for repeated emission.  Figure is from BATSE data web page \cite*{BATSEWeb}.}
\label{BATSEGRBMap}
\end{center}
\end{figure}

The first major result from BATSE was the isotropic distribution of gamma-ray bursts as shown in Figure \ref{BATSEGRBMap}.  Most early GRB models were based on emission from the neutron star population within our galaxy.  However, the distribution of GRBs appears isotropic with no discernible bias towards the galactic plane.  The isotropic distribution suggested a cosmological origin for GRBs, but the issue was still actively debated until the observation of GRB host redshifts.

\begin{figure}
\begin{center}
\includegraphics[width=5.75in]{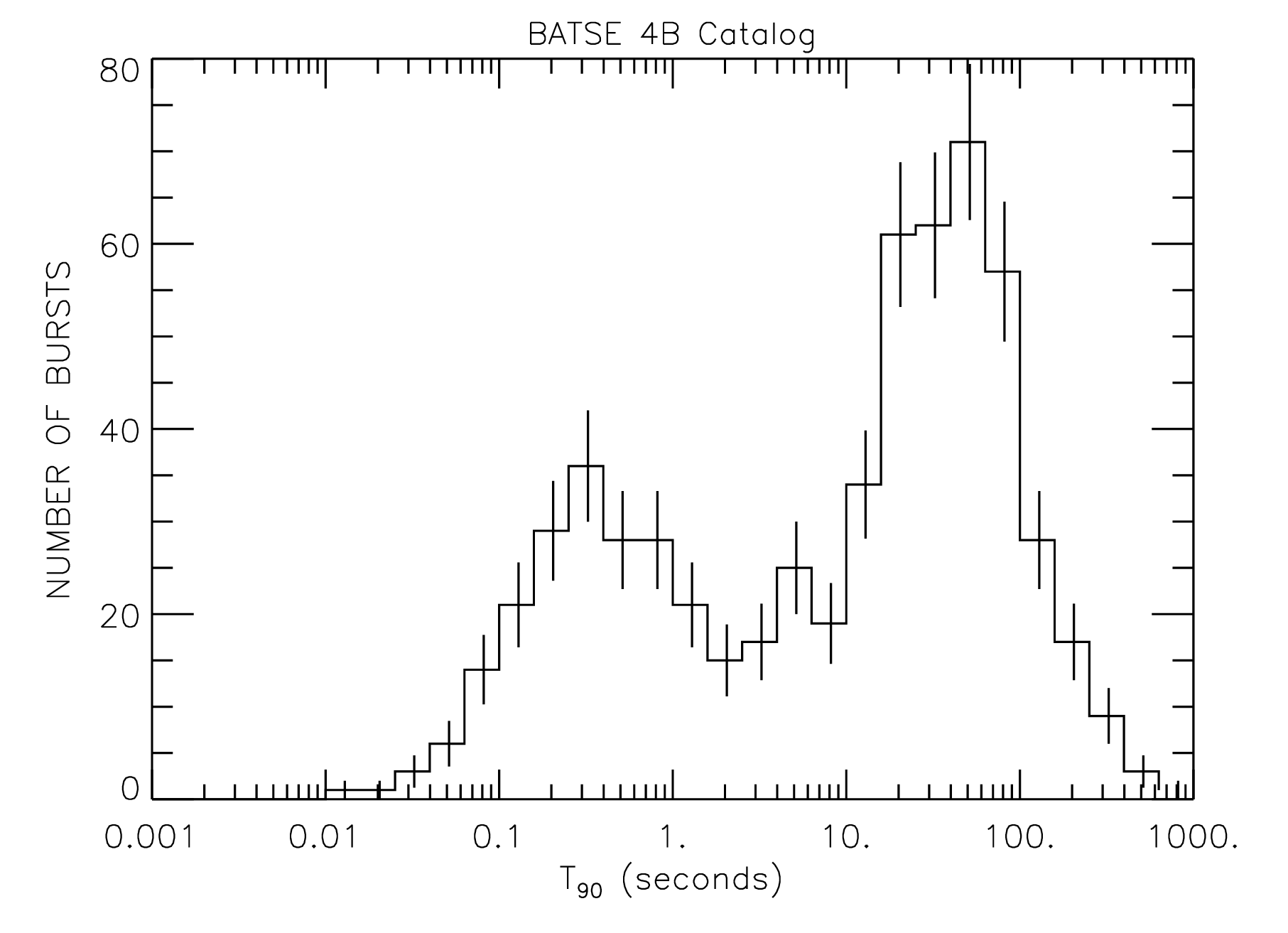}
\caption[Distribution of BATSE $\text{T}_{90}$ distributions]{The distribution of GRB durations listed in the BATSE 4B catalog.  The duration is determined by the time interval in which the central 90\% of the event counts occur ($\text{T}_{90}$).  The bimodal distribution of event times can be clearly seen, with the two types of GRBs descriptively labeled as ``short" and ``long."  Figure from BATSE data web page \cite*{BATSEWeb}.}
\label{BATSEdistributions}
\end{center}
\end{figure}

The second major result from BATSE was the bimodal distribution of GRB durations as shown in Figure \ref{BATSEdistributions}.  This distribution strongly suggested that there are two distinct kinds of gamma-ray bursts --- descriptively named ``short" and ``long" GRBs. However, it is only recently that conclusive evidence has arisen that there are two distinct classes of GRBs \cite{Norris:2GRBs}.  

Towards the end of BATSE's mission GRB science was in a tantalizing holding pattern.  The BATSE data suggested two classes of GRBs at cosmological distances, however there was not enough information to conclusively demonstrate either point.  While BATSE's resolution of $\sim$4 degrees was good for a wide field-of-view gamma ray observatory, it is terrible by optical standards.  The location error boxes provided by BATSE were simply too large to search with standard optical telescopes. BATSE continued to collect locations and spectra about once per day, but the data were largely useless without the key to help interpret it.

The breakthrough came in the winter of 1997 with the first conclusive evidence of an x-ray afterglow by the BeppoSax satellite \cite{BeppoSax1}.  Because x-ray telescopes have a much better angular resolution, they can pinpoint the location of the GRB and allow follow-up observations by optical telescopes.  By the spring of 1997 the first redshift observation was reported with a z $\ge$0.835 \cite{KeckRedshift1}, settling once and for all the local vs. cosmological distance question.

The discovery of transient GRB afterglows at longer wavelengths opened up new avenues of research, and finally put the BATSE data into context.  With redshift measurements the intrinsic gamma-ray luminosity of GRBs could be determined and correlated with the detailed gamma-ray observations of BATSE.  In addition, the multi-wavelength afterglows contained important information about the characteristics of the explosion and the environment surrounding the GRB progenitor.  

The remainder of this chapter will concentrate on the observational advances in the past five years which have been enabled by afterglow observations.  One important caveat to remember in this discussion is that all of the 56 counterparts which have been observed are associated with long GRBs.  At this time very little is known about short GRBs, except that there are no bright afterglows.\footnote{As if to underline how quickly this field is moving, since this was written in mid June 2002 word has come down grapevine that the first redshift for a short GRB has been determined to be $\text{z} \approx 1$.  There are no papers available yet, and it hasn't even made the online lists.  However, as with many significant steps in GRB science, the speed of gossip is infinite.  This is obviously a very important result, but much more data and analysis needs to be done before we can begin to understand short GRBs and how they differ from long GRBs.}

\section{The Redshift Distribution of GRBs}

\begin{figure}
\begin{center}
\includegraphics[width=5.75in]{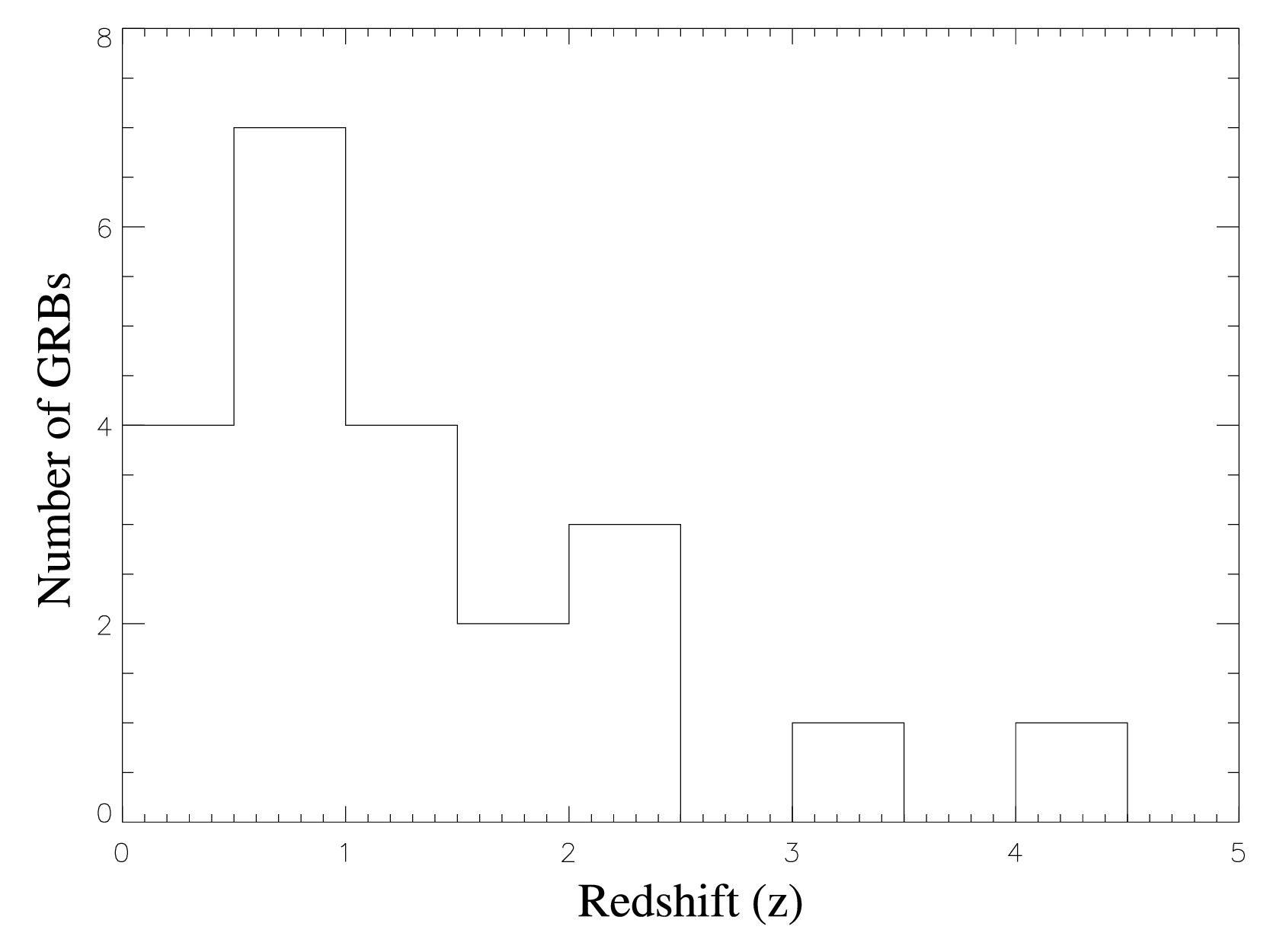}
\caption[GRB Redshift Distribution]{This histogram plots the redshift distribution for the 22 GRBs with reliable redshift determinations.  For context, the era of reionization -- and thus the edge of the visible universe at optical wavelengths -- is believed to be at a redshift near six \cite{SloanReionization}.}
\label{GRBzdist}
\end{center}
\end{figure}

To date, 22 of the 56 observed afterglows have yielded reliable redshift measurements, with the distribution spanning much of the observable universe (see Figure \ref{GRBzdist}).  One host galaxy has been identified at a redshift of 4.5, while one weak GRB has been associated with a supernova at a redshift of only 0.0085. Not only do the redshift measurements conclusively show that long GRBs are cosmological in origin, but they provide the Rosetta Stone for interpreting the gamma-ray spectra measured by BATSE and other experiments.

The first use of the redshift measurements was to convert the fluence measured by BATSE into the gamma-ray luminosity of the source, assuming isotropic emission.  The resulting luminosities are enormous, implying a total energy release of $10^{51}$ -- $10^{53}$ ergs in gamma rays.  This is the equivalent of converting up to $1/10$ of a solar mass into gamma rays in approximately 10 seconds. Even if the gamma-ray emission is highly beamed, GRBs represent the most powerful astrophysical explosions known, and there are very few theoretical scenarios which can satisfy the energetic requirements.

The redshift measurements have also been used to illuminate the spectral and temporal properties measured by BATSE.  Because CGRO was deorbited only three years after the first afterglow observation, there are only eight GRBs with both redshifts and precision gamma-ray measurements by BATSE.  However, numerous groups have used this small subset of GRBs to look for correlations between the gamma-ray data and the isotropic luminosity of GRBs. The goal is to use this very small sample of bursts to help understand the gamma-ray properties of GRBs, then use the knowledge gained to extract science results from the nearly four thousand GRBs detected by BATSE.

\begin{figure}
\begin{center}
\includegraphics[width=5.75in]{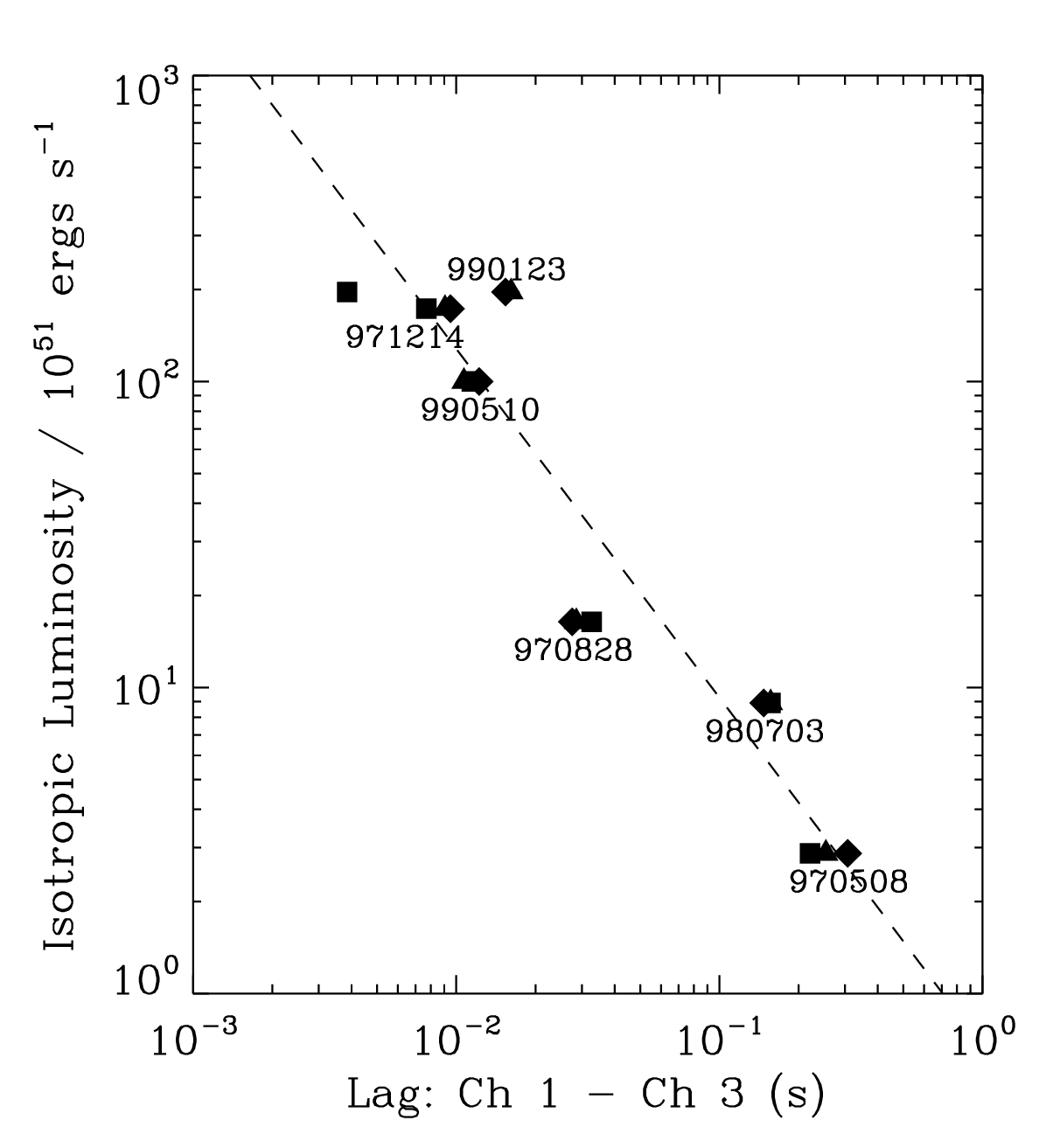}
\caption[Lag-Luminosity Correlation]{The Lag-Luminosity correlation from Norris et al.\ \cite*{Norris:LagLum}. The vertical axis plots the isotropic gamma-ray luminosity while the horizontal axis plots the time lag between BATSE channels 1 (20 -- 50 keV) and 3 (100 -- 300 keV).  For each GRB the lag is plotted for when the count rate is $\ge$0.1 ({\em diamonds}), $\ge$0.3 ({\em triangles}) and $\ge$0.5 ({\em squares}) the peak intensity.  The power law fit shown is for count rates $\ge$0.1 the peak intensity.  The burst 980425 which was associated with nearby supernova 1998bw is four orders of magnitude dimmer than any of the other bursts and was not used.  It generally follows the same trend, but does not fall along the power law shown.}
\label{LagLum}
\end{center}
\end{figure}

Of particular interest are two different features of the gamma-ray emission which appear to correlate well with the isotropic intrinsic luminosity.  The first measure was the lag-luminosity relationship developed by Norris et al.\ \cite*{Norris:LagLum}.  In long GRBs there are many bright pulses of gamma-rays, with the pulses covering the entire BATSE energy range.  However, the high energy photons of a pulse arrive slightly earlier than lower energy photons, creating a lag in the arrival time between different energy channels.  This time lag is anti-correlated with the isotropic luminosity of the GRB (see Figure \ref{LagLum}).  Additionally, Reichart et al.\ \cite*{FenimoreRuiz:Variability} have found a similar relationship between the variability of the GRB signal and the isotropic luminosity.  

Both of these relationships have been studied in depth by many researchers, and despite initial misgivings due to the very small number of bursts used to identify the correlations, there are increasing indications that both relationships are valid.  A number of researchers have used these relationships to determine approximate redshifts for a large number of GRBs in the BATSE catalog and explore the cosmological implications \cite{Lloyd:GRBLumEvolution,Norris:Implications}.  Unfortunately, the current GRB detectors do not have the sensitivity of BATSE, and we must wait for the data from SWIFT and INTEGRAL to increase the sample of bursts and refine the observed correlations. 

One of the side effects of trying to correlate the gamma-ray signal with the isotropic luminosity has been the development of sophisticated measures of the gamma-ray time evolution.  These measures have been used by Norris et al.\ \cite*{Norris:2GRBs} to conclusively show that the time dependent spectral characteristics of short GRBs are distinct from long GRBs, and that there are two classes of GRBs.  In retrospect this separation could have been found without the afterglow information, but the existence of the redshift measurements facilitated this work.

\section{GRB Host Galaxies}

In addition to redshift measurements, optical and infra-red observations have been used to study the properties of GRB host galaxies.  Long exposures of the afterglow positions have identified host galaxies in almost all cases, with the GRB position typically within the galactic disks.  Advanced surveys of the GRB host galaxies are starting to appear, and suggest that the GRB host galaxies are actively forming stars --- though it is unclear how high the star formation rate is \cite{Djorgovski:GRBHost,Holland:3HST,Chary:GRBHost}.  There are also indications that GRBs occur in areas of the galaxies which are actively forming stars, based on their distance from the nucleus \cite{Bloom:GRBDist} and in a few cases by observation of the nearby stellar populations \cite{Holland:3HST}.  

It has also become clear that approximately half of the GRBs with x-ray and/or radio afterglows have no optical counterpart.  The missing optical counterparts were initially thought to be due to the limited sensitivity of the telescopes used.  However, with increasing use of the W. M. Keck Observatory and other large telescopes it has become clear that these bursts must be intrinsically very dim \cite{Reichart:DarkGRBs}.  One possible explanation of the so called ``dark" bursts is the absorption of UV and optical radiation by interstellar dust.  This has been seen as another indication that GRBs may occur in star forming regions where the concentration of dust is very high.

The evidence of GRBs being associated with active star formation fits nicely with the collapsar progenitor model which predicts that GRBs will be associated with the core collapse of very massive stars (see Section \ref{Progenitors}).  Even if the collapsar model does not work, the weight of evidence seems to suggest that long GRBs are associated with star formation.\footnote{It is sometimes hard to judge how good the correlation between GRBs and star forming regions really is because of the popularity of the collapsar model.  In many papers the conclusions seem to overstate the evidence, partly because the authors expect GRBs to be associated with star formation.}  There is also evidence that the gamma-ray luminosity of at least some GRBs is so high that modern GRB satellites can observe all of the events which occur in the observable universe.  This raises tantalizing possibilities of using GRBs to perform cosmological studies extending to the era of reionization and beyond.\footnote{Because gamma rays are not absorbed by neutral hydrogen, in principal GRBs may be visible to very high z.}  If GRBs are associated with star formation and the gamma ray vs. redshift correlations can be refined, GRBs could be used as a tracer of star formation and provide key insights into the early development of the universe.

\section{Gamma-Ray Emission Processes}

Both the prompt and afterglow emission from GRBs can be understood in the fireball shock framework, which has been remarkably successful in explaining both the prompt gamma-ray and the multi-wavelength afterglow observations.\footnote{For an in-depth review of GRB theory see the excellent paper by Meszaros \cite*{Meszaros:Review} which much of this discussion is based on.}  Interestingly, the fireball shock scenario does not posit a GRB progenitor.  It simply assumes certain characteristics for the explosive outflow and leaves the generation of the explosion for others to determine (see Section \ref{Progenitors}). 

The gamma-ray spectra observed by BATSE exhibit a smoothly broken power-law shape, or ``Band function" \cite{Band:Function}, with a characteristic power law of -1 at low energies and between -2 and -3 above a break energy of 0.1 -- 1 MeV.  Because of the small size of the emission region (implied by the short duration of GRB emission) and the extremely high luminosity above 0.511 MeV, the first theoretical difficulty is explaining how to avoid the $\gamma\gamma \rightarrow e^+e^-$ process which would normally absorb the high energy gamma-ray photons.  A natural explanation is that the emission region is moving relativistically with a Lorentz factor of 100 -- 1000, and the observed gamma-ray emission has been blueshifted from longer wavelengths.  

The requisite Lorentz factors can be obtained by depositing a large amount of energy into a small mass and creating a fireball.  The fireball converts the internal energy into an accelerating expansion, forming a relativistically expanding $e^\pm$,$\gamma$ plasma with some entrained baryons.  

Once the fireball has expanded far enough to become optically thin to gamma rays, the question becomes how to convert the kinetic energy of the fireball into the observed gamma-ray radiation.  The answer comes from realizing that shocks are likely to occur either as the fireball collides with the ambient medium (external shock model), or within the fireball itself if the initial power source is variable and creates shells with different Lorentz values (internal shock model).  Charged particles near a shock can be accelerated by scattering off the magnetic fields in the plasma and repeatedly crossing the shock boundary.  Shock acceleration naturally leads to a power-law distribution of high energy particles in a region of high magnetic field, and efficiently converts the kinetic energy of the fireball into synchrotron and inverse Compton radiation.  The Band spectrum observed by BATSE is well explained by synchrotron radiation from a power-law distribution of electrons which is then blueshifted by the relativistic motion to gamma-ray energies.

There is still active debate as to whether internal or external shocks dominate, but for most GRBs the prompt gamma-ray emission is best explained by the internal shock model, with the afterglow emission explained by an external shock propagating into the ambient medium.

\section{Afterglows and Collimated Emission}

The broad-band afterglow emission is remarkably well described by a cooling external shock propagating into an ambient medium.  In general the afterglow spectrum is a broad bump consisting of four power-law segments with three breaks, with the power-law indices and break positions depending on the distribution of electron energies.  As the fireball expands and cools the luminosity decreases and the peak of the emission spectrum slides to lower energies.  The resulting time evolution of the luminosity follows a broken power-law in each energy band, with ``chromatic" breaks as the spectral peak moves through the observation band.

There are also ``achromatic" breaks in the luminosity decay rate which occur simultaneously at all frequencies.  These achromatic breaks are a natural feature of a collimated jet and are the clearest evidence that the GRB emission is not isotropic but instead beamed. At early times the radiation from a fireball is highly forward beamed due to the relativistic boost, and only a small portion of the fireball surface can be seen by an observer.  As the fireball sweeps up more ambient material and slows, more and more of the surface becomes visible --- artificially enhancing the luminosity.   However, this enhancement stops when the entire face of the jet becomes visible, and results in an achromatic break in the luminosity decay rate.

By fitting the data to intensive multi-wavelength observation campaigns, the physical parameters of $\sim$20 bursts have been determined, including the opening angle of the relativistic jets.  Surprisingly, when the gamma-ray luminosity is recalculated to account for the opening angle, the energy release for all the bursts clusters near $5\times10^{50}$ ergs \cite{Frail:StandardEnergyReservoir,Panaitescu:PropJets}.  This result could be extremely important, as it implies that all long GRBs may be produced by the same underlying phenomenon with a well defined energy release similar to the most energetic supernovae.  The fitting of model parameters may also determine the environment near the burst (how dense, how much dust), particularly with the large statistics and very early afterglow observations that SWIFT will provide.

One intriguing side effect of beamed GRB gamma-ray emission is that there should be a large population of ``orphan afterglows."  Since the optical through radio emission peaks after the jet has slowed, the emission should be nearly isotropic.  This means that afterglows at these frequencies should be observable even if we view the GRB off-axis and detect no gamma-ray emission.  A number of researchers are starting to make wide field-of-view transient observations at optical and longer wavelengths, and should be able to set direct limits on the beaming of GRBs and the true GRB rate.

\section{Progenitor Models}
\label{Progenitors}

Despite the success of the fireball shock scenario, models of the GRB progenitor and the creation of the relativistic jet are very speculative.  Because of the enormous energy requirements, most of the models include the formation of a black hole which is rapidly accreting material.\footnote{The recent data on collimation of GRB jets has significantly reduced the energy requirements of the central engine; however, black hole models are still preferred.}  The gravitational binding energy of either the black hole or the disk may then be tapped to power the relativistic jet.  What is not clear is what astrophysical object forms the black hole and accretion disk, or how the gravitational binding energy is transferred to the jet.

The current theory {\em du jour} is the collapsar model by Woosley \cite*{Woosley:collapsars}.  In this scenario the iron core of a very massive spinning helium star collapses to a black hole.  The matter along the polar axis falls into the black hole while the equatorial regions forms a centrifugally supported disk outside the last stable orbit.  Matter continues to accrete through the disk at 0.01 -- 0.1 $\text{M}_\odot$ $\text{s}^{-1}$, a significant fraction of which is ejected as a powerful wind of $\text{Ni}^{56}$.  In this scenario the jet is either powered by neutrinos which are produced in the accretion disk and annihilate in the polar regions, or by magneto-hydrodynamic (MHD) processes powered by magnetic fields in the accretion disk.  In either case, the jet would be naturally collimated by the funnel shaped cavity that forms along the poles of the star.  This theory has had several nice features, including a natural initial population, collimation, association with star forming regions, and a predicted supernova-like light curve at long times from the $\text{Ni}^{56}$.  This theory has been supported by the association of GRB 980425 with supernova 1998bw and several recent GRBs which have shown signs of supernova light curves superimposed on the fading afterglows \cite{Garnavich:GRBSN}.

However, there are many other theories and the collapsar model develops too slowly to easily explain the short GRBs.  Other progenitor models include black hole - helium star mergers (similar to collapsars in behavior), neutron star mergers (for short GRBs), Kerr black holes braked by the Blandford-Znajek process (electromagnetic vacuum breakdown), and $\sim$300 other models \cite{Freyer:GRBformationRate,BlandfordZnajek,Meszaros:Review}.  More data are needed, and extending the gamma-ray observations to TeV energies could provide important constraints on GRB progenitors.

\section{TeV GRB Emission?}

Whether GRBs should emit large amounts of TeV radiation is very model dependent.  If the observed keV-MeV spectrum is due to synchrotron emission, one would expect some of the synchrotron photons to be upscattered by the energetic $\text{e}^\pm$ distribution to create a second gamma-ray peak at TeV energies.  This synchrotron-self-Compton (SSC) mechanism can be very efficient at producing high energy photons and may be responsible for the strong TeV emission of some active galactic nuclei such as Markarian 421 and Markarian 501.  

Models based on both internal and external shocks have predicted TeV emission comparable to, or in certain situations stronger than, the keV-MeV radiation \cite{Dermer:VHEGRB,PillaLoeb:VHEGRB}.  However, TeV emission is sensitive to a number of model parameters.  Because of  $\gamma\gamma \rightarrow e^+e^-$ absorption, TeV radiation is particularly sensitive to the Lorentz factor and the photon density (and thus the distance of the shock from the source) when the radiation is emitted.  The duration of the TeV radiation is also model dependent, with everything from shorter than the keV-MeV emission to extended TeV afterglows.  While these features make the emission of TeV radiation uncertain, they also lend power to the observations.  Because TeV radiation is so model dependent, observations can provide key insights to the emission process and potentially the GRB progenitor.

\begin{figure}
\begin{center}
\includegraphics[width=5.75in]{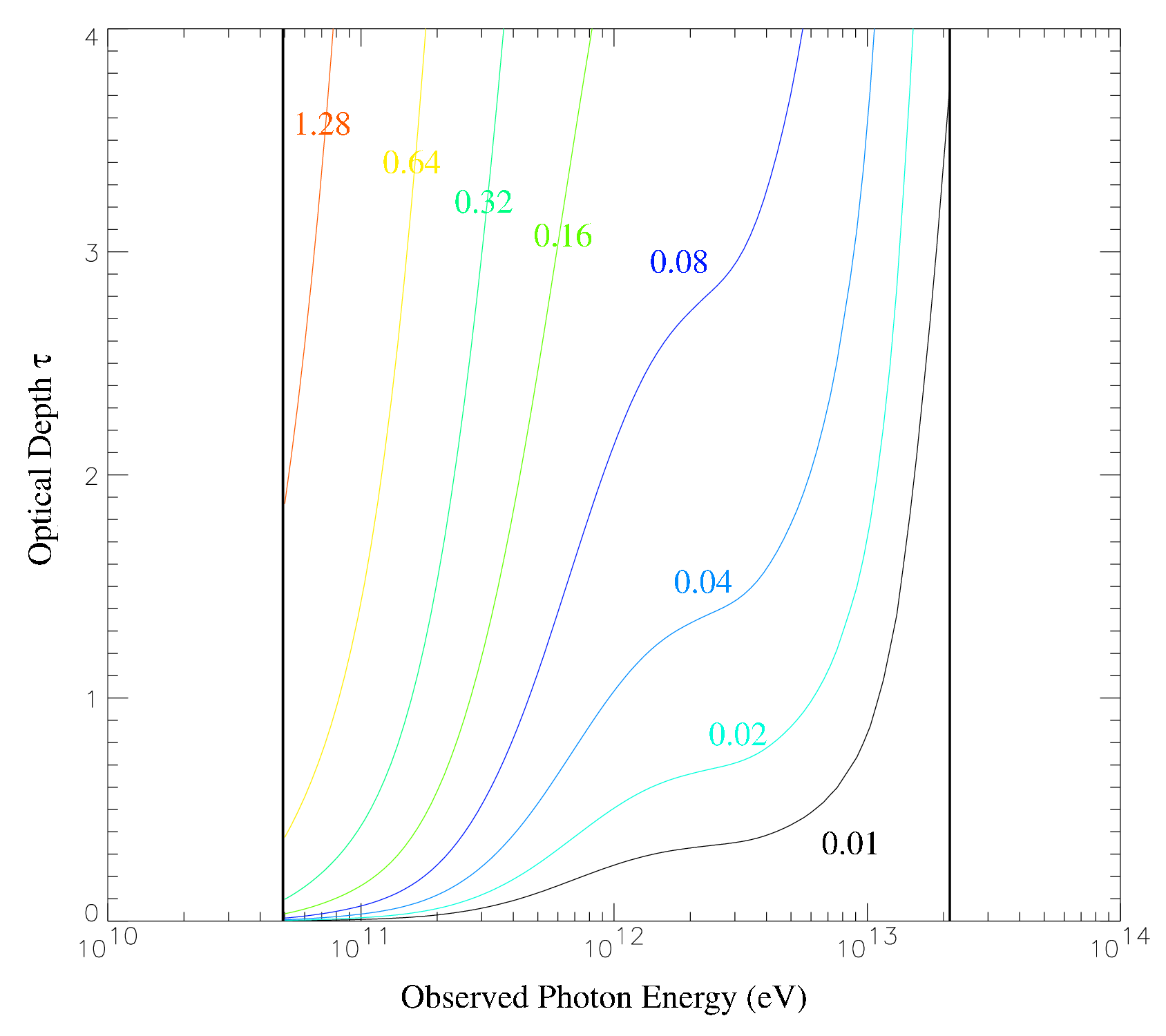}
\caption[Attenuation of Gamma Rays by Extragalactic Background Light]{The optical depth $\tau$ as a function of observed energy and redshift.  The attenuation factor is equal to $e^{-\tau}$, and the redshift of each optical depth curve is indicated by the same color number.  This set of optical depths was obtained by James Bullock using semi-analytic modeling with a flat $\Lambda$CDM universe with $\Omega_\text{M}=0.3$, $\Omega_\Lambda=0.7$, h = 0.65 with a Kennicutt initial mass function.  The black vertical lines show the energy range of the optical depth calculation. Data courtesy Bullock \cite*{James:EBL}.}
\label{JamesAtten}
\end{center}
\end{figure}

An observational complication is introduced by the interaction of TeV photons with extragalactic background light (EBL) \cite{Primack:ProbingGalaxyFormationWGammaRays,Stecker:EBL,Jelley}.  Over cosmological distances space becomes increasingly opaque to TeV photons due to the $\gamma\gamma \rightarrow e^+e^-$ reaction with background starlight (see Figure \ref{JamesAtten}).  EBL absorption depends on the amount of extragalactic light and thus the details of galaxy formation, particularly the star formation history and the effects of dust on the emitted spectrum.  Because the absorption of TeV gamma rays has not been accurately measured, this adds some model dependence to TeV observation of distant GRBs.

Several attempts have been made to detect TeV emission from GRBs.  There have been no clear detections, but several groups have reported hints of TeV emission.  The Tibet air-shower array looked for $\sim$10 TeV emission coincident with BATSE observations from June 1990 -- September 1992.  While no significant excesses were found, adding all burst-like events near the 57 BATSE positions on time scales from 1 s -- 100 s produced a 6$\sigma$ deviation from background \cite{TibetGRB}.  Air-Cherenkov telescopes have also been used to search for extended TeV emission by slewing the telescope to the GRB position within a few minutes of the trigger, but no emission has been observed \cite{WhippleGRB}.  The most interesting TeV result is from Milagrito, the prototype for Milagro, which searched for emission coincident with 54 BATSE detections.  An excess was observed coincident with one of the BATSE triggers, with a chance probability of $1.5\times10^{-3}$ after all trials factors \cite{MilagritoGRB}.  While not strong enough for a discovery, this event provided a tantalizing suggestion of TeV emission.  McCullough also used the Milagrito telescope to search for  bursts of 1 s -- 40 min. duration, and observed no significant excesses \cite{McCullough}.  The thesis by McCullough was unique in searching the entire visible sky for $\sim$ 1 TeV radiation without requiring a satellite-based trigger. 

The initial incarnation of this thesis envisioned using BATSE and other satellites to identify GRBs, then searching the Milagro data for coincident emission.  Soon after being awarded NASA\footnote{I have received NASA Graduate Student Researcher Project Fellowship support for the past two years.} support to perform a ``Multi-wavelength study of GRBs using BATSE, HETE II and Milagro," the Compton Gamma-Ray Observatory was deorbited and the launch of HETE II was delayed.  It became clear that the number of coincident observations between Milagro and the GRB satellites would be too small to be statistically significant, particularly in light of the increasing evidence that most GRBs were very distant and the associated attenuation of TeV signals.  

However, Milagro is a very sensitive wide field-of-view detector in its own right. Not only is the full Milagro telescope more sensitive than Milagrito, it has a significantly lower energy threshold.  The lower threshold is particularly important because of the reduced attenuation at $\sim$100 GeV, which dramatically increases the volume of space observed (see Figure \ref{JamesAtten}).  Milagro has a fluence sensitivity at TeV energies which is comparable to dedicated satellite GRB detectors at keV-MeV energies \cite{Andy:GRB}.  

This thesis was then re-envisioned to use Milagro as the trigger to identify TeV transient emission of 40 s to 3 hours duration.  Smith \cite*{Andy:GRB} was beginning to analyze the Milagro data for TeV transients of 250 $\mu$s to 40 s duration, and this thesis was designed to complement that effort.  In addition, as a part of this thesis a new search technique based on Gaussian weighting was developed to improve the sensitivity of Milagro when performing a real-time all-sky transient search (see Chapter \ref {WeightedAnalysisTechnique:chap}).  Notification of any transients observed by Milagro would be rapidly released to the optical and radio communities through the gamma-ray burst coordinate network (GCN), and follow up target-of-opportunity observations performed using the rapid x-ray timing explorer\footnote{RXTE proposal \#70135, cycle 7.} (RXTE).

One always hopes for detections, but partly because of the model dependence of TeV emission, either a detection or an upper limit can place important constraints on GRB progenitors and emission mechanisms.  This thesis was designed to fit into the broader monitoring goals of the Milagro experiment, and provides the most sensitive search yet performed of 40 s -- 3 hour duration TeV emission from gamma-ray bursts.

\chapter{The Milagro Gamma-Ray Observatory}
\label{Detector:chap}

\section{Introduction}
The Milagro Gamma-Ray Observatory is located at a decommissioned hydro-thermal plant high in the Jemez mountains of New Mexico.  Very high energy gamma rays incident on the earth pair produce in the upper atmosphere and produce extensive air showers (EAS) which propagate to lower altitudes.  Milagro uses the water Cherenkov technique to detect the EASs and reconstruct the directions of the initiating gamma rays.  Because Milagro is observing the EASs when they reach the ground instead of in the atmosphere, it can operate continuously and has an extremely wide field-of-view.  The continuous coverage and wide field-of-view make Milagro ideally suited for observing gamma-ray bursts and other very high energy gamma-ray transients.

This chapter outlines the hardware and software design of the Milagro telescope, and how the detection of individual EASs are turned into astronomical observations.  More detailed discussions of specific parts of the Milagro detector can be found in Sullivan \cite*{GregMilagroReview} and the detailed description of the Milagro prototype by Atkins et al. \cite*{gritoNIM}.

\section{Overview of Extensive Air Showers and the Milagro Detector}
\begin{figure}
\begin{center}
\includegraphics[width=5.75in]{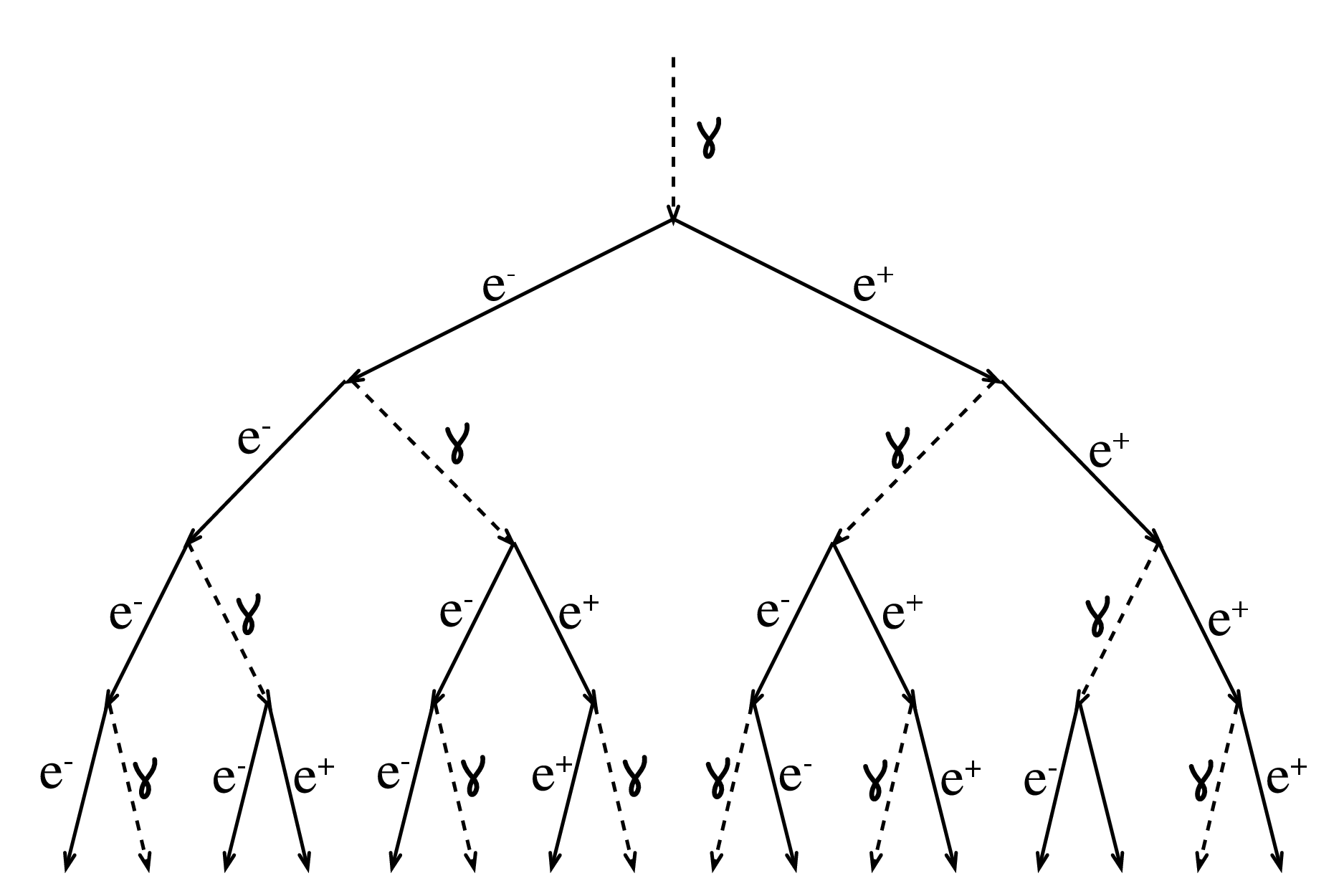}
\caption[EAS Cartoon]{A cartoon of the development of an extensive air shower, showing the particle generations and the iteration of bremsstrahlung and pair-production.}
\label{EMcascade}
\end{center}
\end{figure}

Before describing the Milagro observatory, a brief review of extensive air showers is in order.  In a typical EAS the initiating gamma ray interacts with an atomic nucleus in the atmosphere at an altitude of 10-20 km and pair produces to create a very energetic electron-positron pair.  The electron and positron then bremsstrahlung and deposit $\sim$$1/2$ their energy into secondary gamma rays which subsequently pair produce to continue the cycle.  This process of repeated pair production followed by bremsstrahlung forms an electromagnetic cascade, quickly diluting the energy of the primary gamma ray into a large number of relativistic electrons, positrons, and secondary gamma rays (see Figure \ref{EMcascade}).  Eventually the mean energy of the particles becomes low enough that ionization and excitation effects start to absorb electrons and positrons from the shower.  Shower maximum is when the number of shower particles reaches its highest value just before absorption effects start to dominate.  

The shape of the EAS can be qualitatively described as a very thin pancake of particles normal to the direction of the initiating gamma ray, with a narrow core of relatively energetic particles and an extended skirt of lower energy particles.  The actual size of the EAS depends strongly on the energy of the initiating particle and how many radiation lengths the shower has developed through, but values of a few meters diameter for the core and hundreds of meters for the skirt are typical for Milagro. I have created a number of Monte Carlo based animations of EAS development, and they can be viewed on the web \cite{AnimationWeb}. These movies of EAS are useful for developing a conceptual understanding of the shape and dynamics of extensive air showers.

\begin{figure}
\begin{center}
\includegraphics[width=5.75in]{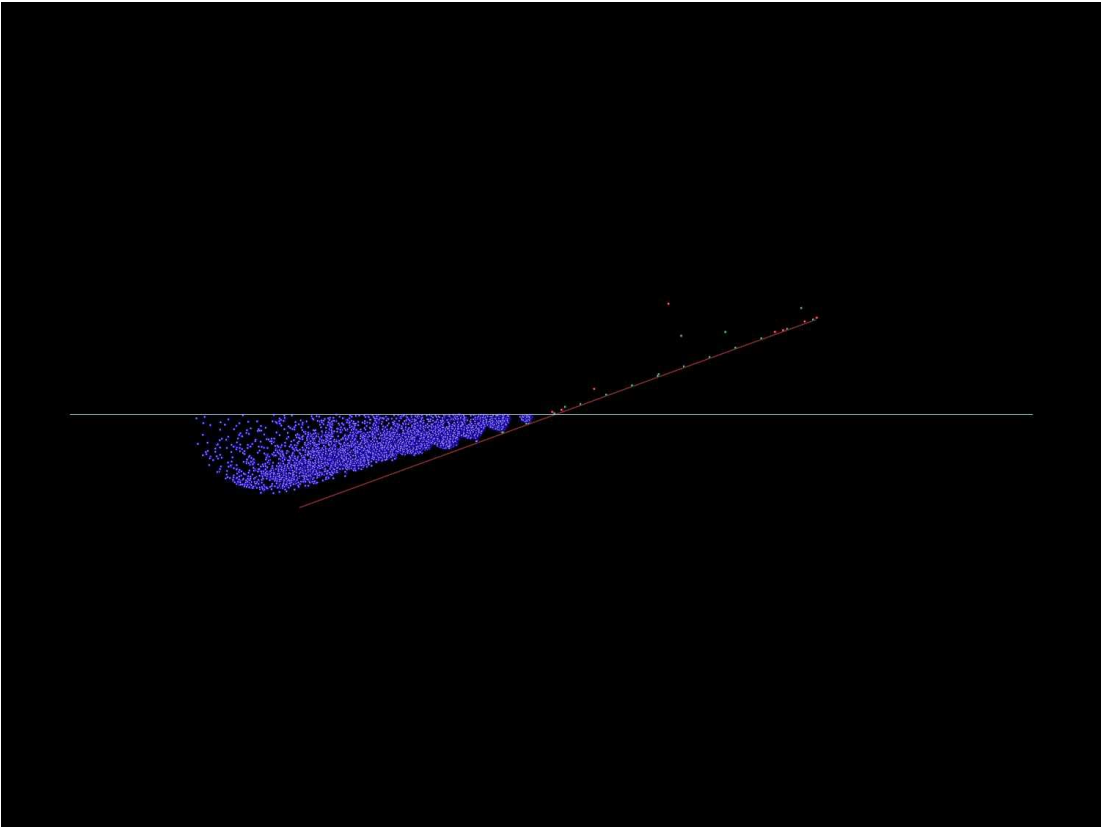}
\caption[Conversion of Toy EAS to Cherenkov Photons]{In this figure a toy model of an EAS consisting of 20 MeV gamma rays arranged on a plane is incident on the Milagro pond.  The white line indicates the water surface. Green is used to indicate a gamma-ray, red an electron or positron, and blue a Cherenkov photon, while the red line indicates the original plane of the EAS as it propagates into the detector.  Note that there is a conversion of the EAS particle front into a front of Cherenkov photons just below the detector surface, with the front of Cherenkov light showing significant broadening and refraction.  This image is from full motion animations that can be viewed at http://scipp.ucsc.edu/milagro/Animations/AnimationIntro.html}
\label{20dG20MeV10C1095}
\end{center}
\end{figure}

The Milagro observatory uses the water Cherenkov technique to identify EASs and reconstruct the direction of the initiating particle.  The central Milagro detector consists of a large reservoir of water instrumented with two layers of photomultiplier tubes, and covered by a light-tight cover.  When the pancake of relativistic particles from an EAS enters the water, the charged particles exceed the local speed of light and produce Cherenkov radiation in the blue and near ultra-violet. In essence, as the water absorbs the relativistic particles, the front of particles from the EAS is converted into a front of visible Cherenkov photons as shown in Figure \ref{20dG20MeV10C1095}, and it is this front of Cherenkov photons which is detected by the photomultiplier tubes.  The conversion of an EAS particle front into a Cherenkov light front is quite subtle and displays rich structure.  I explored this topic in some depth in  Morales \cite*{MMBowlRing} and the accompanying animations which are available on the web \cite{AnimationWeb}.

One of the great advantages of the water Cherenkov technique is the ability to detect both the neutral and charged components of the EAS.  Because Milagro is well past shower maximum for almost all showers, most of the particles in the EAS front are secondary gamma rays. Traditional air shower arrays have relied on detection techniques which are only sensitive the charged component of an EAS.\footnote{Lead or other conversion material is often used above the detectors in traditional air shower arrays to convert some of the gamma-rays to charged particles, but the conversion material also absorbs the charged component and so must be used sparingly.}  In Milagro the water acts as both a converter and the detection medium;  charged particles entering the water immediately radiate Cherenkov light, whereas the secondary gamma-rays produce an electron and a positron which both radiate Cherenkov light.  This allows Milagro to detect nearly all the particles incident on the detector, and significantly reduces the energy threshold of the experiment.

One of the major issues facing all air shower arrays is differentiating the gamma-ray events from the overwhelming background of cosmic-ray events.  EASs initiated by cosmic ray nuclei do have characteristics which can distinguish them from gamma-ray initiated EASs.  In particular, hadron initiated EASs contain a significant number of muons at ground level and have a clumpier shape due to the high transverse momentum of pions and other mesons produced in the hadronic cascade.  These characteristics can be used to partially remove the hadronic background.  

\section{Detector Layout}
The heart of the Milagro observatory is affectionately known as ``the pond" and consists of a 60 m by 80 m reservoir with sloping sides which is 8 m deep in the central region.  The reservoir is located at 2600 m altitude in the Jemez mountains of New Mexico and contains 23 million liters of water.  The floor of the reservoir is covered by a 2.8 m square grid of sand-filled PVC pipe which serves as the structural anchor for all of the equipment in the detector volume, and a flexible light-tight cover floats on the surface of the water to block atmospheric light and environmentally isolate the pond.  This light-tight cover can be inflated to allow access to the pond for routine maintenance using boats and SCUBA divers.  

The detector volume is instrumented with 723 photomultiplier tubes (PMTs) arranged in two layers.  The upper level of tubes is called the air shower layer and consists of 450 tubes 1.5 m below the surface of the water arranged on a square 2.8 m grid.  Signals from the air shower tubes are principally used for their excellent time resolution to reconstruct the orientation of an EAS and thus the direction of the initiating particle.  The lower layer of tubes is called the hadron layer, and consists of 273 tubes on an offset 2.8 m grid near the bottom of the reservoir.  Even though both layers share the same grid spacing, the lower layer is significantly smaller due to the sloping sides of the reservoir. Most of the tubes in the hadron layer are at a depth of 6 m, except the outer ring of tubes which are slightly higher again due to the sloping sides of the pond. The hadron layer is used primarily to identify the position of the shower core, identify penetrating muons, and perform calorimetry.  

The photomultipliers are 20 cm diameter hemispherical tubes made by Hamamatsu (model \#R5912SEL), and have excellent timing and charge resolution.  The base of each tube is encased in a water-tight PVC housing which is attached the electronics building by a single RG-59 coaxial cable which carries both the high voltage DC power and the AC PMT signal.  In addition there is a baffle around each tube made of anodized aluminum and black polypropylene which encircles the top of the photomultiplier tube much like a veterinary dog collar.  
The baffle forms an inverted and truncated cone which encircles the tube.
The reflective aluminum and black polypropylene are arranged in two layers so that the interior of the baffle is reflective to help funnel light into the tube while the outside is black to absorb stray light and keep the tube from seeing horizontal or upwards moving photons.\footnote{The baffles are a major difference in the design of Milagro from the prototype detector Milagrito. During the operation of Milagrito, it was discovered that there are a large number of muons which traverse the pond nearly horizontally.  These muons create upward or nearly horizontal light that can illuminate a large number of tubes in the pond and created a major background for our simple multiplicity trigger.  The baffles have solved these problems, though there have been some issues with long term corrosion.} The tube-base-baffle assembly is buoyant and kept in place by a Kevlar string which anchors it to the PVC grid.

\begin{figure}
\begin{center}
\includegraphics[width=5.75in]{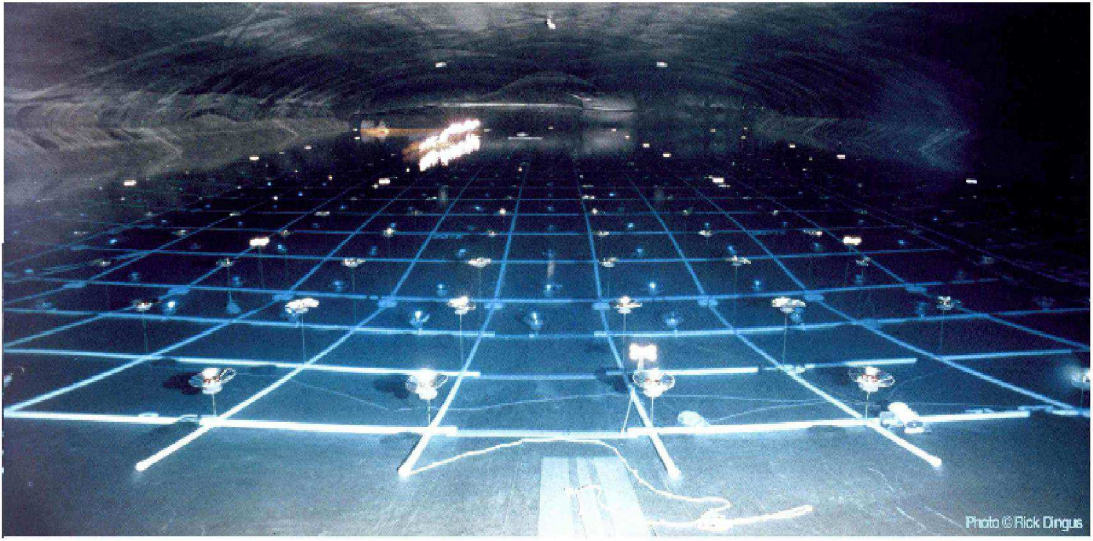}
\caption[Photograph of Milagro from Under the Cover]{This photograph was taken with the light tight cover inflated for maintenance.  The sloped sides and flat bottom of the reservoir can be seen as well as the PVC anchor system, the two layers of PMTs and their baffles. Photograph courtesy Rick Dingus.}
\label{milagro_in}
\end{center}
\end{figure}

Many of the physical features detailed in the previous paragraphs can be seen in Figure \ref{milagro_in} which shows the detector with the cover inflated for maintenance work.  
One of the features not visible in the photograph is the lightning protection system.  The Milagro observatory is located in one of the most active lighting regions in the nation, with the mean waiting time for a lighting strike within the 50,000 $\text{m}^2$ site being about 1 month \cite{gritoNIM}.\footnote{The idea of a $\sim$4,800 $\text{m}^2$ pool of water grounded by 723 high voltage cables directly to our custom front end electronics and the computer racks gave us some pause.}  In response, a large Faraday cage was erected around the central pond and support buildings using telephone poles and large diameter copper cable.  Despite several observations of nearby strikes and discharges from the lighting rods no damage has occurred within the Faraday cage.

\section{Electronic Layout}

The neighboring tubes within the pond are grouped into ``patches'' of sixteen photomultipliers.  The sixteen tubes are gain matched to require the same bias voltage and are driven by one high voltage supply channel. The first stage in the event processing chain is a pair of custom 16 channel boards which are mated back-to-back and process the signals from one patch.\footnote{The boards can be configured so that there are two patches of eight tubes, each with a different high voltage, and this is done in a couple of instances.}  After separating the AC signal from the power supply, these boards divide the signals and send them through two separate gain and discriminator chains which implement Milagro's dual time-over-threshold (TOT) system.  The number of photoelectrons (PEs) produced on the surface of the PMT can be estimated by measuring the total charge in one signal pulse.  This is done by storing the charge on a capacitor and measuring the voltage as the charge bleeds off through a resistor.  In high speed applications like Milagro equipping each channel with a separate analog to digital converter to integrate the voltage measurements and determine the total charge is expensive.  One common technique is to use a discriminator to measure the time over a set threshold --- the more charge in a pulse the longer the signal remains over threshold.  To improve the dynamic range and help differentiate between overlapping pulses Milagro uses a dual time over threshold system, with one discriminator set at $\sim$$1/4$ of a photoelectron and the other at $\sim$5 photoelectrons.

\begin{figure}
\begin{center}
\includegraphics[width=5.75in]{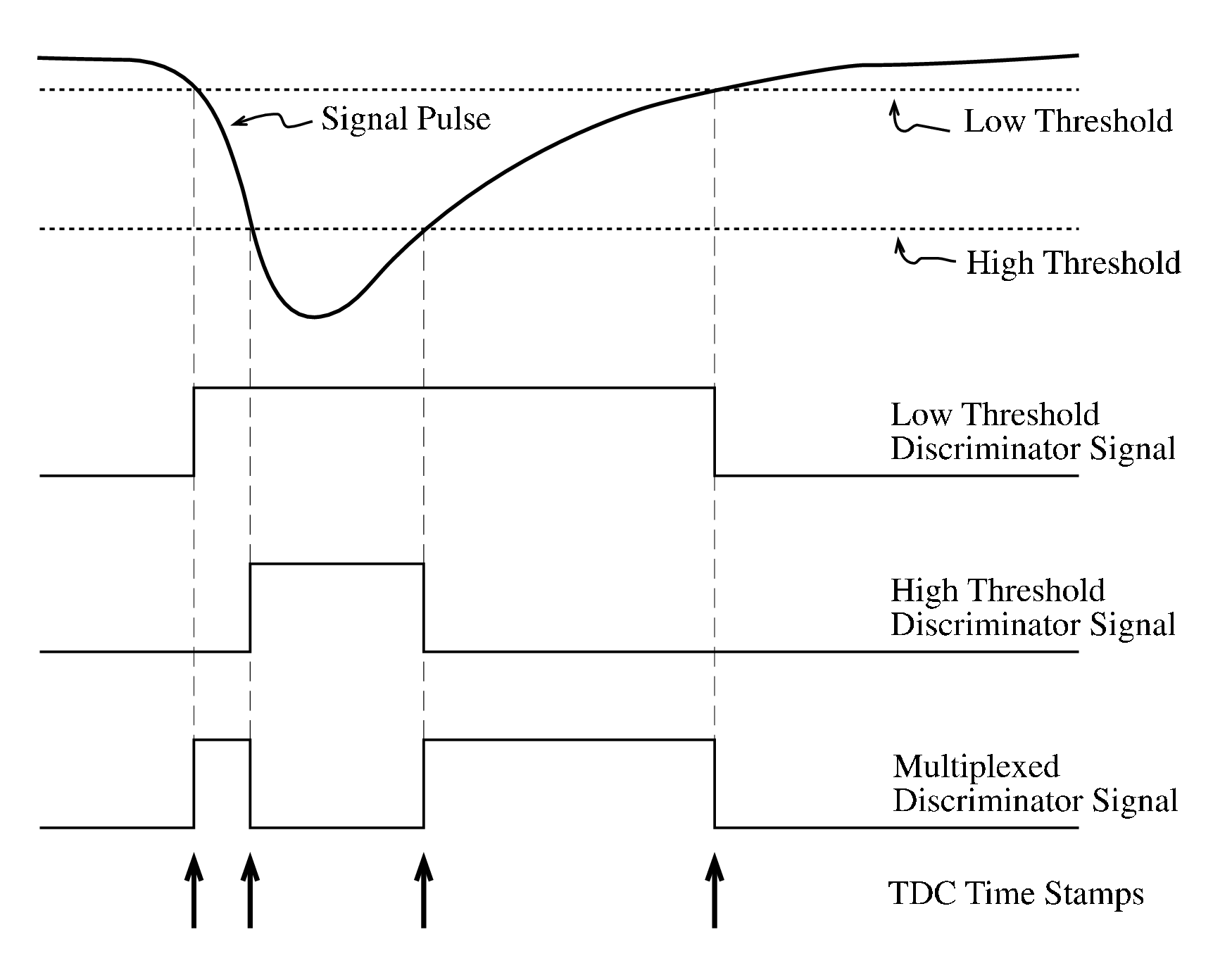}
\caption[Time-Over-Threshold System Diagram]{This diagram shows how the time and pulse height information are digitized in the dual threshold system.  For the diagram voltage is the vertical axis and time is the horizontal axis.  The pulse from a single PMT is shown at top with the dual voltage thresholds indicated.  The middle of the diagram then shows the digital voltage signals of the two discriminators which indicate when the signal pulse was above each threshold. The high threshold discriminator signal is then inverted and added to the low threshold to form the multiplexed signal. The TDC board then measures the time of each edge crossing and creates a digital time stamp.  Note that the separation of the TDC times indicates the size of the pulse while the offset of all the times indicates the arrival time of the pulse.}
\label{TOTDiagram}
\end{center}
\end{figure}

Both high and low threshold discriminators output a signal which is binary in voltage but analog in time, with one voltage for below threshold and another for above.  These signals are then multiplexed together to form a single digital voltage signal which is sent to the time to digital converter (TDC) boards.  The LeCroy 1887 FASTBUS TDCs create a digital time stamp for each edge crossing and can store up to sixteen edge crossings per each channel.  (See Figure \ref{TOTDiagram} for a graphical description of the dual TOT system.)

In addition to forming the multiplexed discriminator signal, the front end boards create a separate 200 ns digital voltage signal for each new PMT signal which crosses the low threshold.  These signals are added on the front end board and then summed across the boards with a fan in to create a single signal which is proportional to the number of tubes in the air shower layer which have triggered within a 200 ns window.  A single discriminator is then used to form a simple multiplicity trigger.  For the year of data used in this thesis the discriminator was set at $\sim$55 tubes.  This spring the trigger system was substantially upgraded to individually read in the tube hits and allow for more sophisticated software triggers \cite{MilagroTrigger}.  Because of the late change, the new very low threshold triggers introduced by the new trigger system were not used in this analysis.

When the trigger condition is met, a common stop is created and the edge time stamps are read out of the TDC boards by the FASTBUS Smart Crate Controller and transfered into a VME dual ported memory module via an Access Dynamics DM115/DC2 smart VME controller. The data is then transferred through a BIT-3 VME interface from the dual ported VME memory module into the data acquisition (DAQ) computer.  For the data used in this GRB search the DAQ computer was a 10 processor SGI Challenge mainframe which performed real-time reconstruction of triggered events.  Though state of the art when purchased, the Challenge has been eclipsed by the computational power of Linux based workstation clusters.  After the end of data taking for this thesis the Challenge was replaced by a clustered system whose computational power can scale to handle more sophisticated reconstruction algorithms.

\section{Reconstruction System}
\label{MilagroReconstruction}
The reconstruction system is a complicated custom software program that is responsible for reading the data from the VME dual ported memory, and performing calibration corrections and event fitting to determine the direction and species of the particle which initiated the EAS. Because of the enormous data rate --- $\sim$1800 triggers per second, or $\sim$5 MBytes per second of raw data --- we cannot afford to save all of the raw data.  The real time reconstruction is often the only opportunity to analyze the raw data and comprises a major piece of the data acquisition system.

The first task of the reconstruction program is to identify and characterize the PMT signals from the TOT stamps recorded by the TDC.  In particular, overlapping signals must be identified and handled correctly.  Once the PMT signals have been identified, the detector calibration is used to estimate the number of photoelectrons in the signal pulse and correct the arrival time for PE dependent risetime effects \cite{MilagroCalibration}.  This produces the calibrated data --- the tube by tube PE and arrival times fully corrected for all detector effects.

The next stage in the reconstruction system is locating the core of the shower.  The EAS particle front is slightly conical in shape and locating the center of the shower is crucial for accurately reconstructing the direction of the initiating particle.  The current core fitting routine uses the average location of the triggered PMTs weighted by the square root of their PE level to locate the shower core on the surface of the pond.  If the core is off the pond --- as is the case for $\sim$$70\%$ of the showers --- then the core fitter uses the average location to determine the direction of the core from the pond, and places the distance of the core at 50 m from the center of the pond.  The 50 m distance is used as a default lacking other information.  Core finding is one of the pieces of the event reconstruction that will be most affected by the addition of the outriggers (see Section \ref{FutureDirections} for more details).

Before determining the direction of the initiating particle, the core location is used to make two additional corrections to the data.  Because of our TOT system, the arrival time of a signal pulse is really the arrival of the first Cherenkov photon to reach the tube.  Near the core there are many more photons than in the extended skirt, and this leads to arrival times near the core being systematically early.  This bias is removed by a sampling correction, which is determined from the data and is a fifth order polynomial of the $\text{log}_{10}(\text{PE})$.  Additionally, while the shower front is very nearly flat, it does have a slightly conical shape.  A curvature correction adjusts for this geometric effect by subtracting 0.07 ns per meter from the core so that the arrival times form a plane. (See McCullough \& Gordo \cite*{MilagroAngleFitter} for complete details on the sampling and curvature corrections.)

The angle fitter uses the corrected times to iteratively fit a plane and reconstruct the direction of the particle which initiated the EAS.  A chi-squared fit is used with each arrival time given a weight determined by the number of PEs.\footnote{The weights are from the average PE dependent time residuals to a plane fit.}  The first iteration uses all hits greater than 2.25 PE to determine an approximate orientation of the shower.  Tubes with time residuals greater than $8.25\sigma$ are then assumed to be noise hits and are removed before refitting the plane with a loosened PE cut of 1.75.  On subsequent iterations the time residual cut is tightened to ($5.25\sigma$, $3\sigma$, $1.5\sigma$)\footnote{While developing the weighted analysis technique, I discovered that the weights used in the chi-squared fit were all high by a factor of $\sim$3.  This does not affect the fitting algorithm, but leads to the reduced chi-squared being low by a factor of $\sim$10, and the value of the actual significance cuts in the fitting routines being different from the numbers listed in the code by a factor of $\sim$3.  Here I have used the actual significance of the cuts, not the values listed in the code.} and the PE cut is loosened to (1.25,  0.75, 0.5).  The final fit determines the reconstructed direction of the initiating particle, and is the end product of the reconstruction system.  The performance of the Milagro reconstruction system is analyzed in detail in Chapter \ref{Characterization:chap}.

\section{Online Analysis}

The real-time reconstructed data produced by the DAQ computer is used for numerous analyses, which can be broadly grouped into ``online" and ``offline" analysis types.   The offline analyses are usually performed on computer clusters at one of the collaborating institutions, and typically involve steady source searches (such as the Crab pulsar, active galactic nuclei, and diffuse galactic emission) or archival analysis of old data.  In contrast, the online analyses are looking for transient signals in real time, with the goal of promptly alerting the astrophysical community to any observed sources.  Because these analyses are time critical, a small cluster of Linux computers has been installed at the Milagro site for the express purpose of performing online analyses.  

The DAQ computer writes the most recent block of reconstructed events, called a subrun, to disk every $\sim$4 minutes.  These subruns are then copied to the online computational cluster, and a link to each subrun file is placed into the incoming data folders of each online analysis.  It is the responsibility of an online analysis to process the links as they appear in the incoming data folders and erase the links once the data has been processed.\footnote{One of the reasons behind this file moving scheme is the very limited network bandwidth from the SGI Challenge computer.  Now that the SGI mainframe has been replaced with a modern Linux cluster we plan to move towards a network port system to eliminate the latency due to waiting for subruns to be completed.  The analysis used in this thesis has been written to easily incorporate reading data off a network port.}

Currently there are three separate online analyses operating at Milagro.  Smith \cite*{Andy:GRB} has developed an analysis which looks for TeV transients from 250 $\mu$s to 40 s duration; this thesis describes the search for transients of 40 s to 3 hours duration; and Elizabeth Hayes has a search which extends from 2 hours to steady state emission.  There are several reasons for the three separate analyses, with background rejection being the most important.  At the shortest timescales Milagro is data limited, and no background rejection should be performed, while at the long timescales Milagro is background dominated and the sensitivity is maximized with very aggressive background rejection.  

Because {\em a priori} we do not know the duration of a TeV transient,  all three online analyses search multiple timescales.  Biller \cite*{Biller:X3} has shown that logarithmically spaced search durations closer than a factor of 3 approach the sensitivity of using a search window with the actual duration of the transient, and allow the detection of transient signals of unknown length.  By searching on many different logarithmically spaced timescales, the three online analyses combine to form a sensitive search of the northern sky for any TeV signal of duration longer than 250 $\mu$s.

\section{Future Directions}
\label{FutureDirections}

\begin{figure}
\begin{center}
\includegraphics[width=5.75in]{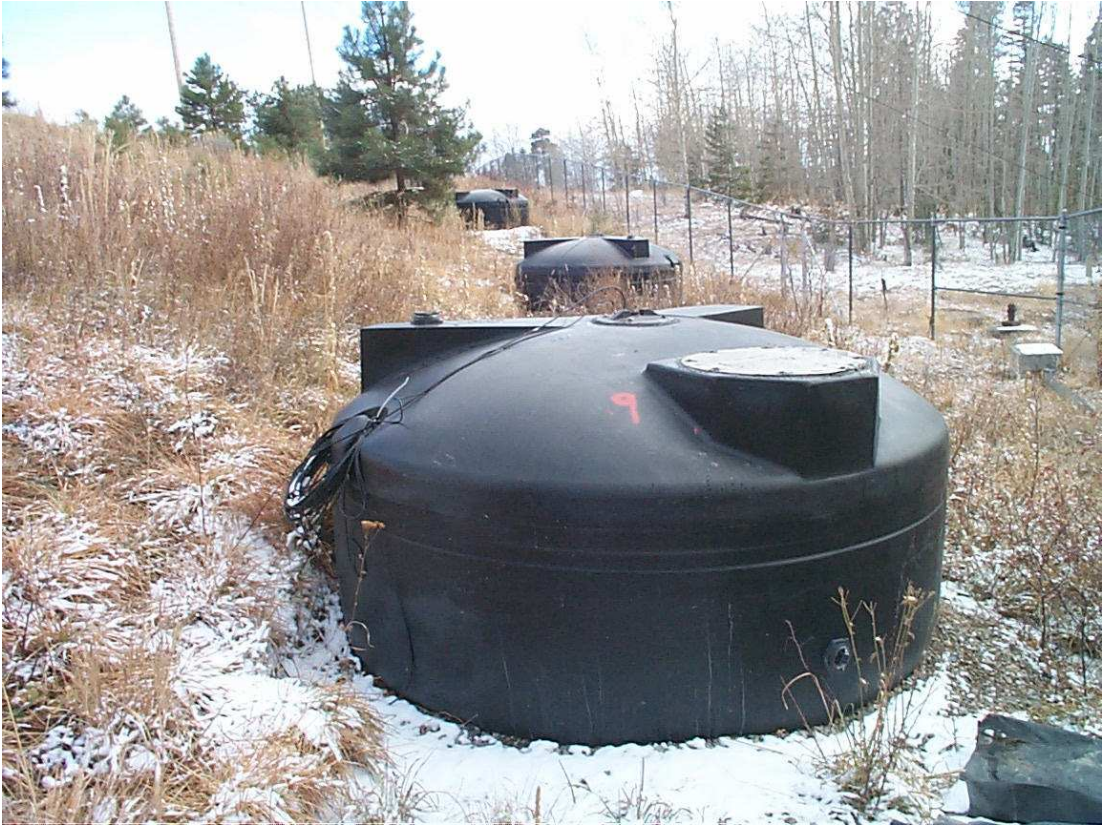}
\caption[Outrigger Photograph]{A row of three outriggers can be seen in the photograph deployed on the shoulder of the central reservoir.  When complete the outrigger deployment will continue into the woods visible in the background.}
\label{Outrigger}
\end{center}
\end{figure}

The full Milagro observatory is currently being completed with the construction of $\sim$170 small satellite detectors called ``outriggers."  The outriggers consist of cylindrical water cisterns 2.4 m in diameter and 1 m deep, which are lined with highly reflective $\text{Tyvek}^\text{TM}$ and instrumented with a single PMT (see Figure \ref{Outrigger}).  
The PMT and front end electronics chain are identical to the system used in the Milagro pond. 

The outriggers are scattered over a 40,000 $\text{m}^2$ area surrounding the Milagro pond, and are used to more fully sample the EAS particle front.  Because the effective area of the Milagro detector is significantly larger than the central pond for most of the energy range, the majority of the EAS cores do not strike the central detector.  Without good knowledge of the core location it is impossible to determine the energy of the EAS or properly correct for geometric effects across the face of the EAS particle front.  The outrigger system will allow Milagro to sample more of the EAS, improving the angular resolution and allowing event-by-event energy determination and the use of advanced background rejection techniques.

\chapter{Weighted Analysis Technique}
\label{WeightedAnalysisTechnique:chap}

\section{Introduction}

This chapter describes the theoretical basis for the weighted analysis technique, which was inspired by the unique requirements of performing a real time transient search in a high data rate gamma-ray observatory like Milagro. The basic requirements for a transient search analysis are to maximize signal sensitivity while keeping the computational cost low enough to enable a real time analysis on many time scales. Any analysis technique involves compromises between sensitivity, model independence, and speed, and I developed the weighted analysis technique as an alternative to the compromises made by the standard binned or maximum likelihood analyses \cite{CygnusTeq}. 

As is common in wide-field gamma-ray observatories, Milagro has a highly variable point spread function (PSF), with an order-of-magnitude difference in width from the best events to the worst. (Please see Chapter \ref{Characterization:chap} where Milagro's PSF is characterized.)  In an optimal binned analysis the sky is divided into equal area bins and the number of events observed in each bin is counted, with the size of the bins chosen to maximize the signal-to-noise ratio for the detector's average PSF.  In essence the binned analysis discards the information on the quality of an individual event, and instead treats every event as if it was drawn from the average PSF distribution. Maximum likelihood techniques, of which there are several approaches, can use the event-by-event PSF information and are the most sensitive methods for analyzing wide-field gamma-ray observations.  However, most implementations are computationally slow because they require fitting model parameters, and this fit usually requires an iterative fitting algorithm such as MINUIT \cite{MINUIT}.

A third analysis method is the Gaussian weighting technique, and has been used by the Fly's Eye and JANZOS experiments as a compromise between the optimal bin and maximum likelihood analyses.  In Woodham's thesis he describes a technique which uses the Gaussian PSFs of individual events to identify excesses, and assumes large statistics so that the central limit theorem can be used to determine the significances of the excesses \cite{Woodhams}. A similar method was used by the Fly's Eye group to analyze data from Cygnus X-3, but they only outlined the technique used and never published the full analysis method.  From the sketch of the analysis provided in Cassiday et al.\ \cite*{FlysEyeTeq} it is not obvious whether the PSF was Gaussian or arbitrary in shape, and they used an extensive Monte Carlo simulation to determine the significance of the excess.\footnote{There is a nice review of point search techniques in Alexandreas et al.\ \cite*{CygnusTeq}, where the authors claim that Gaussian weighting is as sensitive as maximum likelihood in the limit of large statistics and provides a substantial gain in computation time.  Unfortunately, while plausible these claims are not supported by their citations.}

The weighted analysis technique presented in this chapter is an extension of the Gaussian weighting technique as developed by Woodhams \cite*{Woodhams} to PSFs of arbitrary shape and Poisson statistics. To introduce the weighted analysis technique, I return to first principles and build a somewhat idealized sky map, then describe how this sky map can be used as the basis for an analysis.

\section{Building a Sky Map}
\label{WATSkymap}

Let's return to the basics, and imagine making a sky map. An idealized sky map would represent our complete knowledge of the sky --- using all of the available information and adding no spurious or biased information. The question of making a sky map becomes what do we know, and how do we represent that knowledge?  

Surprisingly, the question ``What do we know?" can be quite subtle. Do we use just the information we measure directly (event positions and characteristics), or do we go to the next step and include sources such as the Crab pulsar or characteristics such as an expected source spectrum in the definition of what we know? Including knowledge of sources leads to maximum likelihood analyses, with differences in how much information we include about the sources (spectra, angular extent, etc.) leading to different types of maximum likelihood analysis. Alternately, we could limit the definition of what we know to directly measured quantities --- event positions and characteristics. This approach delays questions of source identification until after the sky map is created. Both approaches are valid, but from different points of view. The weighted analysis technique uses the later viewpoint and includes only directly measured quantities.

The second question is how to represent our knowledge of the event positions and characteristics in a sky map? The PSF is defined as the normalized probability density distribution for the true event position given the measured event position:
\begin{equation}
\label{PSFdef}
PSF(\vec{k}_t-\vec{k}_{m}) = \partial P(\vec{k}_{t}|\vec{k}_{m})/ \partial \Omega,
\end{equation}
where $\vec{k}$ is a vector on the unit sphere and $\vec{k}_{m}$ is the measured location and $\vec{k}_{t}$ is the true location.\footnote{Since in astronomy we are only concerned with the direction of the initial photon, it is useful to represent this direction as a vector on the unit sphere.  The PSF can also be defined by $\partial P(\vec{k}_{m}|\vec{k}_{t})/ \partial \Omega$, which is identical to the form in Equation \ref{PSFdef} by Bayes formula if $P(\vec{k}_{t})$ is uniform on the scale of the PSF.  For studying analyses, the definition in Equation \ref{PSFdef} is more convenient because $\vec{k}_{m}$ is the observed quantity.} For gamma-ray telescopes, the width and shape of the PSF depends on the characteristics $\psi_i$ of the individual event, and may vary considerably from one event to the next. The PSF can also be multiplied by the probability $P_\gamma$ that the event was a photon at all (as opposed to background) to create a value that is the photon probability density for the $i^{\rm th}$ event's true position being at a position $\vec{k}$ on the sky and being a photon:
\begin{equation}
\label{probDensityEq}
p_i(\vec{k}) = p(\vec{k}|\vec{k}_i, \psi_i) = PSF(\vec{k}-\vec{k}_{i}, \psi_i)P_{\gamma}(\psi_i).
\end{equation}

\begin{figure}
\begin{center}
\includegraphics[height=6.75in]{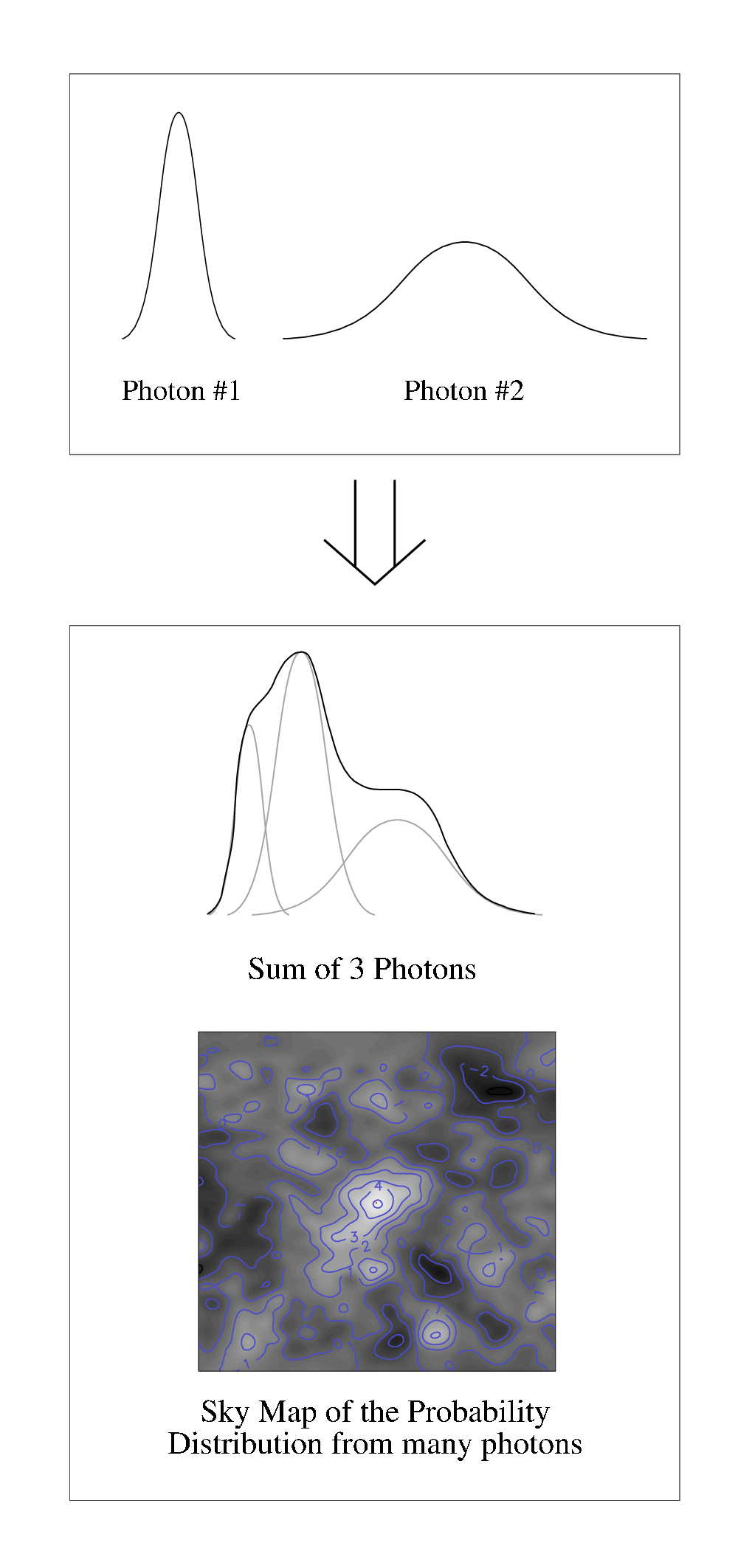}
\caption[PSF Summation Diagram]{This is diagram shows how individual photons are added to form a sky map in the weighted analysis technique.  In the top frame we have events with varying PSFs.  The lower frame shows how these events can be added together to form a probability density map, first as a 1-dimensional example of three events, then a more realistic 2-dimensional example incorporating many events.}
\label{PSFSummationDiagram}
\end{center}
\end{figure}

A sky map can be created by adding together the photon probability distributions of many events to form an overall map of the photon probability distribution (see Figure \ref{PSFSummationDiagram} for a graphical description).  The total photon probability density distribution $w$ is given by the sum of the individual photon probability densities:  
\begin{equation}
\label{weightSum:Eq}
w(\vec{k})=\sum_i^ {\text{all showers}} p_i(\vec{k})=\sum_i^{\text{all showers}}PSF(\vec{k}-\vec{k}_{i}, \psi_i)P_{\gamma}(\psi_i).
\end{equation}
There is some information loss in forming this sky map because the characteristics of the incoming showers cannot be uniquely determined from the sky map.  In essence the sky map represents total photon probability density, but we have lost the individual $p_i(\vec{k})$ values that make up the sum.\footnote{The $p_i(\vec{k})$ information can be retained if  there is a small set of PSFs and $P_\gamma$s instead of continuous distributions.  In this case a separate sky map for each combination of PSF and $P_\gamma$ can be created and this individual information retained.  Sky maps of this type were used in a maximum likelihood analysis by the EGRET collaboration \cite{EgretTeq} for PSFs which were binned in energy and a binary $P_\gamma$ cut. If the PSF or $P_\gamma$ distributions are continuous only the original list of event positions and characteristics retains all of the information, and any sky map is an approximation.} The sky map of the photon probability distribution uses all of the event-by-event knowledge, and represents our knowledge of the spatial distribution of events.

\begin{figure}
\begin{center}
\includegraphics[width=5.75in]{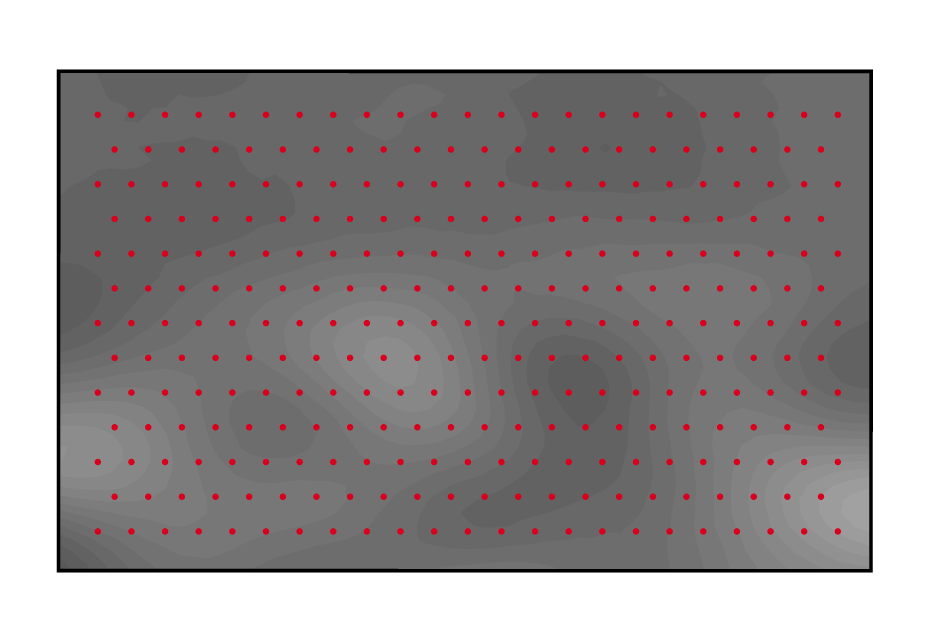}
\caption[Sky Map Sampling Diagram]{The shaded background represents the smooth variation of the probability density seen in an example sky map, with the red dots representing locations where the probability density has been sampled.  The sampling pattern need not be uniform, and as long as the sample spacing is small compared to the narrowest PSF, all the information in the continuous sky map is captured in the sampled map.}
\label{SkyMapSamplingDiagram}
\end{center}
\end{figure}

A continuous sky map created by adding the PSFs of individual photons represents a somewhat idealized map which is difficult to manipulate with a computer. The spatial scale for fluctuations across the sky map is set by the width of  the narrowest PSF. The sky map can be digitized by sampling the total photon probability density $w(\vec{k})$ at individual points on the map surface.  If the spacing of these samples is small compared to the narrowest PSF the information loss can be made arbitrarily small (Figure \ref{SkyMapSamplingDiagram}). There are several nice features of this digitized sky map. Because the value at a location on the sky map represents the photon probability density at that point and is not an integral over nearby locations (as is common in binned analyses), the spacing between the points need not be uniform and tiling problems associated with binning a spherical sky are avoided. Additionally, two sky maps which share a sampling pattern can be summed. An 80 second sky map can be formed by adding two sky maps of 40 seconds duration --- a significant computational advantage when hunting over multiple time scales.

In this discussion spectral information has been ignored, but can be added in a completely analogous manner. The key is determining the normalized energy probability density function $\partial P(E_t|E_m,\psi_i)/\partial E$ --- the one dimensional energy analog to the PSF --- for each event. This energy distribution can then be multiplied by $P_\gamma(\psi_i)$ to form the photon energy probability density $p_i(E)$ and added as a third independent axis of the sky map.  The resulting three dimensional sky map is harder to visualize, but again represents the total photon probability distribution.  Similarly, the probability density of any other parameter of interest (such as polarization) can be added to a multidimensional sky map to enrich the representation of the data.  In conclusion, we can form a digitized sky map that represents our direct knowledge by summing the photon probability density distributions of each event and digitizing the resulting map.

\section{Source Identification}
\label{WATSourceID}

Now that we have a sky map of the photon probability distribution we need to tackle the issue of source identification, and again the analysis approach depends on exactly what question is being asked. A maximum likelihood analysis could be directly applied to the photon probability distribution of the sky map. Maximum likelihood is a very flexible approach, allowing searches for sources of different types and characteristics, and forming the sky map first could lead to significant time savings when the number of photons exceeds the number of sampled map locations (very similar to binned maximum likelihood techniques). However, maximum likelihood techniques tend to be too slow for searching multiple time scales in real time. If we are developing a new technique, the question becomes ``What are we looking for, and what compromises are we willing to make?"

The weighted analysis technique was developed for discovery mode real-time GRB searches in the Milagro experiment. We expect signals to be transient point sources,\footnote{In a transient source the size of the object in light seconds must be smaller than the emission time to allow different parts of the object to be causally connected.  This requirement can be relaxed somewhat in relativistic outflows due to time dilation, but any signal with a duration less than a few hours which originates at galactic or cosmological distances will have an angular extent much smaller than the Milagro PSF, and appear to be a point source.} but the spectrum of a TeV transient is highly model dependent and uncertain. Furthermore, since we don't know the signal duration we need an analysis which is computationally fast enough to handle a real time search over multiple time scales. So we want a search for point sources which is fast, sensitive, and not strongly biased by spectral expectations. The compromise we are willing to make is that this is a discovery mode search:  we just need to identify sources, once sources are identified they can be analyzed at length using slower more precise methods.

Because we are performing a discovery mode search, the relevant statistic is the probability of the background producing the observed signal.  Since we expect point sources, we can look at each of the sampled locations independently and ask ``What is the probability of the background producing the observed photon probability density?" Mathematically, we need to determine the probability that the background could produce a probability greater than or equal to the observed photon probability density  ($w_{obs}$) given the spectrum of probability densities observed in the background ($g(w)$),
\begin{equation}
\label{baseProb}
P(w\ge w_{obs}|g(w,N,\vec{k})).
\end{equation}
Note that in general the probability density spectrum is dependent on both the number of events ($N$) and the position in the sky ($\vec{k}$).  For small $N$, the probability density spectrum is typically very skewed, with many small values from the tails of the PSF and $\text{P}_\gamma$ distributions and only a few large values. However, as $N$ becomes large the central limit theorem comes into play and the probability density spectrum becomes Gaussian-distributed around the mean.  The probability density spectrum can also be spatially dependent if the distribution of event characteristics changes with position in the sky (if the distribution of $\psi_i$ in Equation \ref{weightSum:Eq} depends on $\vec{k}$).

Equation \ref{baseProb} represents the full probability of the background producing an observed probability density if the PSFs used to generate the sky map --- called the weighting functions $PSF'$ --- cover the entire sky and $N$ is deterministic.  In implementing the weighted analysis technique, the weighting functions are often truncated at some angular distance (see Table \ref{Cutoff:table}).  Truncating the weighting function significantly improves the computational speed of the analysis because only sample locations near the position of a new event must be updated, not the entire sky map.  The cost of truncating the weighting functions is a further complication of the statistics.  When the weighting functions are truncated, the number of events summed to form the photon probability density will vary from location to location.  Furthermore, the number of events summed at a location on the sky map ($N_{obs}$) experiences Poisson fluctuations around the expected background value $N_{exp}$, which adds a second observable to the probability.  Determining the probability of the background producing an event which is equal or more signal-like than the current observation is subtle when there are two or more independent variables ($w_{obs}$ and $N_{obs}$ in this case), but can be approximated by\footnote{For a complete explanation of the relevant statistics and this approximation, please see Appendix \ref{ProbAppendix}.} 
\begin{equation}
\label{CorrectProbEq}
\sum_{N=N_{obs}}^\infty P(w\ge w_{obs}|g(w,N,\vec{k}))P(N|N_{exp}).
\end{equation}
If the background probability density spectrum $g(w)$ changes slowly with $N$ ($\partial P(w\ge w_{obs}|g(w,N,\vec{k}))/\partial N \ll \partial P(N|N_{exp})/\partial N$) then equation \ref{CorrectProbEq} can be further approximated by: 
\begin{equation}
\label{Prob:Eq}
P(w\ge w_{obs}|g(w,N_{obs},\vec{k}))P(N\ge N_{obs}|N_{exp}).
\end{equation}
The first term in Equation \ref{Prob:Eq} is just the probability of the observed probability density being produced by the background, and can be easily measured in a background dominated experiment like Milagro. The second term is the Poisson probability of seeing an observed number of events given a background expectation, and is only important if the weighting functions used have a finite extent.  Qualitatively the two terms serve distinct purposes. The first term depends on how gamma-like the events are from the $P_{\gamma}$ values and how clumped the events are from the PSF values (see Equation \ref{weightSum:Eq}), and asks how likely is it that the background could have produced the observed probability density. The second term is looking for a simple excess of events, and becomes important for Milagro only if the PSF is truncated at a given angular distance so that there is an effective bin size for each PSF. Typically $g(w)$ varies slowly with $N$ and a small truncation distance can introduce correlation between the probability terms in Equation \ref{Prob:Eq} which must be accounted for (see Section \ref{WATExamples}).

The probability of the background producing an observation as given by Equation \ref{Prob:Eq} can be used to identify signals in the data.  The typical probability threshold for a source discovery is set at $\sim$5$\sigma$, or a probability less than $2.8\times10^{-5}$.  For a single search location this is exactly the threshold that would be used.  However, in a GRB search we are looking at hundreds of millions of independent locations and multiple time scales, and probabilities less than $2.8\times10^{-5}$ are quite common.  By definition, if a 1000 independent locations are analyzed, $\sim$1 will have a probability $\le1/1000$ and $\sim$10 will have a probability  $\le 1/100$.  A log-log plot of the probability histogram has a characteristic slope of -1 if there are no sources, and this can be used as a diagnostic to ensure that the distributions used to calculate the probability are correct.  For a search involving many locations, the source discovery threshold is set at ``$5\sigma$" below the probability where one background event is expected (for 1000 locations $10^{-3}\times[2.8\times10^{-5}] = 2.8\times10^{-8}$).  For a real-time search, the approximate number of independent locations that will be analyzed is determined in advance, and used to set the probability threshold for the discovery of transient TeV emission.

Signal identification in the weighted analysis technique has several nice features. First, $P(w \ge w_{obs}|g(w,N_{obs},\vec{k}))$ can be stored in computer lookup tables. Because no parameter fitting is required to determine the probability, calculating the significance of a signal is very fast. Second, because we are simply looking for something which does not look like the background, we are less model sensitive than some maximum likelihood implementations. However, the weighted analysis method does sacrifice some features like estimating the observed spectrum in favor of speed. A source identified with this method would need to be reanalyzed with maximum likelihood to obtain all the information from a signal.  

\section{Sensitivity}
\label{WATSensitivity}

We would like to compare the sensitivity of the weighted analysis technique to maximum likelihood and binned analyses. Unfortunately, there is no simple analytic way of comparing the various maximum likelihood analyses to either the optimal binned analysis or the weighted analysis technique. Qualitatively we expect the weighted analysis to do well because it is using all available information, but it is safe to say that it only approaches the sensitivity of a well implemented maximum likelihood analysis. That being said, many common maximum likelihood implementations require a model signal, and their sensitivity can be significantly impacted by mistakes in the original model.\footnote{The oral physics tradition maintains that maximum likelihood is superior to all other techniques.  In the literature there are counterexamples which show cases where maximum likelihood is not ideal \cite{Eadie}, though it may be due to how the maximum likelihood analysis was implemented.  I was not able to find citations proving the superiority of maximum likelihood.}

Comparing the weighted analysis technique to an optimal binned analysis also requires a model signal and a Monte Carlo simulation in most cases. However, we can illustrate the key differences using a few toy models with analytic solutions. In the limit of large statistics, we can obtain analytic solutions for both the binned and weighted analysis techniques for a detector with a single Gaussian PSF, and a detector with events drawn from two Gaussian PSFs of different widths. The limit of large statistics allows us to use the central limit theorem to calculate the variance of $w_{obs}$, and the combination of the large statistics limit, no weighting for background rejection, and Gaussian PSFs reduces the weighted analysis technique to the Gaussian weighting analysis developed by Woodhams \cite*{Woodhams}. Since the Gaussian weighting and weighted analysis techniques are identical in this limit, we will refer to them generically as weighted analyses in the following discussion.

For these toy models, $N$ is the number of signal photons and $b$ is the number of background events per square degree. After background subtraction, the significance of the signal is given by the signal/noise ratio and has the form $AN/\sqrt{b}$, where $A$ characterizes the sensitivity of the search and is the object of the following calculations. Because the PSFs in these models are symmetric, $\vec{k}-\vec{k}_i$ depends only on the angular separation between the source location ($\vec{k}$) and the reconstructed event location ($\vec{k}_i$).  To simplify the equations, $r$ is used to denote the angular separation in degrees between the source and reconstructed positions.

For a single Gaussian, the signal observed in a circular bin of radius $R$ is given by the integral of the PSF 
\begin{equation}
\label{ }
Signal=\int_0^R\frac{N}{2\pi\sigma^2}e^{-r^2/2\sigma^2}2\pi rdr = N(1-e^{-R^2/2\sigma^2}),
\end{equation}
and the noise is given by the square root of the number of background events $\sqrt{b\pi R^2}$.  This ratio is maximized for $R=1.585\sigma$, giving a sensitivity parameter A of $0.255/\sigma$ for a Gaussian PSF. 

For the weighted analyses, the signal from a point source with $N$ photons is the probability distribution of the photon positions (the true point spread function $PSF$) times the weight given to each photon (the weighting function $PSF'$):
\begin{equation}
\label{ }
Signal = \int_0^{\infty} N\ PSF\ PSF'\ 2\pi rdr.
\end{equation}
 Since the weighting function and the PSF are the same Gaussian function in this example, the integral becomes 
\begin{equation}
\label{ }
Signal = \int_0^{\infty} N\Bigl[\frac{1}{2\pi \sigma^2}e^{-r^2/2\sigma^2}\Bigr]^2 2\pi rdr = \frac{N}{4\pi\sigma^2}.
\end{equation}
The noise is given by the square root of the variance of the probability density.  In the limit of large statistics, the variance is given by integrating the flat background distribution by the square of the weighting function:
\begin{equation}
\label{ }
Noise = \Biggl[ \int_0^{\infty} b\ PSF'^2\ 2\pi rdr \Biggr]^{1/2}.
\end{equation}
Since the weighting function is the same Gaussian as the true PSF, this is the same integral as used  for the signal with $b$ replacing $N$. The signal to noise ratio becomes $\frac{N}{\sqrt{4\pi\sigma^2}\sqrt{b}}$, giving a sensitivity parameter A of $0.282/\sigma$.  This implies that the weighted analyses are $\sim10\%$ more sensitive than an optimal bin analysis.  Woodhams \cite*{Woodhams} argued that this 10\% improvement should be a lower limit, and that detectors which have a spectrum of PSFs should benefit even more from a weighted analysis. 

The next toy model has two Gaussian PSFs, with 25\% of the events coming from a PSF of width 0.33$\sigma$, and 75\% from a PSF of width 1$\sigma$. Following the previous calculation, the optimal bin size is $0.764\sigma$ and the sensitivity parameter is $0.312/\sigma$ for the optimal binned analysis.  For the weighted analyses the sensitivity parameter is $0.489/\sigma$, or a $\sim 56\%$ improvement in sensitivity over the binned analysis. This is the kind of improvement we expected from a weighted analysis technique. 

However, the improvement depends very much on the spectrum of PSFs, and in special circumstances the improvement can be zero. To show that the 10\% improvement from a single Gaussian PSF is not a lower limit, consider a spectrum of PSFs given by $g(\sigma)$.  The general problem of finding the signal in a round bin becomes
\begin{equation}
\label{ }
N\iint_0^R PSF(\sigma,r)g(\sigma)2\pi r \, dr \, d\sigma,
\end{equation}
and the signal in a weighted analysis becomes 
\begin{equation}
\label{ }
N\iint PSF^2(\sigma,r) g(\sigma)2\pi r \, dr \, d\sigma.
\end{equation}
For a flat spectrum of Gaussian PSFs from width 0.1$\sigma$ to width 1$\sigma$, the weighted analysis gives less than a 7\% improvement over the binned analysis despite the wide range of PSFs used. In retrospect, this can be explained by reversing the order of integration. By integrating the spectrum of PSFs first (over $d\sigma$), a composite PSF can be obtained which has a distinctly non-Gaussian profile. By choosing the appropriate PSF and spectrum, a composite PSF with a top-hat profile could be generated, and in this extreme case the optimal binned analysis would be just as effective as the weighted analyses. This can be seen by realizing that the weighted analysis technique with a top-hat weighting function
\begin{equation}
\label{tophat:Eq}
 \Theta(\overrightarrow{\triangle k}) = \begin{cases}
      a & \overrightarrow{\triangle k}<R \\
      0 & \overrightarrow{\triangle k}\ge R\ ,
\end{cases}
\end{equation}
is identical to a binned analysis.  In Equation \ref{tophat:Eq} $R$ is the size of the bin and $a$ is a constant. Returning to the probability of a background fluctuation producing the observed signal as defined in Equation \ref{CorrectProbEq}, the top-hat weighting function leads to $g(w)=\delta(aN_{obs})$ since all the events with a non-zero probability density have a probability density of $a$.  Consequently, the total observed probability density $w_{obs}$ is deterministic and the first probability term is always equal to 1.  The total probability of the background producing the observed signal is solely determined by the second term which is simply the Poisson probability of seeing $N_{obs}$ events inside a bin of radius $R$ --- exactly the same result as a binned analysis.  It can also be shown that the optimal weighting function to use in the weighted analysis technique is the true PSF \cite{Woodhams}. Since the optimal weighting function is the true PSF, and the weighted analysis with a top-hat weighting function is identical to a binned analysis, it follows that the sensitivity of a weighted analysis is never worse than a binned analysis, and would only be equal for a detector with a top-hat composite PSF.  In general, the less square the composite PSF is, the more effective a weighted analysis will be.

One final topic we can explore with simple examples is model sensitivity.  Returning to the example with two Gaussian PSFs, we can compare the sensitivity of both analyses to signals where all the signal events come from either the narrow or wide PSFs while the expectation is still for a 25\% -- 75\% division between the PSFs. In these examples, the bin size or background distributions will be wrong, and we can explore how errors in the expected PSF affect the sensitivity of the analysis.  If the PSF of the signal is $0.33\sigma$ (all narrow PSF events), the weighted analyses are more than twice as sensitive as the binned analysis (114\% improvement).  At the opposite extreme, if the PSF of the signal is $1\sigma$ (all wide PSF events), then the binned analysis is nearly 13\% more effective than the weighted analyses.  This surprising result is because the weighted analysis techniques only use information from a single position on the sky map to identify excesses. An excess in photon probability at one location can be produced by either a few high quality photons with narrow PSFs, or a larger number of poor quality photons with wide PSFs.  The weighted analyses determine the significance of a signal by looking at only one position on the sky map and implicitly assuming that the excess has the expected spatial distribution. Another way of looking at this is that the power of the weighed analysis techniques comes from weighting the events with the expected PSF.  However, if the expected PSF is wrong, there can be times when the expected optimal bin/top-hat PSF from a binned analysis happens to be more accurate than the expected PSF.  This shows that there is some model dependence in the weighted analysis technique which can be detrimental in certain specific scenarios.

In the preceding examples the $P_\gamma$ term from Equation \ref{probDensityEq} has been assumed to be one.  This is equivalent to a hard background cut which treats all events passing the cut identically ($P_\gamma = 0$ or 1).  The weighted analysis technique can use an analog $P_\gamma$ value instead of a hard cut, and this will magnify the sensitivity advantage of the weighted analysis technique over a binned analysis. This can be seen by observing that a background cut is equivalent to a 1-dimensional bin in the cut parameter, and the same argument which showed that the sensitivity of the weighted analysis technique is greater than or equal to that of the binned analysis applies (if the correct PSF and $P_\gamma$ distributions are used).  In effect, background rejection adds a third dimension to the analysis, and an optimal binned analysis with a background cut uses a step-like probability distribution in all three dimensions, whereas the weighted analysis uses the expected probability distributions.

In general, the sensitivity of two analyses can only be compared using Monte Carlo simulation and an expected signal. There are a number of subtleties which have been masked by the simplicity of these examples, including the effect of fluctuations (on all parameters) in the limit of low statistics. For GRB searches, the limit of large statistics does not hold and the similarity between Gaussian weighting and the weighted analysis technique is broken. Gaussian weighting as developed by Woodhams \cite*{Woodhams} can only be used in the limit of large statistics, and the weighted analysis technique can be seen as an extension of Gaussian weighting to arbitrary PSF and the regime of Poisson statistics. Alexandreas et al.\ \cite*{CygnusTeq} performed a Monte Carlo simulation to compare the sensitivity of  maximum likelihood and optimal binned analyses, and for the simple case of a single Gaussian PSF (see the first example in this section) they also observed a $\sim10\%$ improvement with maximum likelihood.  This implies that the weighted analysis technique is similar to the sensitivity of maximum likelihood in this limit. The weighted analysis technique is more sensitive than the binned analysis for much but not all of the possible phase space, and should approach the sensitivity of well implemented maximum likelihood searches for at least some of the phase space.

\section{Summary}

The weighted analysis technique fits a particular analysis niche. For discovery mode GRB searches with variable PSF instruments, we want an analysis which is fast and uses all available information. Binned analyses are very fast, but sacrifice sensitivity by ignoring the variable PSF typical of wide-field gamma-ray telescopes. Maximum likelihood techniques use all available information, and can give valuable information like the estimated spectrum, but are computationally slow.  The weighted analysis technique is a compromise between binned and maximum likelihood techniques, landing somewhere in the middle on computational speed, but like advanced likelihood techniques uses the event-by-event PSF and $P_\gamma$ information for source identification.

\chapter{Characterizing the Milagro Detector Response}
\label{Characterization:chap}

\section{Introduction}

The sensitivity of the weighted analysis and maximum likelihood techniques depends on how accurately the PSF of an individual event can be characterized (see Section \ref{WATSensitivity} and Appendix \ref{appendixA}).  Though there is some sensitivity to be gained from determining the cumulative PSF of all events, this is not optimal because the PSF varies from one event to the next.  In sophisticated analyses such as the weighted analysis and maximum likelihood techniques, the more event-by-event information that can be supplied the more sensitive the analysis becomes, particularly in the Poisson limit when the shape of a signal PSF will display significant statistical variations.  In this chapter I characterize the shape and variations in the PSF and $P_\gamma$ distributions seen in the Milagro observatory. This characterization could be used in implementations of either the weighted analysis technique or maximum likelihood, but was made with an eye towards the weighted analysis implementation detailed in Chapter \ref{ImplementingWAT:chap}.

\section{The Problem}

While the cumulative PSF can be easily determined from either Monte Carlo simulations or data, characterizing the event-by-event variations in the PSF is much more challenging. In particular, how do we know if we have an optimal characterization of the event-by-event variations?  For example, there may be an observable (such as Cherenkov photon distribution in the bottom tube layer) which would allow us to pick out a class of events with a very narrow PSF or sharp $P_\gamma$ distribution.  By identifying this subset of events, we would have improved the sensitivity of our analysis.  But how do we go about formulating an optimal (or at at least good) characterization of the event-by-event variations without knowing which observables are important? 

Obtaining an optimal detector characterization is particularly difficult for the Milagro observatory because the detector response is driven by characteristics of the shower which are not observed.  In a satellite gamma-ray telescope such as GLAST, the variations in the PSF are largely determined by the energy of the incoming photon and the position in the detector of the first interaction, both of which are well measured.  Similarly, the PSF variations for Milagro are dominated by the photon energy and position of first interaction in the atmosphere (both height and distance of the core from the detector), but in Milagro's case neither of these shower characteristics are well determined.  

The conclusion is that the characterization of Milagro will change as the detector is upgraded (see Section \ref{FutureDirections}) and our understanding of detector response characteristics improves. If we could observe the physical characteristics of a shower which drive the detector response, we could formulate a near optimal characterization of the detector. Barring this, we are limited to trying to determine the best characterization that we can from our current understanding of the detector.  This characterization will not be optimal and will evolve as our understanding of the Milagro response characteristics improves.

\section{Identifying Regions}
\label{RegionIdentification}

In general the PSF shape should be a continuous function of the event characteristics.  However, the implementation of the weighted analysis technique for which we are developing this characterization uses PSF lookup tables to improve computational speed (see Section \ref{Skymap}). Thus we need to group events into a small number ``regions," with a composite PSF for all the events in the same group.  To be effective, all the events in a given region should be very similar to minimize the difference between the discrete regions and a continuous characterization of the PSF shapes.

In determining the PSF regions, I decided to rely upon the data and not the Monte Carlo simulations.  Using the data has two advantages.  First, because we have a large quantity of Milagro data, statistics are not a problem as is often the case with Monte Carlo simulations.  Second, there have been concerns about the accuracy of the Milagro simulations --- particularly the early versions that existed when this work was started --- and data-based region finding is not biased by problems in the Monte Carlo simulations.  The disadvantage of using data is that almost all of the Milagro data consists of background proton-initiated showers, which may not have the same characteristics as the gamma-initiated showers that we are really interested in.  However, the uncertainties in the Monte Carlo simulation outweighed the potential benefits during the region finding phase.  Subsequently, the gamma-based Monte Carlo simulation was compared the to the data to ensure that the regions found by analyzing proton-dominated data samples form a sensible way of grouping gamma-ray initiated showers.

To identify groups of similar events, I looked through the data in search of patterns in the PSF.  The measure I chose to characterize the PSF is called deleo/2.  In this measure, a shower is split into two separate events, one with the even numbered tubes, and another with the odd numbered tubes.\footnote{In practice, the even/odd division explains the conceptual process and not the actual computer code.  The actual routine divides events into two even/odd groups in a checkerboard pattern based on position rather than the channel number.} These two events are then separately fit by the Milagro reconstruction algorithms, and the angular difference between the two fits is called deleo.  For normally distributed errors with no systematic effects, deleo/2 is approximately equal to the PSF.  Neither of these assumptions are true for Milagro, but they are close enough that this is still a useful measure.

The best observables for capturing the variations in the PSF came from the angle fitting routine, and are the reduced chi-squared of the fit, and a variable called nFit.  In the shower fitting routine currently in use, there is an iterative fitting process. After an initial fit, tubes with times far from the initial fit are removed from the event.  The shower is then iteratively refit, with the criteria for including a tube being gradually adjusted.  The number of tubes used in the final fit is called nFit, and serves as a measure of how many good tubes were available to the shower fitting routine.  The shower fitting routine works by performing a chi-squared fit to a plane, and as such assumes symmetric errors, where the errors are related to the number of photoelectrons observed by the tube (see Section \ref{MilagroReconstruction} for a full description).  It was discovered during this analysis that the errors given to the fitting routine are artificially high by approximately a factor of three.  This does not affect the fitting in any way (chi-squared is still minimized), but the reduced chi-squared values are an order of magnitude lower than expected.  Since the PSF of Milagro is symmetric, $\vec{k}-\vec{k}_i$ in Equation \ref{probDensityEq} depends only on the angular separation of the vectors.  To aid the readability of the following discussion $\vec{k}-\vec{k}_i$ is replaced by the variable $r$, where $r$ is the angular separation of $\vec{k}-\vec{k}_i$ in degrees.

\begin{figure}
\begin{center}
\includegraphics[width=4in]{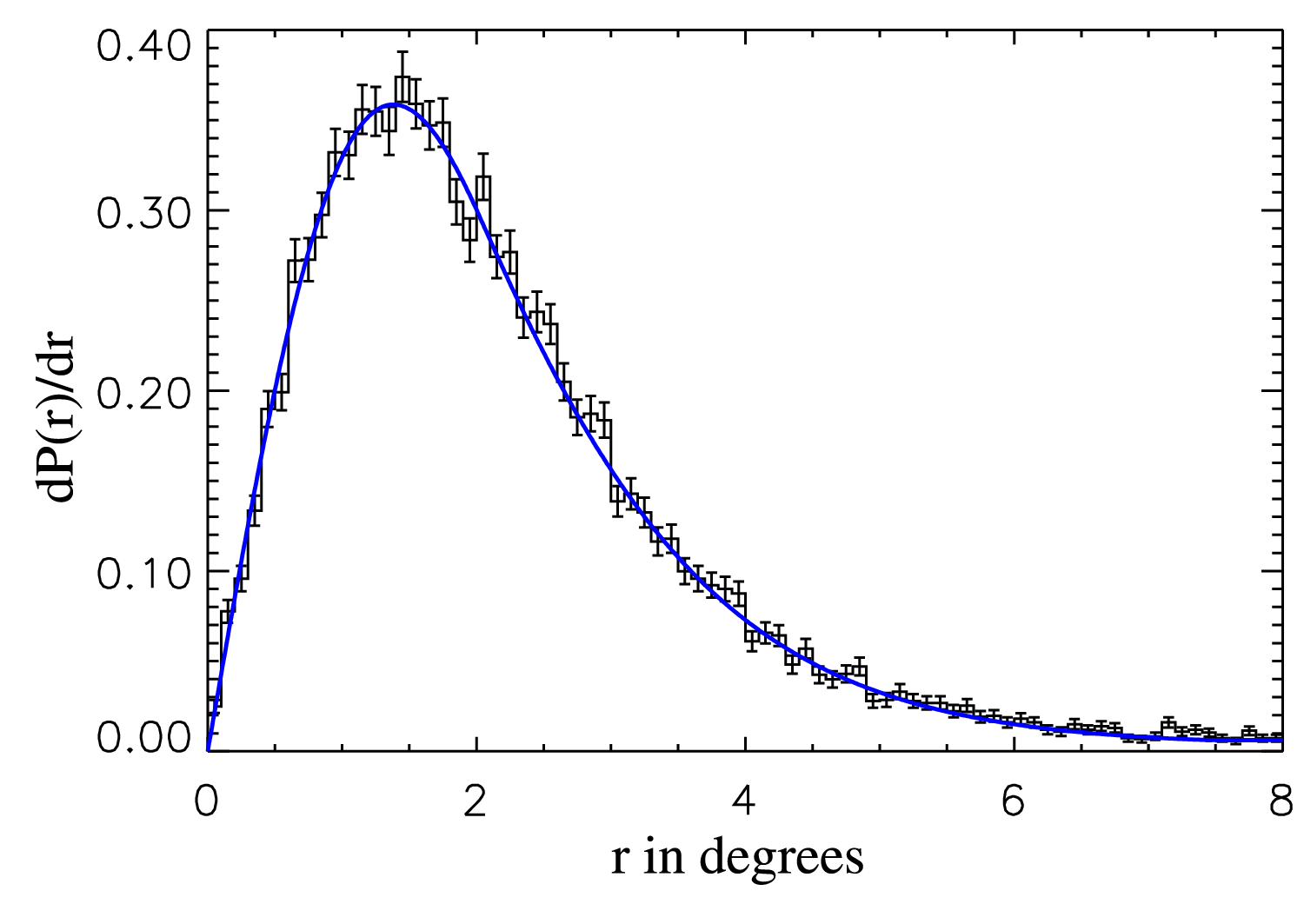}
\caption[Example PSF Fit for a Group of Similar Events]{An example fit to $\partial P(r)/ \partial r$ for a group of similar events.}
\label{ExamplePSFFit}
\end{center}
\end{figure}

The deleo/2 distributions for any group of similar events had a characteristic shape that was well described by the following function: 
\begin{equation}
PSF(r) = \frac{dP(r)}{d\Omega}= \left\{ \begin{array}{ll}
	\frac{a_0}{2\pi}e^{-r^2/2a_1^{\,2}} + \frac{a_3^{\,2}}{2\pi} & r<a_2a_1^{\,2}   \\
	\frac{a_0}{2\pi}e^{a_2^{\,2}a_1^{\,2}/2}e^{-a_2r} + \frac{a_3^{\,2}}{2\pi} & r \geq a_2a_1^{\,2}
	\end{array} \right.
\label{PSFfit:Eq}
\end{equation}
Qualitatively this function is a Gaussian which is smoothly joined to an exponential at $r=a_2a_1^{\,2}$, with an additional ``random" component ($\frac{a_3^{\,2}}{2\pi}$) for the small subset of showers the reconstruction had difficulty fitting. An example of the fit can be seen in Figure \ref{ExamplePSFFit}.  Using this functional fit was useful because the fit parameters captured the key characteristics more accurately than the median and mode of the deleo/2 distributions.

\begin{figure}
\begin{center}
\includegraphics[width=5.75in]{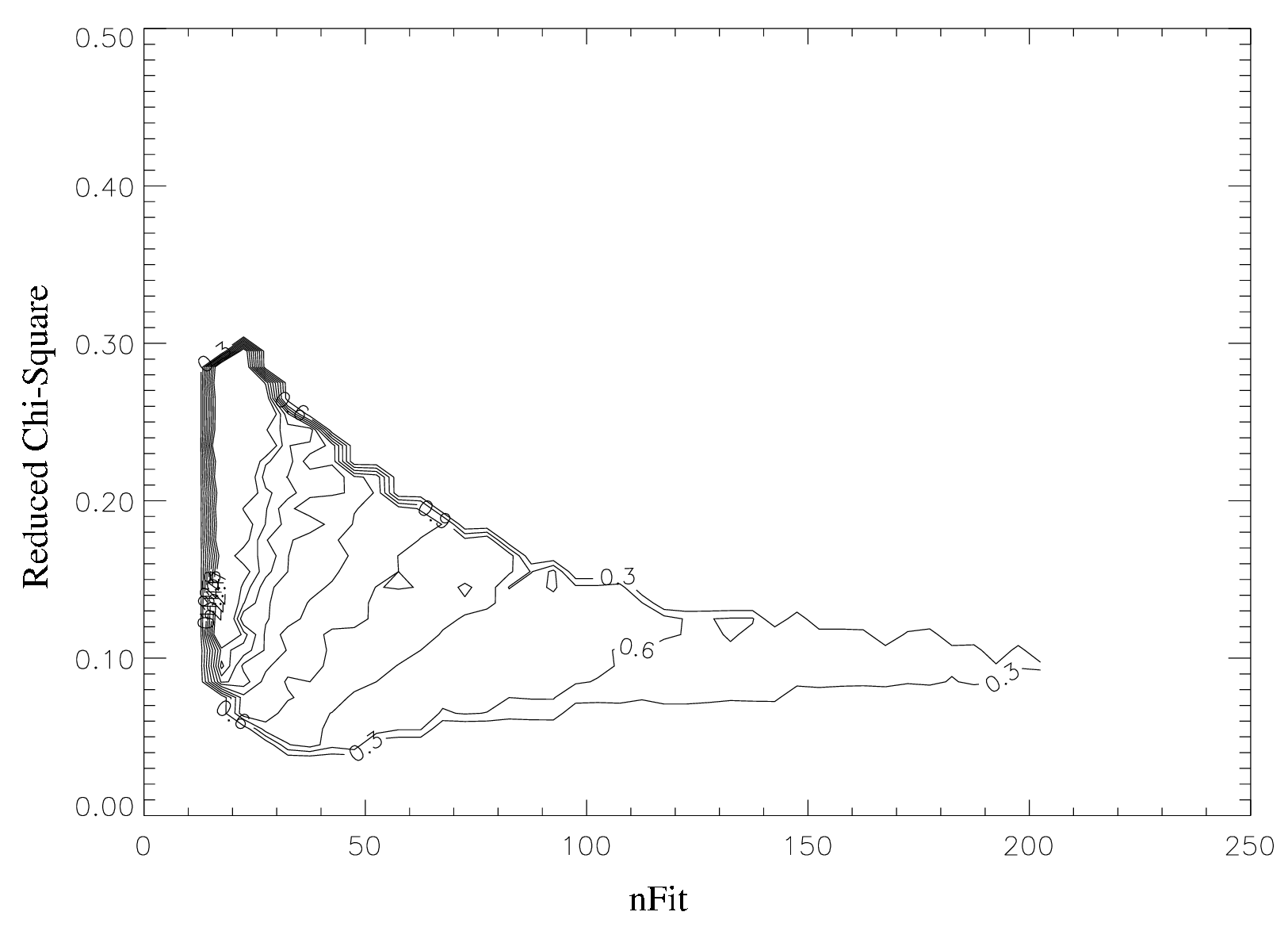}
\caption[PSF Width Contour Map]{A contour map of the PSF width (peak of $\partial P(r)/ \partial r$) in nFit vs. reduced chi-squared space. Note the bands of events with similar PSF width.}
\label{PSFContour}
\end{center}
\end{figure}

\begin{figure}
\begin{center}
\includegraphics[width=5.75in]{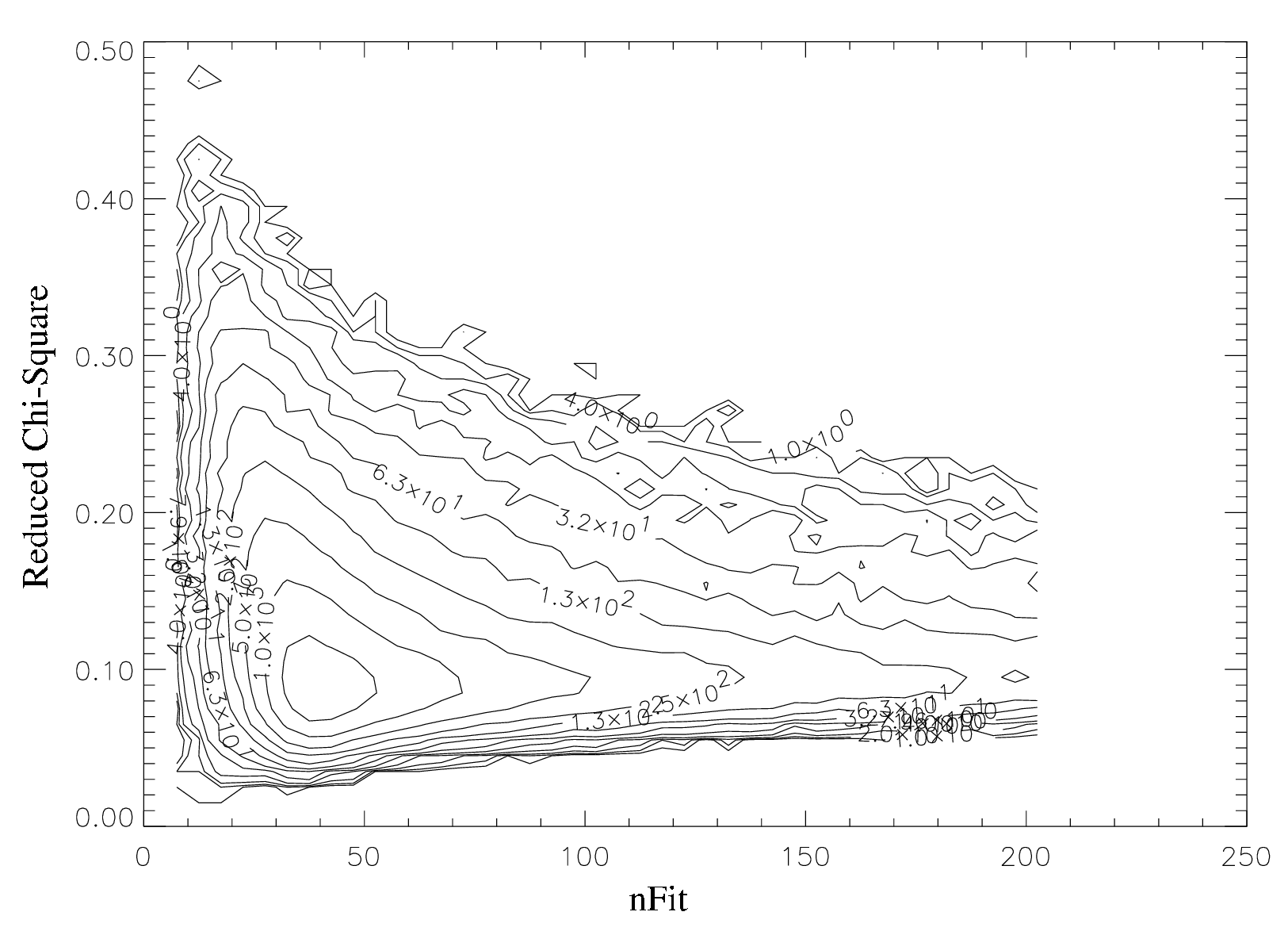}
\caption[Contour Map of Event Distribution]{A contour map of the number of events in nFit vs. reduced chi-squared space, where the contour lines follow a $\text{log}_2$ spacing. Note the ``double wing" structure, with ridges of high numbers of events running along reduced chi-squared $\sim 0.10$, and another ridge at very low nFit. }
\label{Num_chiSqvsnFit}
\end{center}
\end{figure}

In the reduced chi-squared vs.\ nFit space, the PSFs were determined by making a histogram of deleo/2 for many very small areas, then fitting with the function \ref{PSFfit:Eq}.  Note that because the fits are performed as a function of  $\partial P(r)/ \partial r$ not  $\partial P(r)/ \partial \Omega$, the peak of the function corresponds to the one sigma position of the space angle PSF.  Figure \ref{PSFContour} shows the contour plot of the peak position of the functional fit.  There are clearly strong trends in the PSF width as a function of nFit and reduced chi-squared, but neither nFit nor reduced chi-squared captures the variations on its own. Figure \ref{Num_chiSqvsnFit} shows the overall distribution of event frequency.  Note the ``double wing" structure, with ridges of high numbers of events running along reduced chi-squared $\sim 0.10$, and another ridge at very low nFit.

\begin{figure}
\begin{center}
\includegraphics[width=5.75in]{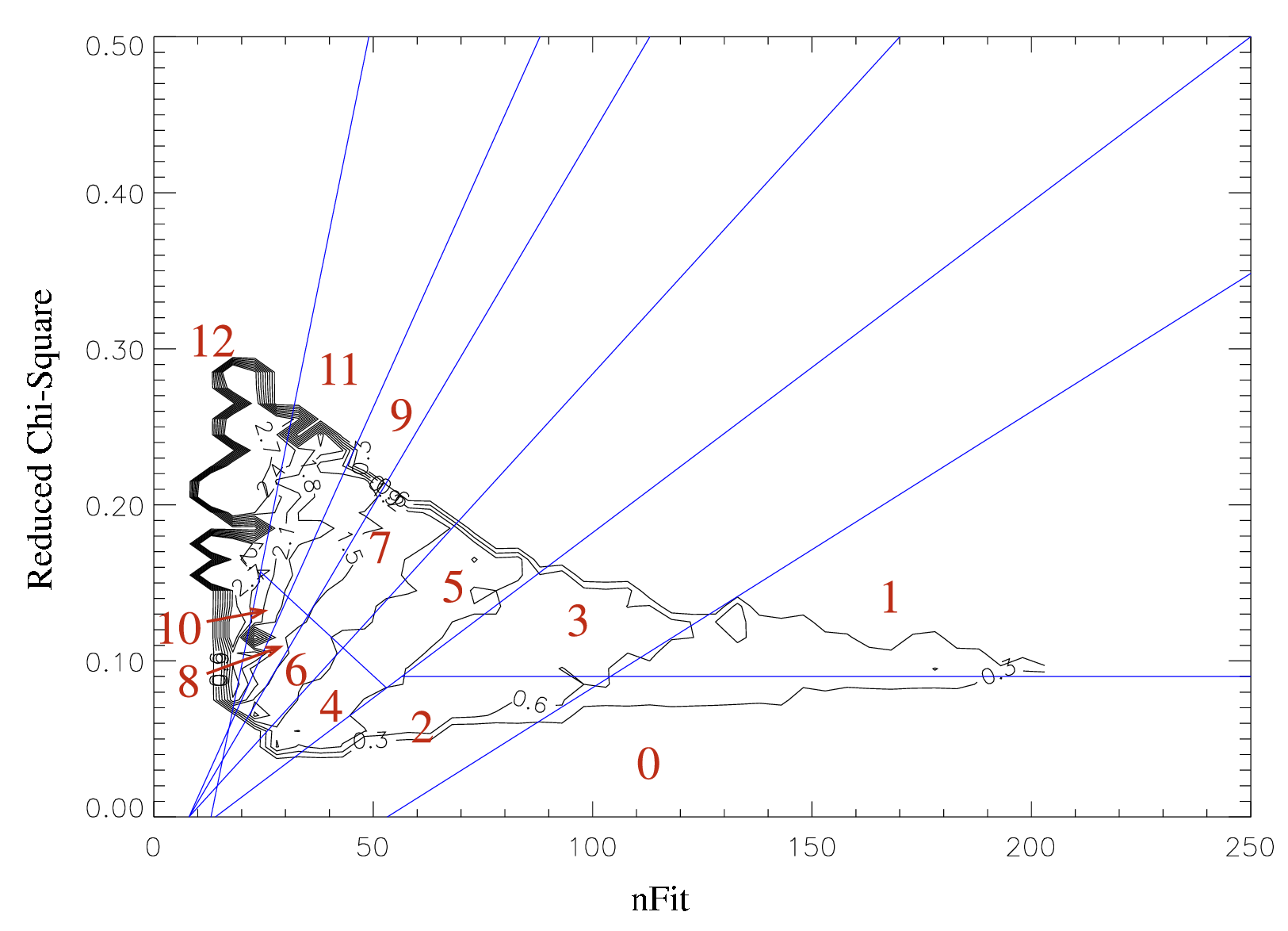}
\caption[Region Definitions, Original Calibration]{Definition of the 13 regions of events with similar PSF (numbered 0-12) shown with the peak position contours from the original calibrations.}
\label{nFit_chiSqRegions_v3b}
\end{center}
\end{figure}

From examining the contour graph of the peak positions, there are natural groups of events with similar PSF that run parallel to the contour lines.  After initial partitioning into regions by the contour lines, it was noticed that areas with lower reduced chi-square tended to have smaller tails at high $r$ values.  These large regions were better fit if they were divided into two separate areas.  The locations for these separations were determined by hand, but the best area for separation was quite clear in most cases.  Figure \ref{nFit_chiSqRegions_v3b} shows the final region definitions. 

\begin{figure}
\begin{center}
\includegraphics[width=5.75in]{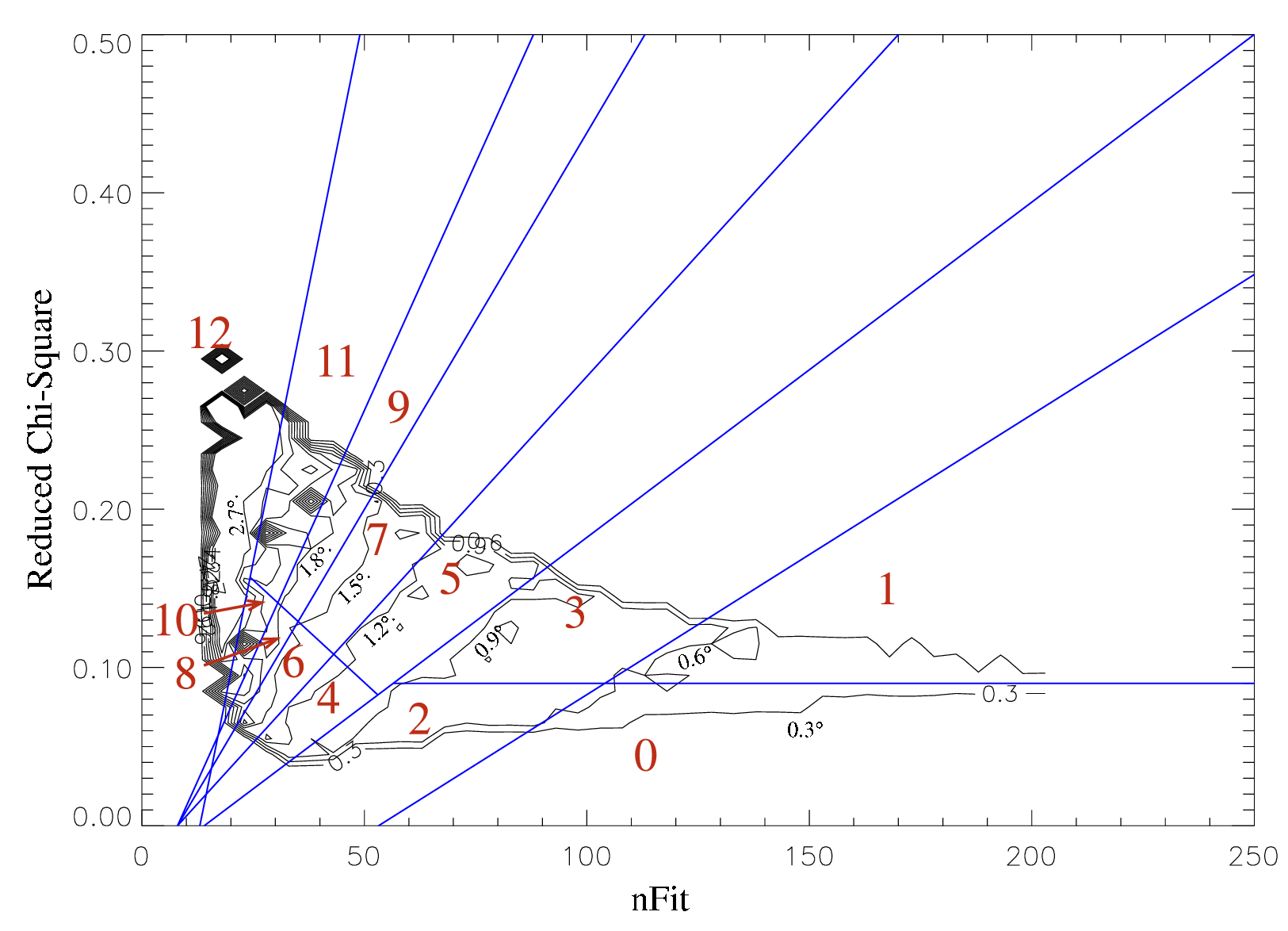}
\caption[Region Definitions, New Calibration]{Definition of the 13 regions of events with similar PSF (numbered 0-12) shown with the peak position contours from the new version 53 calibrations.}
\label{nFit_chiSqRegions_v3_53}
\end{center}
\end{figure}

After the region finding was done, the calibration for Milagro went through a major upgrade.  This did affect the distributions, but not in a meaningful way.  I thus decided to leave the region definitions the same.  Figure \ref{nFit_chiSqRegions_v3_53} shows the same regions, but with contour lines from the new calibration.  All of the results presented from here onwards will be in reference to the new set of detector calibrations.  

\begin{figure}
\begin{center}
\includegraphics[width=5.75in]{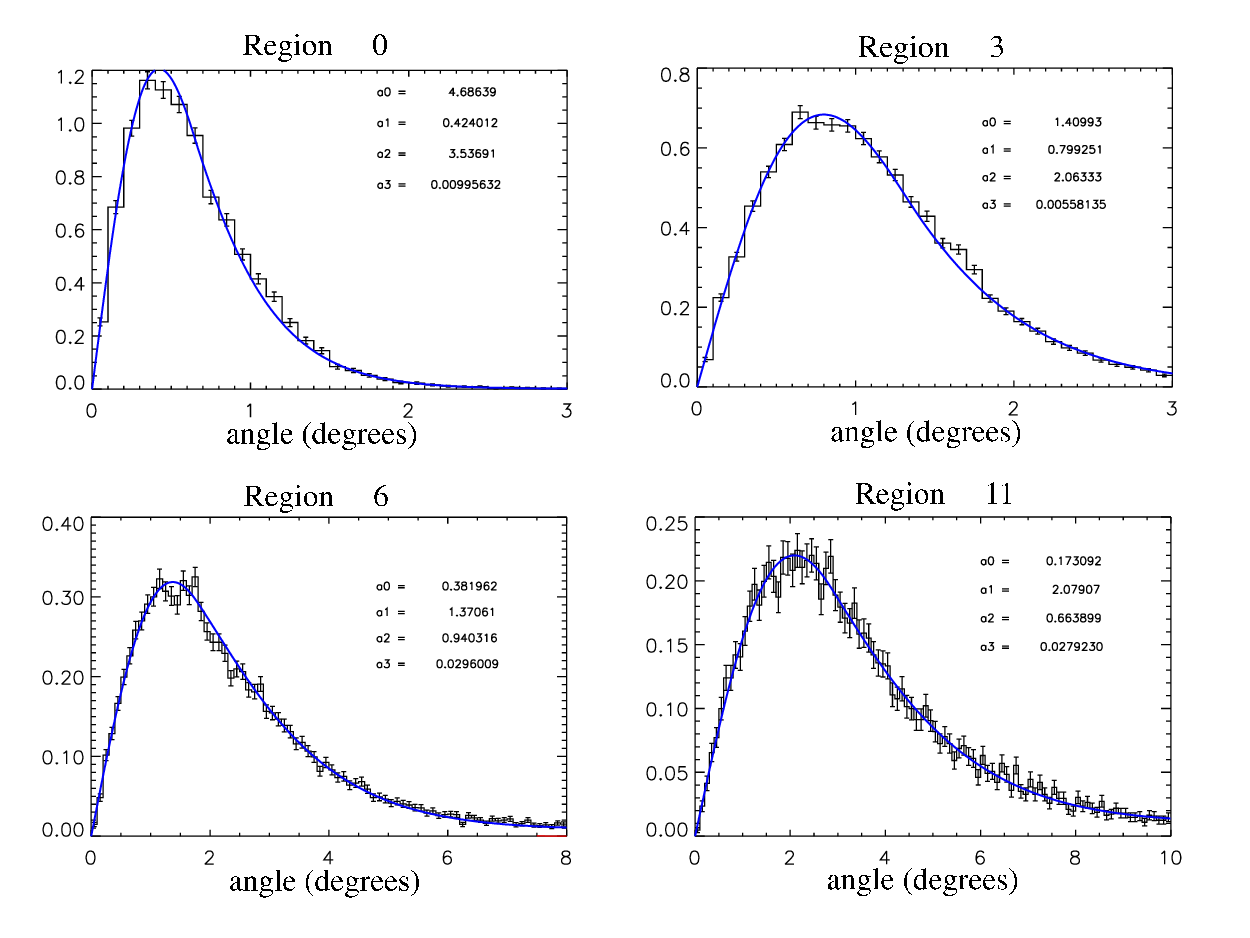}
\caption[PSF Fits To Deleo/2 Distributions]{PSF fits to deleo/2 distributions for 4 of the 13 regions. The fit parameters are displayed in the corner of each graph.  The fits are not constrained to be normalized, so that they accurately represent the PSF at each $r$.  Note the different scales on the horizontal axis.}
\label{v3DataRegionFits6Ex}
\end{center}
\end{figure}

Figure \ref{v3DataRegionFits6Ex} shows the function fits to the deleo/2 distributions for a few example regions using the functional form of Equation \ref{PSFfit:Eq}.  (For all regions see Figure \ref{v3DataRegionFits6} in appendix \ref{PlotAppendix}.)  As can be seen here, the functional fits nicely match the data within each region, and there are clear differences in the PSF from one region to another.

\section{Determining the PSF Fits}

Now that the regions characterizing the PSF have been chosen, we need to analyze the behavior of the detector with respect to these regions and determine the final PSFs to be used in the analysis.  In particular, we need to compare the results from the data-based region finding with the gamma-based Monte Carlo simulation.  This ends up being a sensitive test of the Milagro simulations and the differences between proton and gamma initiated showers. 

\begin{figure}
\begin{center}
\includegraphics[width=5.75in]{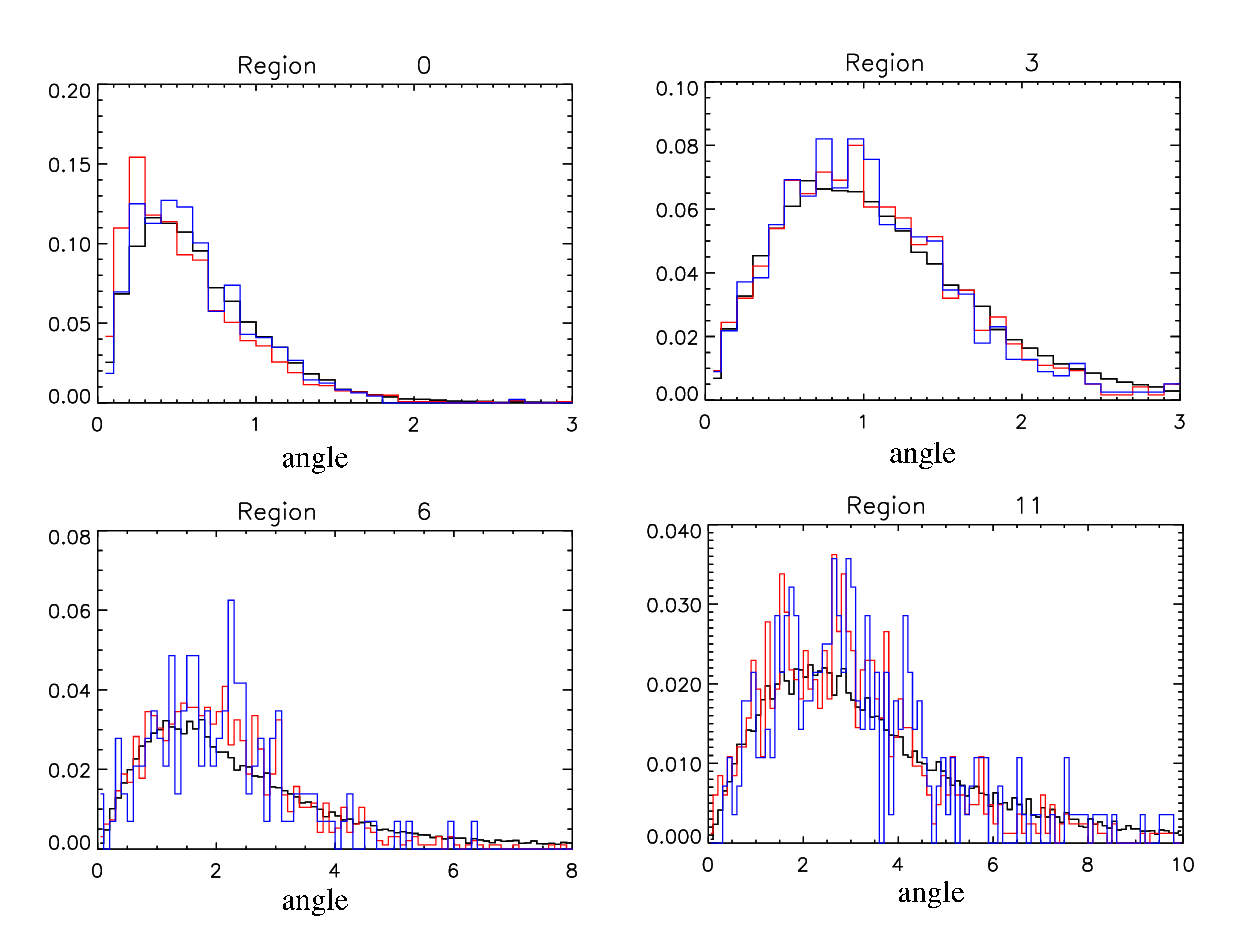}
\caption[Deleo/2 Distributions for Data and Monte Carlo Simulations]{Examples of the deleo/2 distributions for 4 of the 13 regions. The data distribution is in black, the Monte Carlo gamma initiated showers in red, and the same Monte Carlo showers with a very hard compactness cut of 3.0 in blue.  The plots for all 13 regions can be seen in Figure \ref{v3dat53mcgmcdhdeleo}.}
\label{v3dat53mcgmcdhdeleoEx}
\end{center}
\end{figure}

Since the data is dominated by background cosmic rays, the first concern is that the regions found using the data are not effective for classifying gamma rays.  Unfortunately we do not have a strong gamma-ray signal to use, but must instead rely on the Monte Carlo simulations to provide gamma-ray events.  In addition, most Milagro data analysis is done with a ``compactness" requirement to help separate gamma-ray and cosmic-ray-induced showers \cite{Gus:X2}.  This raises the concern that the gamma-hadron separation could adversely affect the PSF grouping developed in the last section.  Figure \ref{v3dat53mcgmcdhdeleoEx} shows the deleo/2 distributions for the data (black), Monte Carlo gammas (red), and the same Monte Carlo gammas after a very hard compactness cut of 3.0 (blue).  Though some of the regions have very low statistics, the agreement between all three distributions is quite good.  This reassures us that the PSF regions depend principally on the number and quality of data points used in the reconstruction, and not the species of the initiating particle.  Since the compactness cut had a minimal effect on the deleo/2 distributions, and we can continue to treat the PSF classification and gamma-hadron separation as independent problems. 

\begin{figure}
\begin{center}
\includegraphics[width=5.1in]{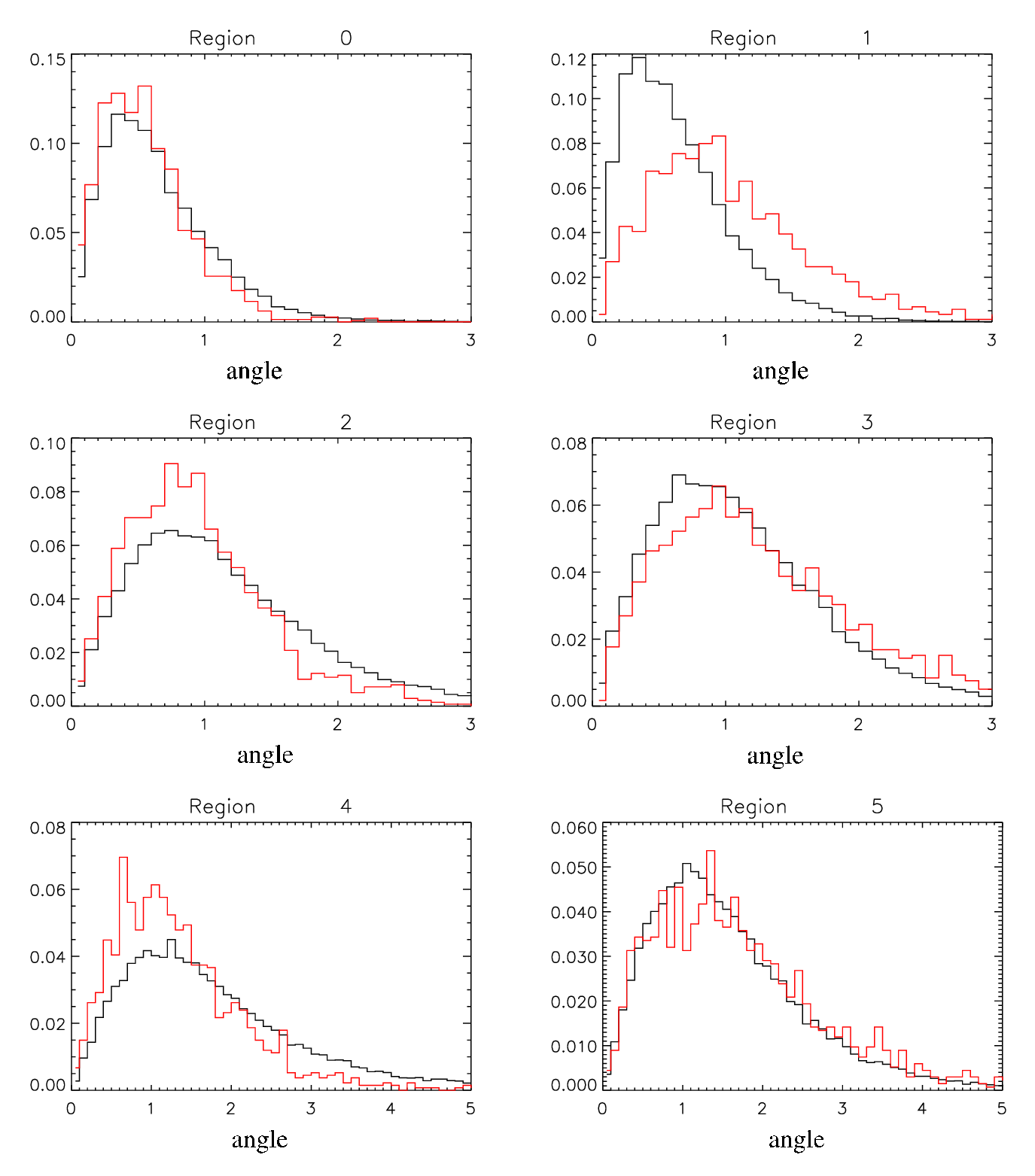}
\caption[Data Deleo/2 vs. Monte Carlo PSF Part 1]{Deleo/2 for the data (black) vs. the angular reconstruction error for Monte Carlo gamma-rays (red) for regions 0 -- 5. Note that in the even regions (with lower reduced chisquared) the Monte Carlo PSF is narrower than the data deleo/2, while the Monte Carlo PSF is wider than the data deleo/2 for the odd regions (having higher reduced chi-squared).}
\label{v3dat53_vs_mc30g}
\end{center}
\end{figure}

\begin{figure}
\begin{center}
\includegraphics[width=5.1in]{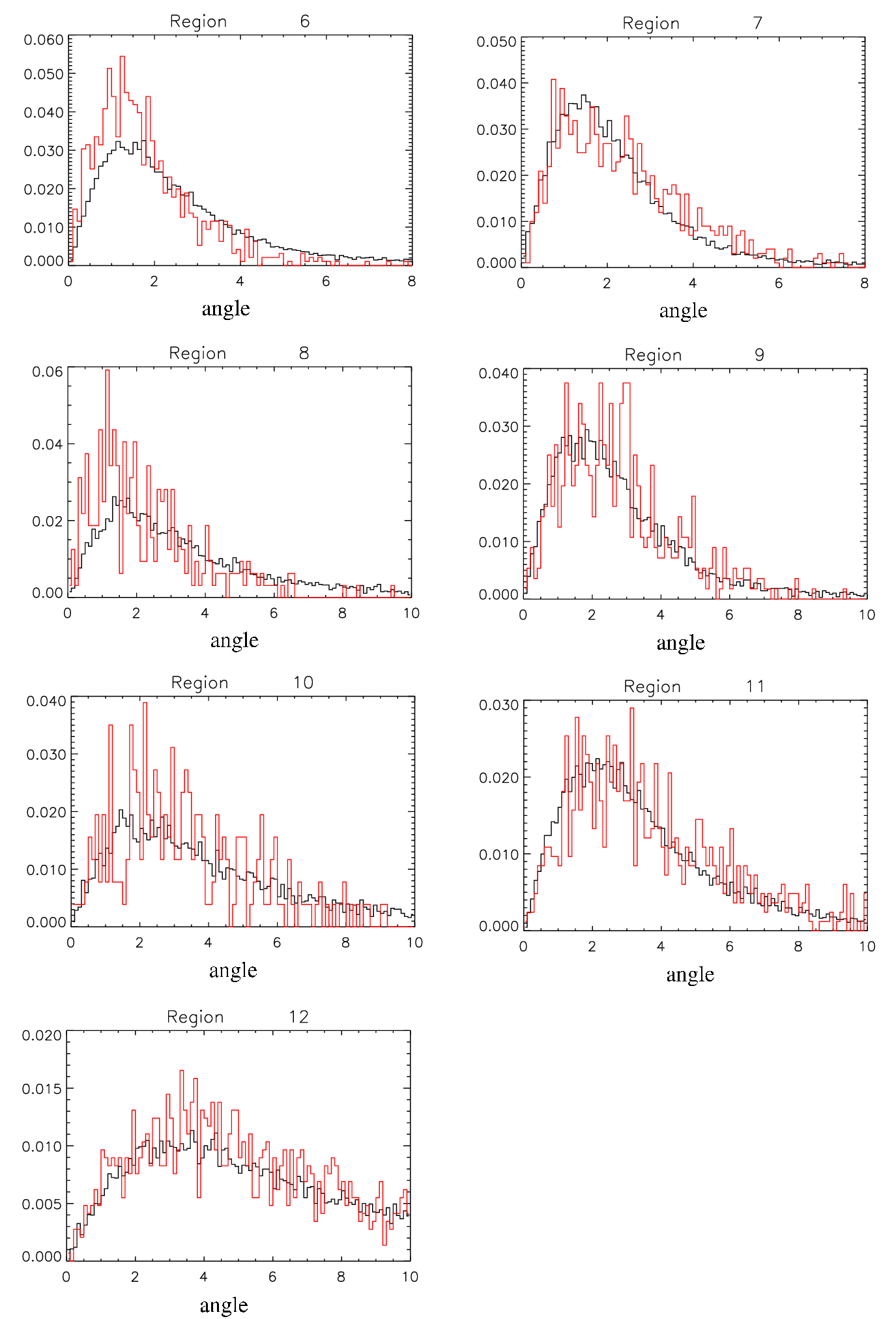}
\caption[Data Deleo/2 vs. Monte Carlo PSF Part 2]{Same as Figure \ref{v3dat53_vs_mc30g} for regions 8 -- 13.}
\label{v3dat53_vs_mc30gB}
\end{center}
\end{figure}

So far we have used the measure deleo/2 to classify the PSF.  However, because the timing distributions are asymmetric and systematic effects are not included, we don't expect deleo/2 to be the true PSF.  The reconstructed direction for the Monte Carlo data can be compared to the true direction to get a direct determination of the PSF.  Figures \ref{v3dat53_vs_mc30g} and \ref{v3dat53_vs_mc30gB} shows the data deleo/2 distributions in black, and the Monte Carlo gamma-ray PSF distributions in red.  Region 1 shows a significant discrepancy, where the deleo/2 peaks near 0.35 degrees and the true angle difference peaks closer to 0.9 degrees.  Region 1 events have a large number of tubes in the fit, but the higher than average chi-squared value indicates that the angle fitter is having trouble with these events.  In contrast, the PSF distribution for region 0 events is slightly narrower than the deleo/2 distribution. This general trend continues though the other regions.  Remember that the even numbered regions have low reduced chi-squared values, while the next higher odd numbered regions have a similar deleo/2 distribution but higher reduced chi-square (see Figure \ref{nFit_chiSqRegions_v3_53}).  Looking through the regions, the PSF is typically better than deleo/2 for the even regions, and worse than deleo/2 for the odd regions.  

The argument can be made that even better regions could be found by tuning on the Monte Carlo gammas. However, significantly better statistics would be needed, with separate tuning and testing samples to avoid bias.  We currently do not have the computational resources needed to make a Monte Carlo sample of the necessary size.  It is encouraging that the regions based entirely on the deleo/2 distributions from data are finding clear differences in the Monte Carlo sample.

\begin{figure}
\begin{center}
\includegraphics[width=5.75in]{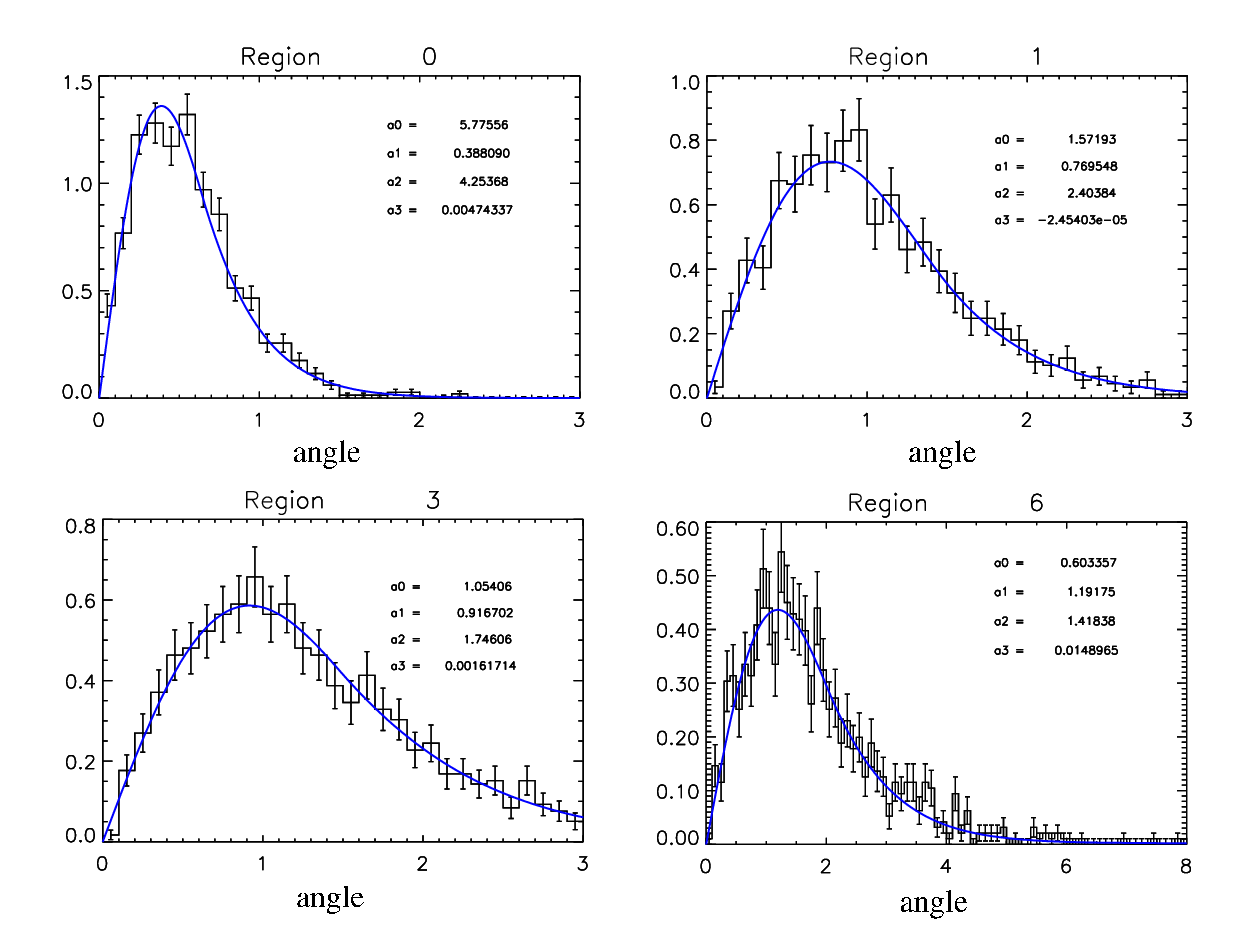}
\caption[PSF Fits to Monte Carlo Distributions]{PSF fits to the gamma-ray Monte Carlo angle difference distributions for 4 regions. The fit parameters are displayed in the corner of each graph.  So that they accurately represent the PSF at each distance, the fits are not constrained to be normalized. Figures \ref{v3mcgRegionFits6} and \ref{v3mcgRegionFits6B} show the PSF fits to the gamma-ray Monte Carlo angle difference in all 13 regions.}
\label{v3mcgRegionFits6Ex}
\end{center}
\end{figure}

The final PSFs used in the analysis of Milagro data come from a mixture of Monte Carlo and data distributions.  For regions 0 -- 7 the Monte Carlo distributions were used because they correctly capture the systematic errors.  However, as the number of tubes used by the angle fitter decreases the statistical errors begin overwhelm the systematic errors.  For regions 8 -- 12, the Monte Carlo distributions agree well with the distributions from data (see Figure \ref{v3dat53_vs_mc30gB}) and the data-based distributions have much better statistics.  Figure \ref{v3mcgRegionFits6Ex} shows examples of the Monte Carlo based PSF distributions and the functional fits used to characterize the final PSFs in regions 0 -- 7, where Figure \ref{v3DataRegionFits6B} shows the fits based on the deleo/2 distributions of the data and used for the final PSF distributions in regions 8 -- 12.  

\section{Determining the $P_\gamma$ Distributions}
\label{Pgammasection}

In Milagro analyses to date, gamma-hadron separation has consisted of a global cut, where some events are tagged as gamma-like and others as hadron-like.  This binary logic is necessary for binned analyses, but does not use all of the available $P_\gamma$ information (see Section \ref{WATSensitivity}). Because of Milagro's poor signal-to-background ratio, this binary logic tends to favor a very hard cut that throws out most of the gamma-ray events along with most of the hadron events.  In a weighted analysis, we want the probability that a given event is a gamma ray --- this uses the quantitative information on how gamma-like events are and in theory allows use all of the events.  

Currently the most established gamma-hadron separation technique used in Milagro is based on the ``compactness" parameter.\footnote{This is an active area of research, and there are several competing techniques.}  This parameter compares the number of tubes in the muon layer with more than two photoelectrons to the highest number of photoelectrons observed in any single phototube in the muon layer.  A simplified way of thinking about this parameter is that it looks for penetrating particles that deposit a large amount of light into a spot in the bottom layer. A significant amount of the compactness parameter's gamma-hadron separation comes from the identification of muons which often accompany hadron initiated showers.  

\begin{figure}
\begin{center}
\includegraphics[width=5.75in]{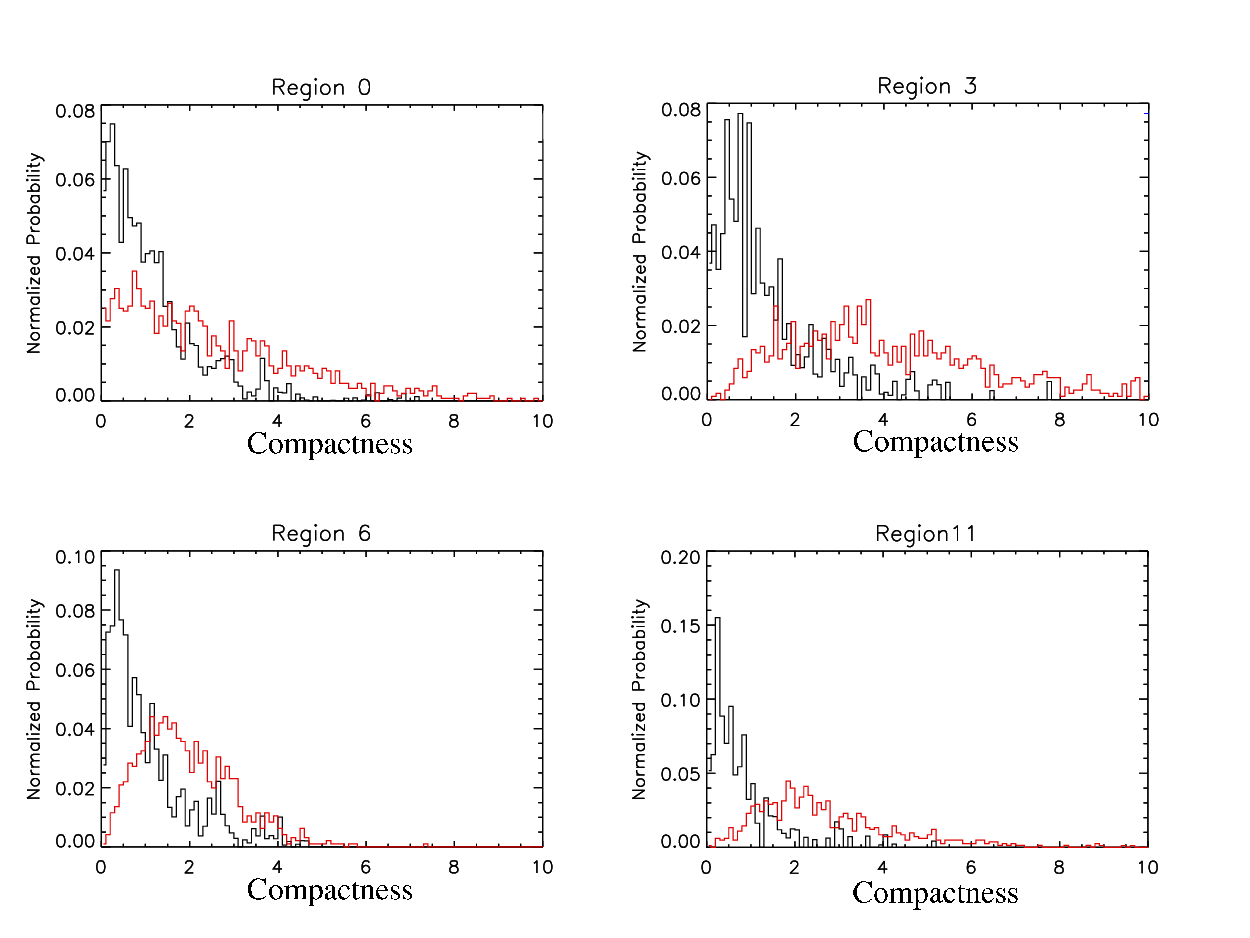}
\caption[Compactness Distributions by Region] {The probability distributions for 4 example regions plotted as a function of the compactness parameter for Monte Carlo protons (black) and Monte Carlo gamma rays (red).  The probability distributions are derived from the event distributions by globally normalizing both the protons and gamma rays to one.  (See text for discussion of this normalization factor.)  The distributions for all 13 regions can be seen in Figures \ref{X2_RegionSimplified} and \ref{X2_RegionSimplifiedB}.}
\label{X2_RegionSimplifiedEx}
\end{center}
\end{figure}

Figure \ref{X2_RegionSimplifiedEx} shows the number of events as a function of the compactness parameter for four of the regions defined in Section \ref{RegionIdentification}, with gamma-ray events histogrammed in red and proton events in black.  While the region-by-region PSFs were largely independent of the compactness parameter, the compactness distributions are clearly different from one region to the next.  This indicates that we need to determine the gamma probability separately for each region in order to maximize the sensitivity.

One important subtlety in Figure \ref{X2_RegionSimplifiedEx} is the normalization.  Regardless of the compactness parameter, what is the probability that an event is a gamma ray?  If we are looking at a weak but steady source like the Crab pulsar, the signal-to-noise ratio is very low and gamma rays are relatively unlikely.  However, for a millisecond gamma-ray burst, the time window is so short that background noise is insignificant and proton initiated showers are very rare.  In Figure \ref{X2_RegionSimplifiedEx} the gamma ray and proton showers have been globally normalized to one, so that there are equal numbers of gamma-rays and protons.  Equal weighting was chosen in an effort to remain as unbiased as possible over intermediate time scales (40 sec. -- 3 hours).  In these intermediate time scales we expect the signal to be strong, but there is still significant background.  A truly optimal source search would tune this normalization to the expected signal strength.

\begin{figure}
\begin{center}
\includegraphics[width=5.75in]{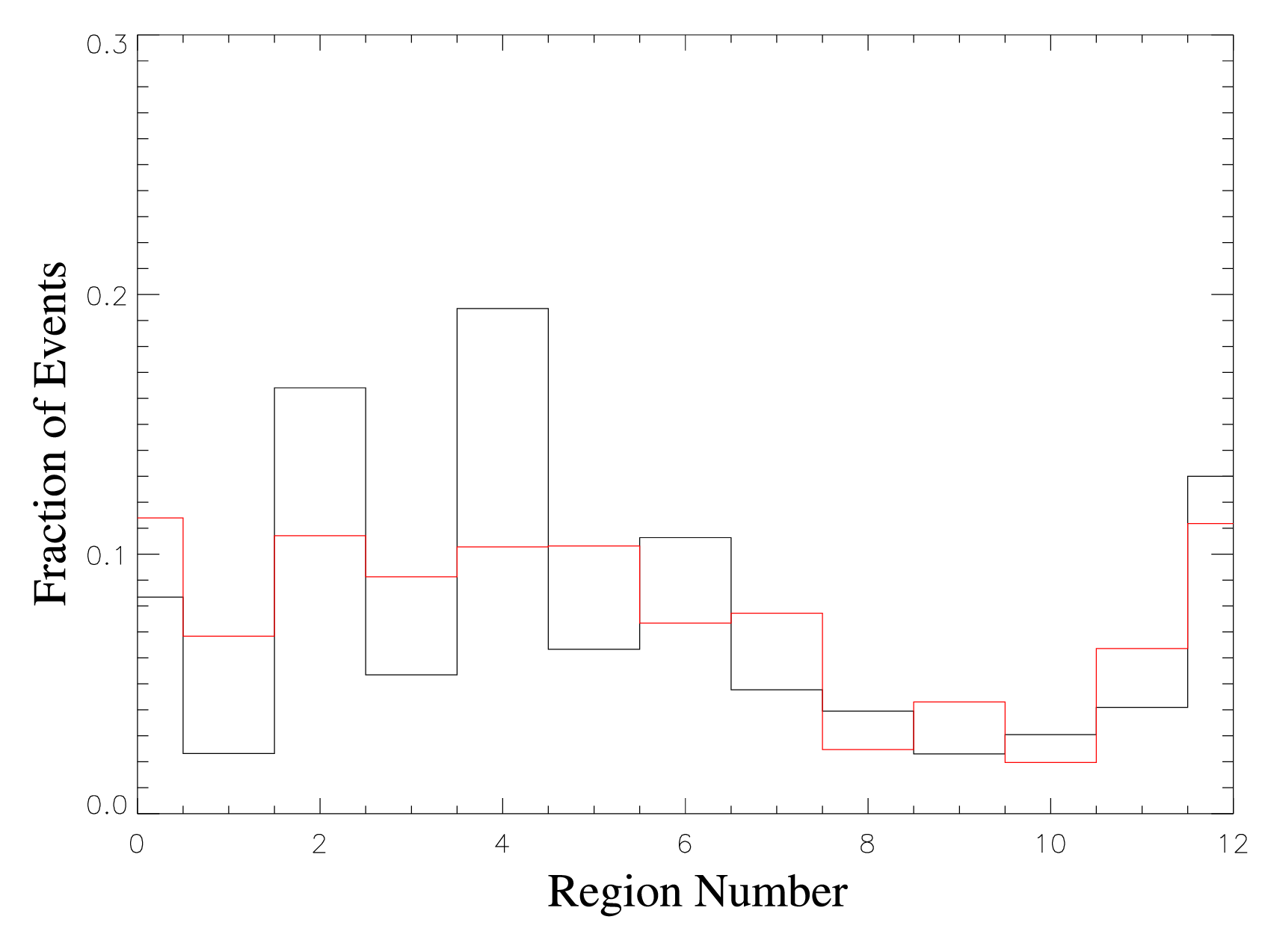}
\caption[$P_\gamma$ by Region]{The fraction of the total number of gamma rays in each region is indicated in red with the fraction of protons in each region indicated in black.  Note that gamma rays are relatively more common in the odd numbered regions.}
\label{X2_RegionFraction}
\end{center}
\end{figure}

An additional decision had to be made whether to normalize the distributions globally or region-by-region.  Global normalization was chosen because gamma-ray events are more likely to appear in certain regions.  Figure \ref{X2_RegionFraction} shows the fraction of gamma rays in each region.  Notice that there are relatively more gamma-ray induced showers in the odd numbered regions.  This implies that the gamma-ray events have characteristically higher reduced chi-squared values than the hadron induced showers.  Why the purely electromagnetic gamma-ray showers appear noisier to the reconstruction algorithms is not understood.

\begin{figure}
\begin{center}
\includegraphics[width=5.75in]{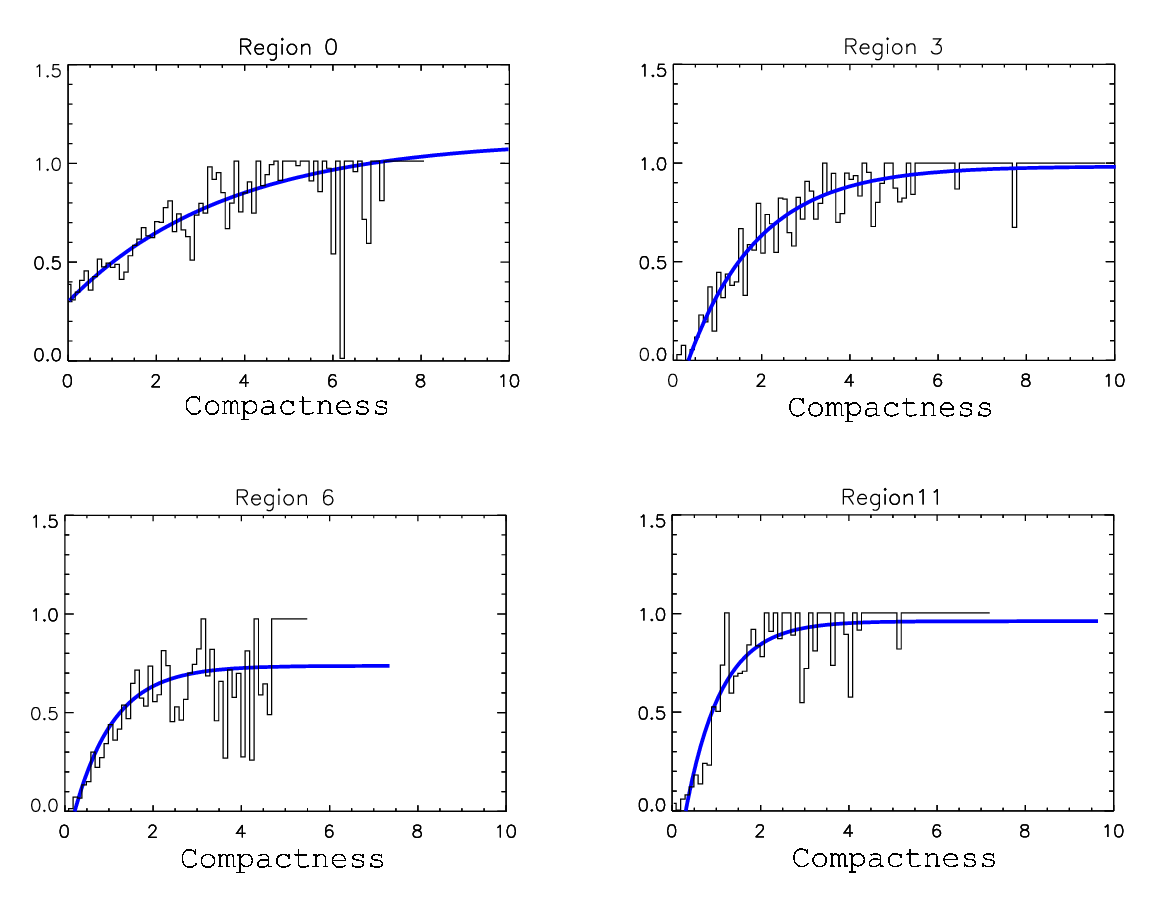}
\caption[$P_\gamma$ vs. Compactness Fits by Region]{The $P_\gamma$ distributions for four example regions with the associated fits in blue. While the error bars are not show for legibility, they were used in the fit.  All 13 $P_\gamma$ distributions, fits and fit parameters can be seen in Figure \ref{X2_RegionFits}.}
\label{X2_RegionFitsEx}
\end{center}
\end{figure}

The probability $P_\gamma$ is the relative probability that a particular shower was initiated by a photon as opposed to a proton.  The data in Figure \ref{X2_RegionSimplifiedEx} is used to calculate the relative probability by determining the fraction of showers initiated by gamma rays as a function of the compactness parameter. The resulting distributions are shown in Figure \ref{X2_RegionFitsEx} and were fit to the function $P_\gamma=a_2-e^{-a_1(compactness)+a_0}$ as indicated by the blue lines.  It is this fit to the distributions which is used in the analysis.  Note that the relative probability does not reach one when separation is hard to achieve, as in region 6.  In some cases the fit exceeds one for high compactness values, and is truncated to a maximum of one in the code.

Another subtlety arises due to the spectrum of the incoming gamma-rays.  The Monte Carlo files were generated using an $E^{-2.4}$ spectrum for the gamma-rays and an $E^{-2.7}$ spectrum for the protons.  The proton spectrum is from measurement, but \em a priori \em we do not know the spectrum of a signal. $E^{-2.4}$ was chosen as being ``reasonable" and similar to the observed Crab spectrum and near the observed MeV GRB spectrum of  $\sim E^{-2.2}$ \cite{BATSEcatalog}. This choice of spectrum in the Monte Carlo simulation leads to a small spectral dependence in the source search.  A very different gamma-ray spectrum could affect both the number of gamma-rays expected in each region (the normalization), and to a lesser extent the compactness distributions in each region. This means that the search is optimized for the Monte Carlo spectrum, though it is still sensitive to all incoming spectra.

\section{Future Directions}

It is expected that the details of the characterization of the Milagro observatory will change, but the techniques for characterizing the detector will remain relatively constant.  In particular the completion of the outrigger field will significantly improve both the angular reconstruction and the background rejection capabilities of the Milagro detector.  There are also a number of planned software and hardware advances that should lower the energy threshold and improve gamma-hadron separation.  These changes in the performance of the detector must be mirrored by updates in the detector characterization.  

This characterization of Milagro should apply equally well for both the weighted analysis technique and a maximum likelihood analysis.  With an adjustment of the expected number of signal events used in the $P_\gamma$ formulation, this characterization could be used by a maximum likelihood analysis to extract the spectral parameters of a steady source and other precision measurements.

\chapter{Implementing the Weighted Analysis Technique in Milagro}
\label{ImplementingWAT:chap}

\section{Introduction}

Implementing the weighted analysis technique for use as a GRB search in Milagro became a significant programming task, with more than seven thousand lines of code. When implementing any analysis, there are always trade-offs, and the quality of an implementation is often in the details of the code and which trade-offs were made. Because a real-time GRB search needs to be computationally fast, there are many areas where code simplicity was sacrificed in favor of speed. The next two chapters detail my implementation of the weighted analysis technique, and the approximations that were made.

The weighted analysis technique can be applied to both transient and steady source signals, and many parts of the code are common to both types of analyses.  For this reason the code base is partitioned into two separate pieces. The weighted analysis framework contains all of the generic code that is common to all analyses such as building sky maps and background maps, adding maps together, identifying signals, etc. Individual analyses must still determine which events and PSF distributions to use, and when to add maps or search for signals. The weighted analysis framework contains about 3700 lines of code, and does most of the hard computational work. The amount of code needed to convert this framework into a full analysis depends on the type of analysis to be performed. A steady source analysis using cleaned data may be a single page of code --- just enough to organize the adding of events and when to perform source searches. In contrast, the online GRB search program with its real time requirements, on the fly event selection, and multi-threaded concurrent searches is nearly three thousand lines of additional code. This chapter describes the implementation of the analysis framework, with the next chapter concentrating on turning this framework into a real-time GRB search.

\section{Introduction to Object-Oriented Programming}

All of the code discussed in this chapter and the next uses advanced object-oriented programming (OOP) techniques and is written in the Objective-C computer language. Objective-C is compiled by the standard gcc compiler and simply adds some OOP abilities to the standard C programming language. I chose Objective-C because there is a sophisticated standard library for Objective-C which greatly reduced the amount of code that needed to be written. 

There is one idea from OOP that needs to be briefly explained before the implementation of the weighted analysis is discussed --- the idea of an ``object." A programming object can be though of as a little autonomous mini-program, which is designed to mimic some real world object. The idea is to create many little programs, and then tie the pieces together with ``messages" which are sent from one object to another.  For example, I can write an object (also called a class) called photon that is designed to represent one shower. Now in my main program I can create a photon object (give me a new photon with right ascension...), ask it questions (what is your declination?), give it commands (save a copy of yourself to disk), or even destroy the photon object. Programming becomes a task of managing the communication among many small independent objects. For a large program like the GRB Search, I have a different object for every important concept, such as sky maps, background maps, or photons. I can then read in a data file and create a few hundred photon objects (one for each event), put them into an array object (a useful object from the standard library), then hand the array of events to the sky map object with the directive ``add this array of photons to the sky map," at which point the sky map object will ask each photon for its position, PSF, and other characteristics, and add the appropriate weights to the sky map. One advantage of this kind of programming is that an object can be thoroughly tested in a simple test program, and when it is used in the main program it is guaranteed to behave in exactly the same way. (This is code proving.) The framework that I describe in this chapter covers the objects which are needed to perform a weighted analysis (sky maps, background maps, etc.).  These objects can then be tied together in different ways, and in the next chapter I describe how they are assembled to perform an advanced real-time GRB search in Milagro. For an superb conceptual introduction to object-oriented programming (for any language) read the second chapter of the online book Object-Oriented Programming and the Objective-C Language \cite{Objective-C}.

\section{Building a Sky Map}
\label{Skymap}

The heart of the weighted analysis technique is building a digitized sky map from a sum of PSFs, and the skyMap object is designed to perform this task (see Section \ref{WATSkymap}). There are two required pieces of information for every sampled sky map location -- the sum of the weights (Equation \ref{weightSum:Eq}), and the number of events used to make the sum (see Equation \ref{Prob:Eq}). In addition, it is useful to retain the exposure as a function of sidereal time, again as both a weighted sum (sum of $P_\gamma$) and as a simple event count.  The key information for the sky map is contained in a two dimensional array of sky map locations --- with each array entry containing a weight and number of showers --- and an array of the sidereal exposure  --- again with each entry containing a weight and number of showers.

In this implementation the sampling pattern of sky map locations is equally spaced in right ascension and declination. This sampling pattern was chosen to simplify the implementation of the direct integration background method, where a distribution of events in local coordinates and the sidereal exposure are used to create an expected background distribution (see the next section). Sampling in equal right ascension and declination steps leads to regions near the pole containing much denser sampling than near the equator in the right ascension dimension. Since we are sampling at points, as opposed to integrating over pixels, this oversampling near the pole does not hurt us as long as we make sure the sampling distance near the equator is small compared to the smallest PSF. The skyMap object is hard wired to use the equal spacing in right ascension and declination, but the size of the spacing is a variable that can be set by the user to match the PSFs under consideration.

In general, for each shower we need to determine the distance from the reconstructed position to each sampling location, determine the weight of the PSF at that distance, and add it to the sky map. In practice, the PSFs are truncated at some distance so that only nearby sampling locations need to be considered.  Table \ref{Cutoff:table} lists the truncation distances for all 13 PSF regions used in the current detector characterization (see Chapter \ref{Characterization:chap}). When a PSF is truncated it is not renormalized.  Renormalizing the PSF would incorrectly give too high of a weight to events of very poor resolution --- the PSF distribution should accurately represent the probability density of an event's true position being at that distance.  Not renormalizing, but simply setting the probability density to zero at some distance is equivalent to only keeping showers within a predetermined radius. Even with this truncation it is typical to have $\sim 400$ locations which need to be updated for each incoming shower. Clearly, repeatedly calculating the distance and PSF weight for each location would present an enormous computational overhead. To mitigate the computational load, a series of small approximations has been made which allow the tabulation of the distance and PSF values into lookup tables.  These lookup tables can then be used to rapidly add weights to the sky map.

\begin{table}
 \centering
\begin{tabular}{|c|c|c|c|c|c|c|c|c|c|c|c|c|c|}
	\hline 
	Region & 0 & 1 & 2 & 3 & 4 & 5 & 6 & 7 & 8 & 9 & 10 & 11 & 12\\ 
	\hline 
	Cutoff  & 2.0 & 3.0 & 3.0 & 3.0 & 3.5 & 3.5 & 3.5 & 4.0 & 4.0 & 4.0 & 4.0 & 4.0 & 4.0 \\
	\hline
\end{tabular}
  \caption[PSF Cutoff Distances]{The distance in degrees at which each PSF is truncated in the current Milagro characterization (see Chapter \ref{Characterization:chap}). The truncation was chosen to be where the probability density became negligible or four degrees, whichever was smaller. }
\label{Cutoff:table}
\end{table}

The first approximation is to quantize the shower arrival positions in right ascension and declination. Instead of using the full floating point value of right ascension and declination provided by the angular reconstruction, the value is rounded to some set precision (user determined). For a event of a given declination, the distance to a sampling point can now only take a small set of values instead of an infinite number, and this set of possible distances can then be used in building lookup tables. Digitizing the possible arrival positions is equivalent to adding a small amount of noise and slightly widening the PSF distributions.  The amount of PSF widening can be minimized by making the digitization of arrival positions very small compared to the smallest PSF. Digitizing the arrival positions enables an enormous increase in computational speed, with the precision only being limited by the available computer memory for storing lookup tables. A finer position digitization makes for a larger lookup table, but does not impact the speed of the skyMap algorithm. 

For a given declination, there is a predetermined set of distances to the nearby sampling locations. However, because the sampling becomes denser approaching the pole, a different set of distances is needed to for each declination. Using the azimuthal symmetry of the PSFs, a given distance corresponds to a particular PSF weight. We can then envision making a 4 dimensional lookup table that contains templates for all of the weights to be used. Given a declination and a PSF (2 dimensions), there would be a corresponding two dimensional template of the PSF weights that can be directly added to the sky.  Basically, for a given declination and PSF we know the pattern of weights that must be added to the sky, and this information can be stored in a lookup table.

The four dimensional lookup table presented above can become enormous, so in practice there are several tricks used to reduce memory usage.  Because the PSFs are azimuthally symmetric, the templates are symmetric around their centers in right ascension. In the lookup tables only half of the template is stored, then mirrored on the fly when added to the sky map. Additionally, changing the declination by one sampling spacing introduces a negligible change in the distances used to create the templates. Allowing for sparse filling of the lookup table in declination can greatly reduce the memory footprint. Using a template for a nearby declination adds a very small amount of noise and is equivalent to adding a small additional widening of the PSF. Again, the sparseness of the declination sampling can be set by the user. A standard analysis of Milagro data with typical values of photon position digitization and declination undersampling will produce a $\sim100$ MByte lookup table. Given that the computational speed increase, the cost of $\sim100$ MByte memory and the associated code complexity is a small price to pay for the performance gain.

The basic layout of the skyMap object consists of three memory structures: the sky map, the sidereal exposure, and the lookup table. When the skyMap object is first created, the user must determine the spacing of the sampling locations, the digitization of the photon positions, and the declination undersampling to be used in the lookup table.  The user then gives the skyMap object a set of PSFs to be used. The skyMap object uses this information to allocate the sky map and sidereal exposure arrays and build the lookup table (this table can be given to another skyMap object, so it only needs to be created once). Once this initialization is complete, the sky map and sidereal exposure can be formed by adding events. For every shower, the declination and PSF region is used to select the appropriate template from the lookup table. This template is then multiplied by the probability that the event really is a photon ($P_\gamma$) and added to the sky map. $P_\gamma$ is also added to the sidereal exposure, and then the process is repeated for the next event.  Adding showers to the sky map is the most time consuming part of the weighted analysis technique, and is why the weighted analysis technique is 2 -- 8 times slower than a comparable binned analysis.

\section{Building a Background Map}
\label{backgroundMap}

The background calculation used in this implementation of the weighted analysis technique is a variation on the background map method \cite{CygnusTeq} called direct integration.  The background map methods of background determination are based on the idea that if the data is background dominated, a long exposure in local coordinates will accurately represent the average distribution of the background seen by the detector. Any signal contribution to the local coordinate map should be minimal because steady sources are diluted by their motion across the sky and transient sources will be buried in the long exposure time. This allows us to assume that the expected probability distribution of the background in local coordinates is proportional to the event distribution observed in local coordinates $E_{b}(\theta,\varphi,t)$. The method of direct integration can be used when the time variation of $E_{b}(\theta,\varphi,t)$ is due almost entirely to changes in the rate $R_b(t)$, with the local event distribution $E_{b}(\theta,\varphi)$ remaining nearly constant with time. In this approximation, the probability distribution for a single background event can be found by normalizing the local event distribution (${P_{b}(\theta,\varphi)} = {E_{b}(\theta,\varphi)}/N_{b}$ where $N_b=\int R_b(t)dt$).  The expected background distribution in celestial coordinates $P_{obs}'(RA,DEC)$ for an observation of any length can then be determined by integrating the normalized local probability distribution times the observed all-sky rate during the source interval $R_{obs}(t)$, and the transfer function that maps the local position and time to right ascension and declination $T(\theta,\varphi,t \rightarrow RA,DEC)$:
\begin{equation}
\label{DirectInt:Eq}
P_{obs}'(RA,DEC) = \int \frac{E_{b}(\theta,\varphi)}{N_{b}}R_{obs}(t)T(\theta,\varphi,t \rightarrow RA,DEC)\,dt
\end{equation}
In essence, the local map is used to determine the probability distribution for a single background event, which is then integrated with the all-sky rate observed during the source time to produce the expected background in celestial coordinates.  This is equivalent to smearing the expected spatial distribution of events in local coordinates across the observed sky in RA and DEC, with the normalization of the local distribution tracking the observed event rate.  This process works equally well for both a smooth probability distribution and a binned analysis --- the definitions of $P_{b}(\theta,\varphi), N_b, R(t)$ just change slightly.  Because the observed sky map contains both the smooth probability distribution and a binned component from the number of events used in the sum, Equation \ref{DirectInt:Eq} is used twice --- once for the photon probability density and once for the number of showers.

In practice the angular distribution of events ``breathes" with diurnal variations \cite{MilagroAsymmetry} and changes in overburden. However, these changes are slow, and the $P_{b}(\theta,\varphi)$ distribution is constant on time scales of 2 hours or less. When used in an analysis, this time variation in the local distributions must be taken into account and the $P_{b}(\theta,\varphi)$ distribution recalculated at least every 2 hours. 

In this computational framework the background calculation is performed by the localMap object and represented in the backgroundMap object. When events are added to the localMap object to form an $E_{b}(\theta,\varphi)$ distribution, the addition of the events is identical to the process described for the skyMap object, except that the sidereal time in degrees is subtracted from the right ascension. In essence this makes a sky map in right ascension and hour angle coordinates which represents what the sky looks like in coordinates local to the detector.

The local and sky maps are digitized in right ascension and declination to create a discrete symmetry in right ascension. Rotating the sky one right ascension sampling distance causes the sampling pattern on the sky to repeat, and this symmetry is used to efficiently build the expected background distribution from the local coordinate map. To exploit this symmetry, the sidereal exposure is digitized into the same number of bins as the number of sky map samples in right ascension. Thus every sidereal time bin corresponds to the right ascension of detector zenith at that time and is aligned with one of the sky map sampling positions. This allows equation \ref{DirectInt:Eq} to be integrated by summing over the sidereal time bins and adding the result to the background map.

\section{Identifying Signals}
\label{SignalIdent}

The backgroundMap object is the last major piece of the weighted analysis framework and is used for source identification. The key to source identification is to compare the signal and background maps, and calculate the probability that the observed signal could be produced by the background.  For this implementation of the weighted analysis technique this probability is given by equation \ref {Prob:Eq}:
\begin{equation}
\label{}
P(w\ge w_{obs}|g(w,N_{obs},\vec{k}))P(N\ge N_{obs}|N_{exp}).
\end{equation}
(see Source Identification, Chapter \ref{WeightedAnalysisTechnique:chap}).

The Poisson probability term can be calculated in a number of ways, and in this implementation a slightly modified algorithm from ``Numerical Recipes in C" \cite{Num_Recipes} is used to determine an iterative approximation. Calculating the integral Poisson probability is a relatively slow process and could be accelerated by using a large two dimensional lookup table, but for simplicity this implementation currently recalculates the Poisson probability for every position. The value calculated is actually $\text{log}_{10}$ the integral Poisson probability to speed computation and avoid the floating point errors that creep in with very small values.

\begin{figure}
\begin{center}
\includegraphics[width=5.75in]{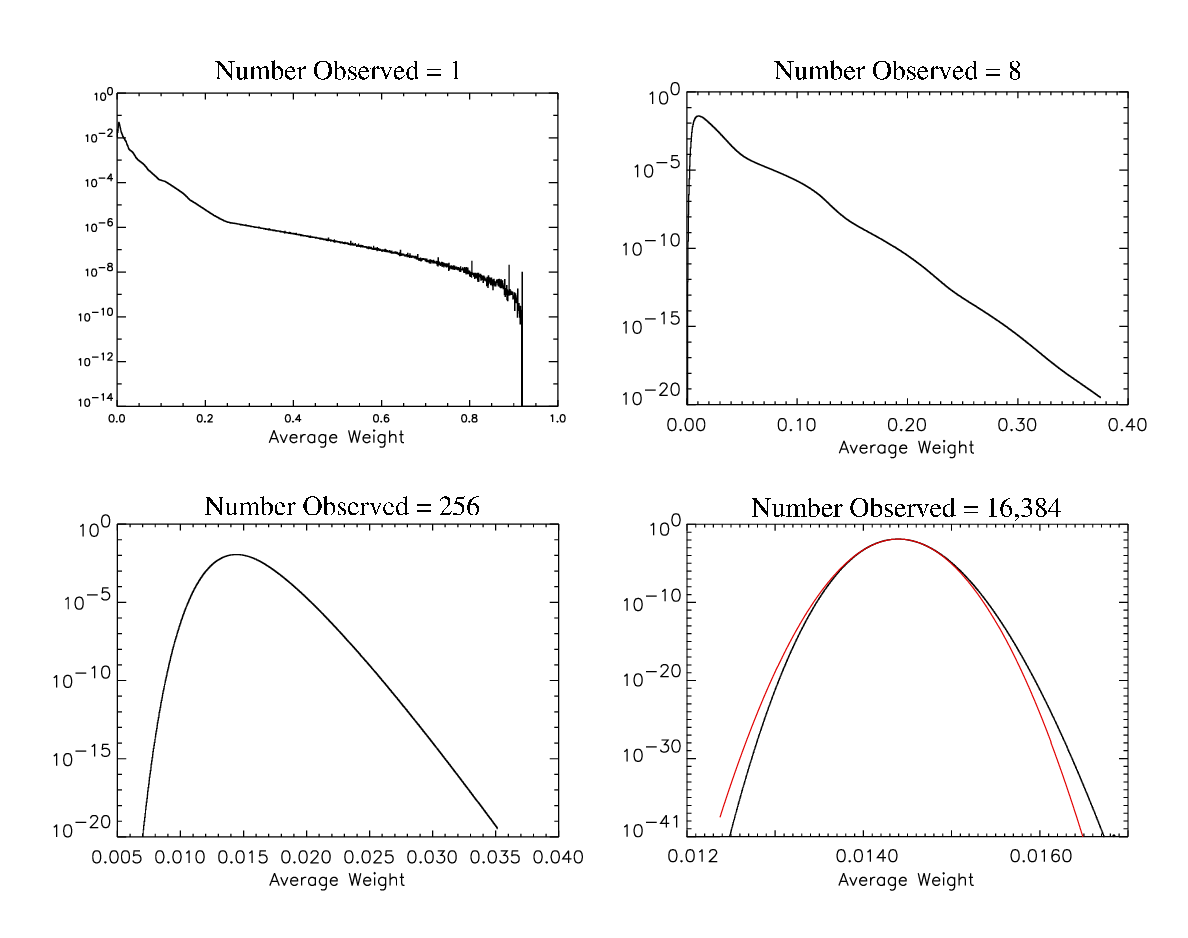}
\caption[Example Average Weight Probability Distributions]{These plots are examples of the $P(\bar{w}|g(w,N_{obs}))$ distributions for four selected $N_{obs}$ values from a representative set of background data. Note that both the horizontal and vertical scales change dramatically from one plot to the next as the distribution of $\bar{w}$ narrows, and the red line for the $N_{obs}= 16,384$ plot shows a Gaussian fit covering 41 orders of magnitude in probability.  The Gaussian approximation is not used until $N_{obs} > 30,000$ to allow the approximation to become more exact.}
\label{weightProb_Examples}
\end{center}
\end{figure}

The weight probability can be measured from the observed background distribution, though several subtleties must be considered. For a given spectrum of weights, the probability distribution of the average weight will narrow with increased $N_{obs}$.  As the number of events becomes very large the central limit theorem applies and the distribution becomes Gaussian distributed around the mean. However, for small $N_{obs}$ the probability distribution is far from Gaussian and must be calculated with care (see Figure \ref{weightProb_Examples} for example distributions). For Milagro the observed spectrum of weights is very constant over the field of view and steady in time.  This allows $P(w\!\ge\!w_{obs}|g(w,N_{obs}))$ to be calculated once and loaded into a lookup table that can be used for all zenith angles and observation times. The average weight is used instead of the total weight to simplify a number of computations in the code, though the two are statistically identical ($\bar{w} \equiv w/N_{obs}$ and $P(w\! \ge\! w_{obs}|g(w,N_{obs})) \equiv P(\bar{w}\!\ge\!\bar{w}_{obs}|g(w,N_{obs}))$ ).

A pair of auxiliary programs are used to measure the weight spectrum and calculate the  $P(\bar{w}\!\ge\!\bar{w}_{obs}|g(w,N_{obs}))$ distribution. The weight spectrum $g(w, N_{obs}=1)$ is measured using a modified version of the analysis and the SkyMapStat object. The SkyMapStat object is identical to the skyMap object discussed earlier in this chapter, but instead of adding weights to the sky map, records these weights in a histogram. Running over a few hours of data using the main analysis but with SkyMapStat substituted for skyMap produces an extremely accurate measurement of the weight spectrum used. This weight spectrum is proportional to the weight probability distribution for a single event ($P(\bar{w}|g(w,N_{obs}=1)$)). 

In order to convert this weight probability distribution into $P(\bar{w}|g(w,N_{obs}))$ for all $N_{obs}$ we need to iteratively convolve the observed spectrum. The process is exactly analogous to calculating the probability distribution for the sum of $N$ dice. The probability of throwing any given sum (or average) is the sum of all permutations which lead to this sum, weighted by the probability of each permutation. It is straightforward to calculate this distribution for successive values of $N$, and this is done in the probConvolution program. The probability that we are interested in is really the integral probability of seeing a weight greater than the weight observed ($P(\bar{w}\!\ge\!\bar{w}_{obs}|g(w,N_{obs}))$). The probConvolution program outputs files with the $\text{log}_{10}$ of the integral probability --- with one file for each $N_{obs}$ --- and these files are used to build a lookup table used by the backgroundMap object for source identification.

To save memory and computational time the $N_{obs}$ of $P(\bar{w}\!\ge\!\bar{w}_{obs}|g(w,N_{obs}))$ is sparsely sampled in regions where the probability distribution is changing slowly. In current usage, all $N_{obs}$ from 1-300 are calculated, then every 10th $N_{obs}$ from 300 -- 3,000 and every 100th $N_{obs}$ from 3,000 -- 30,000.  After $N_{obs}$ = 30,000 we have entered the Gaussian regime, and probabilities are determined from the measured width of the Gaussian distribution (see Figure \ref{weightProb_Examples} for example distributions).

The basic search algorithm compares the weight and number of events observed in the signal map to the expected values in the background map for each sampling location. The integral log probability of observing the signal weight is obtained from the internal lookup table, and the integral log Poisson probability of seeing $N\!\ge\!N_{obs}$ given the $N_{exp}$ in the background is determined using the Numerical Recipes algorithm. The sum of these log probabilities gives the total log probability of a background fluctuation producing the observed signal. 

One complication that arises for large $N_{obs}$ is that the shape of the probability distribution is very stable but the mean wanders slightly with changes in the detector. As the probability distributions become very narrow any error in the mean becomes significant. For this reason when $N_{obs}>300$ the weight observed in the background map is used as the true current mean of the distribution in order to correct for minor temporal or spatial variations in the mean.  This is accomplished by adding the difference between $w_{obs}$ and $w_{exp}$ to the mean $w$ from the lookup table to form $w'_{obs}$, which is insensitive to changes in the observed mean.  For $N_{obs}>300$, $w'_{obs}$ is used to determine the probability from the internal lookup table.

A technique that can be used to speed up the search process is sparse searching of the sky map. Because the sky map is densely sampled, any signal appears in several neighboring sample locations. The sky map can be searched on a sparse grid eliminating the large areas of the sky map where nothing of interest appears, with complete sampling of areas which show significant positive fluctuations. The backgroundMap object can be told what probability the user considers ``interesting."  It will search every third location in both right ascension and declination, then analyze all locations which are near a position within four orders of magnitude of ``interesting." This algorithm finds all significant signals while substantially increasing the speed of the search. The backgroundMap object will also search all locations upon user request.

\section{Examples}
\label{WATExamples}

Though the full performance of the analysis will be analyzed in Chapter \ref{Discussion:chap}, it would be nice to know that the weighted analysis framework is identifying signals as expected. To that end, I will briefly review the results of two analyses which have used this weighted analysis framework.

\begin{figure}
\begin{center}
\includegraphics[width=5.75in]{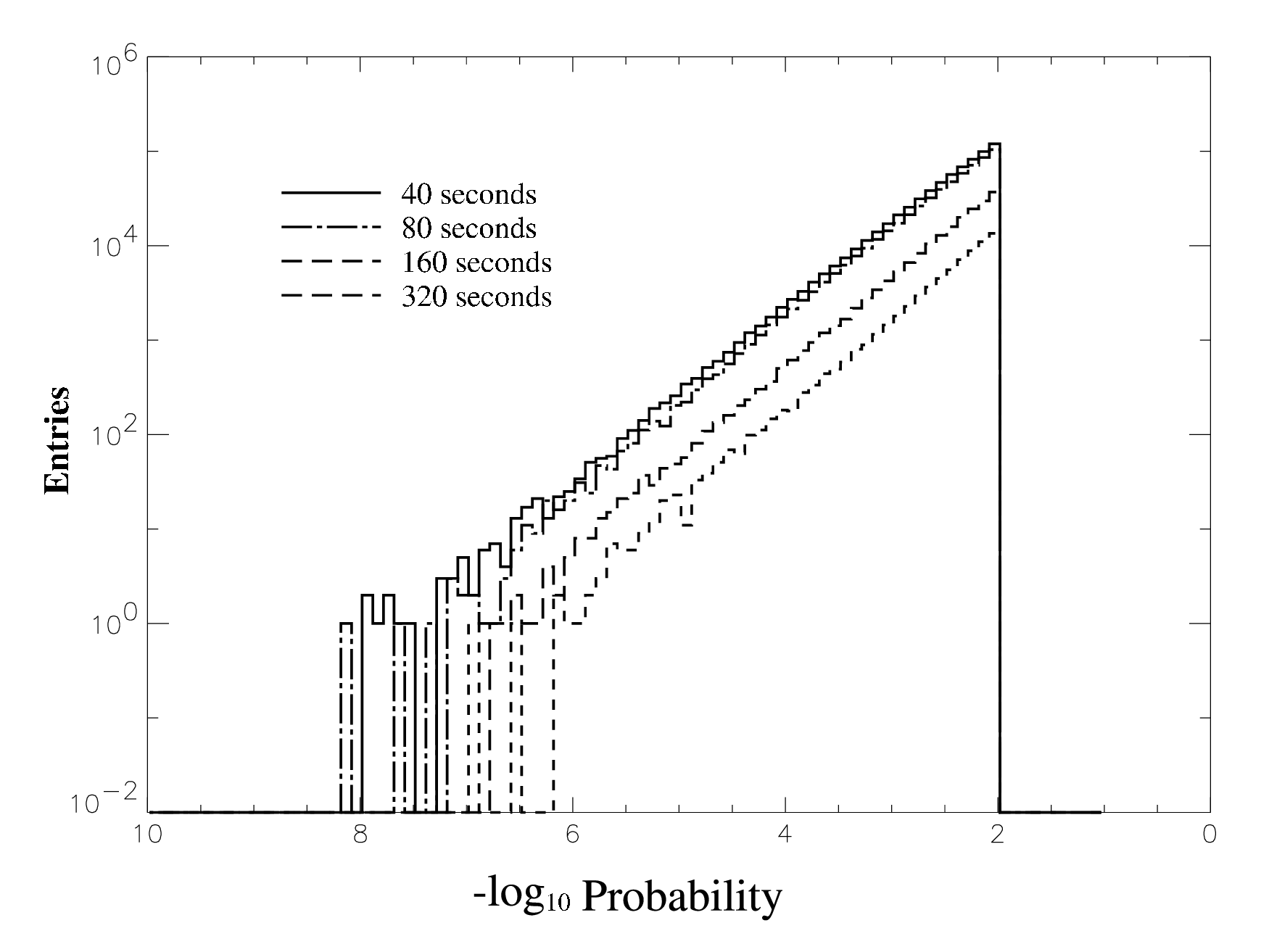}
\caption[Example Background Distributions]{Histogram of the probability distribution of background fluctuations observed in four time scales of the GRB search algorithm over $\sim16$ hours of searching.}
\label{backdist2}
\end{center}
\end{figure}

Figure \ref{backdist2} shows the results for the background fluctuations seen in four time scales of the GRB search algorithm over $\sim16$ hours of data. Note the expected power-law distribution of fluctuations and the similarity of the distributions for all four timescales. However the slope of the distributions is not 1, as expected but $\sim0.90$. The flattening of the probability distribution is caused by the approximations made in Equation \ref{Prob:Eq}, and can be easily corrected by multiplying the $\text{log}_{10}$ probability by the observed slope of the background distribution (see Appendix \ref{ProbAppendix} for details).

The background distributions in Figure \ref{backdist2} show that the probabilities are reasonably calculated, but a signal is required to test the sensitivity of the technique and assure ourselves that gamma-ray signals will be correctly identified. To this end an analysis for long duration sources using the weighted analysis framework was developed and used to analyze 2.5 months of data during the winter 2000 flare of Mrk421. This analysis used the $P_\gamma$ distributions developed for the GRB search (see Chapter \ref{WeightedAnalysisTechnique:chap}), and thus was optimized for short bright sources instead of the long relatively dim source analyzed here, and should be more sensitive if an optimized $P_\gamma$ distribution was used. Nevertheless, the weighted analysis detected a 4.3 sigma signal\footnote{The significance has been corrected for the approximations made in Equation \ref{Prob:Eq}.} at the position of Mrk421. This can be roughly compared to the standard optimal bin search which detects a 4.1 sigma signal when using a compactness cut optimized for DC searches (compactness $> 2.5$). Because of the disparate techniques, the two analyses are largely independent and the fluctuations should be weakly correlated. Thus the two analyses give surprisingly close agreement -- well within the 1 sigma errors -- and no conclusions can be drawn as to which analysis is more sensitive. However, we can conclude that this implementation of the weighted analysis technique is working as expected, and is sensitive to gamma-ray sources.

\chapter{Gamma-Ray Burst Search Procedure}
\label{GRBSearchProcedure:chap}

\section{Introduction}

The complete Milagro transient observation program consists of three separate analyses which are used to identify all TeV signals from 250 $\mu$s duration to steady state. On the short time scales Milagro is signal limited and the apparent motion of the sky becomes negligible. Andrew Smith has developed a very fast optimal bin analysis for the 250 $\mu$s -- 40 s region which uses no background rejection and stationary sky and background maps \cite{Andy:GRB}. The search described in this thesis uses the weighted analysis framework to identify moderate duration TeV signals of 40 seconds to 3 hours duration. In this middle region the background is comparable to the signal strength for a threshold detection and the apparent motion of the sky must be taken into account. For long time scales Milagro becomes background dominated and the science focus shifts from GRBs to flaring active galactic nuclei (AGN). Elizabeth Hayes is analyzing signals from 2 hours duration to steady state using an optimal bin search and aggressive background rejection techniques. All three searches overlap to ensure complete coverage from 250 $\mu$s to steady state and to allow cross-checking on unusual fluctuations.

\begin{figure}
\begin{center}
\includegraphics[height=7in]{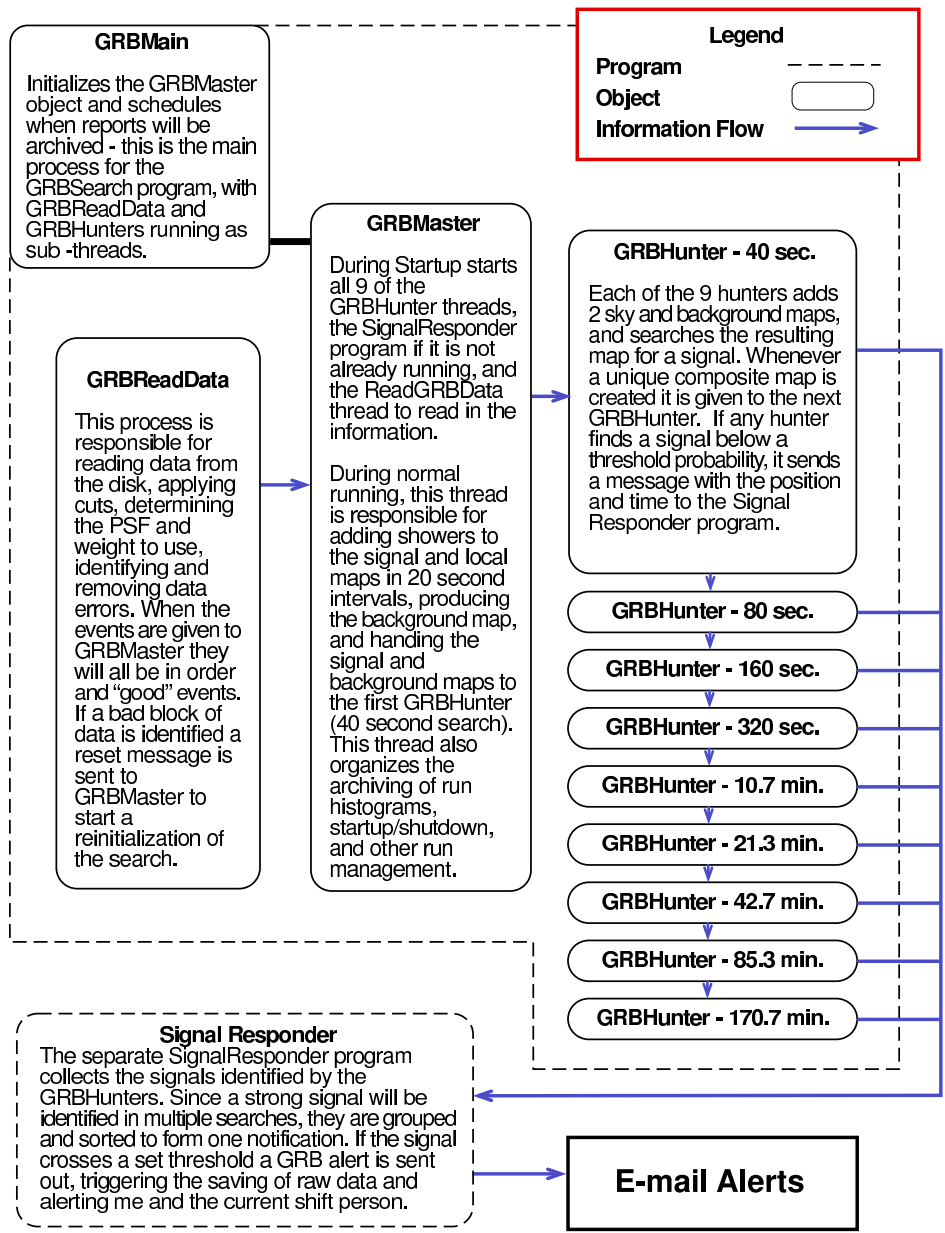}
\caption[GRB Search Diagram]{Conceptual diagram of the programs used to perform the 40 s -- 3 hour TeV transient search.  The dashed boxes surround the two independent programs, with the solid boxes indicating separate threads and the blue arrows showing the flow of information through the analysis.}
\label{GRBSearchDiagram}
\end{center}
\end{figure}

This chapter describes the 40 second to 3 hour search and the use of the weighted analysis technique to search for TeV transients. This search consists of a large number of independent pieces, each performing a specific task in the GRB search process. Performance was a principal concern in developing this analysis, and so there are several advanced computing techniques which are employed. Using the strength of the Objective-C language allows different programs to easily communicate with one another. This can speed computation by allowing two separate programs to work on different CPUs or even separate computers, while synchronizing the work they perform. A related idea is the concept of a programming ``thread.'' A thread acts just like an independent program, except that it shares the memory space with the parent program and any sibling threads. By sharing the memory space, look-up tables and other memory intensive storage can be easily shared between threads of the same program, even as the threads are performing their work on different CPUs within the computer. In this analysis there are two independent programs, and including the threads, twelve separate execution loops all working simultaneously through the data.  
Figure \ref{GRBSearchDiagram} shows a schematic layout for the search programs, which divide into four functional parts. The first thread to analyze the data is ReadGRBData which is responsible for reading the data as it becomes available; determining the PSF region and weight for each shower; checking, cleaning and applying cuts to the data; and then giving it to the GRBMaster thread in 20 s blocks.  The GRBMaster is responsible for maintaining the local map, managing the starting/stopping and reporting of the other threads in the program (thus the GRBMaster name), and building 20 second sky and background maps which are given to the third step in the analysis --- the chain of GRBHunter threads. The GRBHunter threads form long duration maps by adding together shorter maps before searching for transient signals, with each consecutive thread looking for transients of successively longer duration. If any of the GRBHunter threads identifies a candidate signal, the position and time of the signal is sent to the separate SignalResponder program for final processing. The SignalResponder compiles the candidate signals and manages the e-mail reporting process. In each section of this chapter I detail one of the four pieces of the analysis chain and the duties and design of that stage of the analysis.

\section{Data Reading and Event Selection}
\label{EventSelection}

Because this is a real-time analysis with rapid notification, particular care must be given to event selection and identifying unusual detector conditions. In many archival analyses, times when the detector was working poorly or in an unusual state (calibration or engineering modes, equipment failures, DAQ bugs, {\em etc.}) are filtered out before the analysis is performed. However, to identify and report transients as rapidly as possible we need to filter out these conditions as they occur. As my office mate Lowell Boone says, ``Computers combine absolute stupidity with absolute patience," and are perfectly content to send thousands of GRB alerts if defective data is handed in. The main jobs of the ReadGRBData thread are to read the data, identify defective data, and prepare the events for further processing.

A significant upgrade of the Milagro DAQ system is planned, and the data reading routine has been designed with this upgrade in mind.  Currently data is written to disk in $\sim$4 minute segments called subruns.  Once a subrun data file is complete, it is copied from the DAQ computer to the online analysis cluster, and a link to the subrun is put in the data folder of each of the online analyses. ReadGRBData polls the data folder for the 40 second -- 3 hour search, and identifies and starts analyzing a subrun within 30 seconds of it being copied to the analysis cluster. In total there is a typical delay of 2 -- 5 minutes until the data is analyzed online. After the DAQ computer is replaced we plan to go to a port system where the data is passed directly to the online analyses through the network, cutting this latency to under one second. Though this cannot be fully implemented until the DAQ changeover is complete, changing a few lines of code will enable ReadGRBData to pull the data from a port instead of a file.

Once the data is read in, it passes through two tiers of quality filters. The first set of filters analyzes individual showers and is designed to provide a uniform set of events while removing showers with obvious defects. The second set of filters looks at groups of events, and is designed to identify more complex problems such as bad DAQ reads and unstable detector modes.

The event level filters select showers with zenith angles less than 45 degrees, no GPS read errors, and a multiplicity of  $\ge$53 tubes in the air shower layer. In addition the PSF region and $P_\gamma$ values are determined, and region 12 events (very wide PSF) and showers with a $P_\gamma < 0.5$ are removed. The $P_\gamma$ cut serves only to reduce the computational load by removing events which are almost definitely background events. All of the event cuts must be included when the weight spectrum is calculated, because they have a significant impact on the weights used by the analysis (see Section \ref{SignalIdent}).

The second tier of event filters is significantly more complex, and tries to catch some of the odd errors that are seen in the data. These filters work by analyzing the global properties of a group of showers, such as the event rate and time order. To catch errors that happen to coincide with the 20 second block boundaries (such as data gaps, out of order events, {\em etc.}), the initial analysis is performed on approximately 80 seconds worth of data in a revolving buffer.  Once the events in the $\sim$80 second buffer have been checked for time errors, 20 seconds of the oldest events are removed to form a data block and more data is read in to refill the 80 second buffer.

Though the frequency of problems has dropped with successive DAQ software upgrades, there have been times when events have had erroneous times, are out of time order, or repeated. To try and filter out these errors it is assumed that most of the showers are good, and that the errors will typically be large. All the showers in the $\sim$80 second buffer are sorted into time order, and any events with duplicate times or times farther than 60 seconds from the median are removed. This  removes events which have gross time errors, while ignoring possible time errors of a few tens of seconds.

Most major problems with the detector can be identified by an unusual event rate or sudden step in the event rate. After sorting and removing the outlying events, we have a nice clean set of events covering an 80 second interval. Now the event rate can be accurately calculated without fear of contamination from defective events. After a 20-second data block is formed, the rate for the data block is calculated and used in the final set of filters.

The first cut in this filter set requires the event rate within a 20-second data block to be within the 1200 -- 2200 showers/second range typical of normal operation.  This filter is fairly loose to allow for the large but slow rate fluctuations caused by variations in atmospheric pressure, snow and rain accumulation, or the formation of reflective ice under the Milagro cover.  Despite the wide range of accepted rates, this cut is quite effective at filtering out many unusual problems, such as test runs or low voltage power supply failures.  Since small data dropouts of a few seconds duration are fairly common, one data block with a bad rate only causes the 80-second data buffer to be discarded.  If 10 or more bad blocks occur in succession, we have definitely entered a bad data taking mode and a full reset of the entire GRB search is performed, including all sky maps and background information.  

Most detector changes are not severe enough to fall outside of the rate window used above, but instead cause a step in the event rate.  These steps are usually due to a sudden change in the detector configuration --- such as the loss of a patch of 16 tubes\footnote{Typically this is caused by water leaking into the connector between the photomultiplier base and the RG-59 cable on one tube and tripping the high voltage for that patch of 16 tubes.  Depending on the time of day, it is usually a few hours until the bad tube can be disconnected and the high voltage reset.} --- and since the detector configuration has changed the background distributions need to be updated. To identify these steps, the average rate observed during the last seven 20-second data blocks is compared to the data rate in the current 20-second data block, and a deviation greater than 7.5$\sigma$ triggers a full reset of the GRB search.  The bar for a reset is set as high as it is to make accidental resets rare, since a full reset requires rebuilding the background distributions and the long timescale sky maps.

\section{Building Sky and Background Maps}

The GRBMaster thread is responsible for building 20-second sky and background maps from the data blocks generated by ReadGRBData (see Figure \ref{GRBSearchDiagram}). This is in many ways the heart of the analysis, with the sky and background maps produced here being handed to the GRBHunter chain for source identification.

One of the most important jobs of the GRBMaster thread is to manage the local map representation which is used in determining the background map. The local map must contain enough data to accurately represent the local event distribution, but have a duration short enough to allow for slow variations in the distribution (see Section \ref{backgroundMap}). The first 10 minutes of data is used to initialize the local map. Because the local map is needed to generate the background maps, we cannot effectively analyze data until the local map is formed, and 10 minutes is deemed the minimum amount of data needed to make a reasonable determination of the background event distribution.  As data continues to come in it is added to the local map until 2 hours of data has accumulated. At this point we need to limit the integration time of the local map so that it can respond to slow changes the in the experiment. After 2 hours, the local map exposure is kept constant by removing showers more than 2 hours old while continuing to add the new events.  

On initial startup of the GRB search, or after a reset, the first 10 minutes of data is used to build up an initial local map, and no sky or background maps are generated. After this initialization period, each 20-second data block is used to generate a new sky map and to update the local map.  The current local map is then used to generate a background map to match the sky map, and the 20-second sky and background maps are given to the 40-second GRB hunter to search for excesses.  As more data comes in the local map representation improves until 2 hours of data has accumulated, at which point the 2-hour-old data block is subtracted for each new data block which is added.

Ideally, the local map exposure would be centered around the signal exposure to allow linear changes in the local event distribution to average out.  However, in the GRB search we need to analyze the data immediately to generate prompt GRB notifications.  The GRBMaster local map contains the past 2 hours of data, so the local map exposure is not centered around the signal map exposure and linear changes in the local event distribution do not average out. In practice the detector changes are small enough that this has not been an issue, however, a follow-up analysis would almost certainly want to use centered local maps to limit this effect.

\section{Searching For Transient TeV Signals}

The sky and background maps generated by the GRBMaster thread are given to a chain of GRBHunters as shown in Figure \ref{GRBSearchDiagram}. The GRBHunter threads are arranged in series, with the first hunter searching for signals on 40-second time scales, and each of the subsequent 8 GRBHunters analyzing a time scale exactly twice as long (80 s, 160 s, 320 s, ...).  Sky and background maps of the appropriate exposure length are created by adding together shorter duration maps, then searching the composite maps for a signal. When a signal is identified by a hunter, the position, time, and duration of the candidate signal is sent to the SignalResponder program, which handles the GRB notification process. 

The first task of a hunter is to create a composite map of the appropriate time scale.  The maps handed in to a GRBHunter are half the length of the timescale to be searched, so 2 sky maps and 2 background maps are stored in first-in first-out (FIFO) arrays, with one array each for the sky and background maps.  The more recent maps (\#2) are then added to the older maps (\#1), so that the older maps now contain twice the exposure.  These composite maps --- one signal and one background --- are searched by the hunter to identify candidate signals, then handed to the next GRBHunter in the analysis chain and removed from the FIFO arrays (the \#2 maps become the new \#1 maps), and the process is repeated.  

This chain is designed to allow efficient summing of maps and an oversampling of 2 at each time scale. Note that each incoming map is used in two separate source searches (once as map \#2, then as map \#1), so the time window is shifted by half the search window on every iteration. Additionally, the maps given to the next GRBHunter in the chain are twice as long as the incoming maps --- in essence a GRBHunter makes use of the map summations created by all the previous hunters in the chain.  

One subtlety is in the passing of maps from one hunter to the next. A GRBHunter is expecting the input maps to be contiguous in exposure, but the composite maps created by two consecutive GRBHunter searches are not contiguous, but instead overlap in time because of the oversampling.  Hence, every other map is passed to the next GRBHunter to ensure that the composite maps received by the next hunter are independent and contiguous.

The actual work of identifying signals is performed by the sparse search method of the backgroundMap object as outlined in Section \ref{SignalIdent}. In addition, each hunter keeps a probability histogram of all locations with probability less than $10^{-2}$.  These histograms are snapshots of the observed background fluctuations (see Figure \ref{backdist2}), and help monitor the stability of the search and define the significance of any detected transients. The probability histograms are archived daily at 6 a.m. UT and whenever a full reset of the GRB search occurs.

The spacing of the time scales and the oversampling was chosen as a compromise between sensitivity and computational demands. It has been shown by Biller et al. \cite*{Biller:X3} that any search with a time scale spacing of 3 (40 s, 120 s, ...) or less, and an oversampling of 2 (50\% overlap) or more approaches the sensitivity of a massively oversampled search.  The algorithms used in the GRBHunters can perform searches on any time or oversampling scale, and the factors of 2 were chosen to give good sensitivity while fitting within the CPU and memory constraints.

\section{Generating GRB Alerts}

The oversampling in both duration and time means that a strong signal (or unlikely fluctuation) will be identified in several time windows by multiple GRBHunters. This generates a flurry of identifications for each signal, and the job of the SignalResponder program is to collect these identifications and manage the e-mail alert process.

The collection of signal identifications is performed by grouping events in time and position.  Two candidate events are considered members of the same group if they are separated by less than 9 degrees and their start times are closer than three times the duration of the longer event.  When an candidate position comes in, its position and time is compared to a table of active groups.  If it qualifies as a member of one of the active groups, it is added to the group, otherwise it becomes the first member of a new group.

All the candidate positions in a group are considered multiple detections of the same underlying signal, and are reported together. Determining when a group is complete --- and a composite alert should be sent --- can be tricky. Currently the closing time for a group is calculated by selecting the most significant event in the group and multiplying the signal duration by 10 and adding the start time.   If a more significant event is later added to the group, the closing time is recalculated and more events may be added.  After the closing time, the most significant detection --- presumably the true signal location/duration --- is used to generate the e-mail alert.

Once the most significant event in has been identified, a series of filters are used to determine whether an alert should be generated, and the type of alert. If the probability of seeing a random fluctuation of greater significance is less than $\sim$20 per year for the observed time interval, an e-mail alert is sent to me.  I receive approximately 1 of these e-mails per day (there are 9 independent time scales), and they help monitor the performance of the search program but are assumed to be background fluctuations. 

If the probability of seeing a random fluctuation of greater significance is less than $\sim$2 per year for the observed time interval, an e-mail alert is sent to burst@kahuna.lanl.gov.  The alert details the position, time, duration, and probability of the signal, and how many searches it was identified by.  This e-mail sets in motion a series of actions, including paging the shift person and archiving the raw data surrounding the candidate signal.  For particularly strong candidates the shift person, in consultation with the Milagro collaboration, may issue a full GCN \cite*{GCN} alert and ask for target-of-opportunity observations by the rapid x-ray timing explorer (RXTE) satellite. 

In many ways the SignalResponder program is a work in progress, and is expected to change as we gain experience with the online search.  Currently the e-mail alert system is very conservative, and only generates a few e-mail alerts which are scrutinized by the collaboration.  None has yet been forwarded to the GCN network. Once we have seen at least one signal and have confidence in the search program, new e-mail alerts with faster or automated reporting may be implemented, but for now we want to minimize the chance of sending false alerts.

\section{Summary}

Logically, the search for 40 second -- 3 hour transients is divided into 4 distinct pieces: selecting events and identifying changes in the detector, managing the background calculation and creating sky and background maps, adding maps and searching for signals, and finally collecting and managing e-mail alerts for the candidate events.  The current GRB search is able to process events five times faster than the current data rate on one computer with dual 866 MHz Pentium III$^{\rm TM}$ processors and 1 GByte of memory. The ability to analyze data faster than it is created allows for efficient processing of archival data and quick failure recovery. When working on archival data (where we don't have to wait for data to appear), the CPU usage is dominated by building the sky and background maps in the GRBMaster routine and takes all of one processor ($>90\%$), followed by event selection in ReadGRBData ($\sim30\%$ of one processor), and the GRBHunters ($\sim10\%$ for the 40 second search, with each subsequent search dropping by a factor of 2 in CPU usage). In many ways the analysis is limited by memory usage, which averages $\sim700$ megabytes, and there is evidence that the processing speed is limited by the memory bandwidth and not the CPU speed.

\chapter{Results}

\section{Introduction}

The Milagro data taken between May $2^\text{nd}$, 2001 and May $22^\text{nd}$, 2002 was searched for unidentified transients of 40 s to 3 hours duration.  Since January 2002 the analysis has been operating in real time with the capability of rapidly alerting the community to any observed TeV gamma-ray bursts.  No evidence for TeV emission of 40 s to 3 hours duration has yet been observed.  This chapter details the history of the online analysis and presents the observed probability distributions.

\section{History of the 40 s -- 3 hour Online Search}

The GRB search program analyzed all of the Milagro data taken between May $2^\text{nd}$, 2001 and May $22^\text{nd}$, 2002 for transients of 40 s to 3 hours duration.\footnote{Dates and times used in the text are local to New Mexico.} The search interval of slightly over one year was chosen to coincide with significant changes in Milagro's data acquisition system.  Before the $2^\text{nd}$ of May, 2001, the reduced chi-squared from the angle reconstruction was not recorded due to an error in the data archiving code.   Since the reduced chi-squared value is crucial in determining the PSF of an event, data before this date was excluded from this study.  On the $22^\text{nd}$ of May, 2002, the Silicon Graphics Challenge data acquisition computer was replaced with a cluster of Linux machines.  Though this is a positive change for the Milagro experiment, the transition was not trouble free and provided a convenient date to close the data analysis window for this thesis.

Though this thesis covers more than a year of Milagro data, the real-time analysis code was not operational until the fall of 2001.  On the $17^\text{th}$ of October, 2001, the online analysis was started in an engineering mode searching only three of the nine time scales.  Between the middle of October and early January the code remained in an engineering mode, with numerous bug fixes and improvements.  Some of the highlights were:

\begin{itemize}
\item October $19^\text{th}$ -- Running with all 9 time scales.

\item November $30^\text{th}$ -- Using crontab, search is online all the time.  Numerous code improvements.  Real-time filtering of odd experiment states much improved.

\item December $12^\text{th}$ -- Automatic email reporting of significant events enabled.  After this date any observed transients save the raw data, page the shift person, and set in motion possible notification of the GRB community of any observed events. Stability of the search program is still an issue.

\item January $8^\text{th}$ -- After hunting down a Linux kernel problem and an error in the Objective-C standard library the online search becomes stable --- I didn't realize how hard this program would push modern software.  From this date forward all but one death is due to power outages or Linux networking lockups, and the transient search program is extremely stable.

\item January $11^\text{th}$ -- Update email reporting, code officially leaves engineering mode. 

\end{itemize}

After the $11^\text{th}$ of January, 2002, the online GRB search is officially in data mode.  Over the next few months there are a few small updates to correct minor problems, with all changes being tested offline before they are introduced to the online analysis.  On the $30^\text{th}$ of January these small changes  culminate in the release of version 1.0 of the online search.

On the $1^\text{st}$ of April, 2002, a final code update was applied to keep track of the total exposure in each time scale. April $4^\text{th}$ marks the last bug fix to the GRB search code:  if the search remained running longer than $\sim$1 month an integer overflow occurred in a counter.  The GRB search has been running online continuously without modification since this date.\footnote{The only intervention since April $4^\text{th}$ has consisted of rebooting the online computer approximately once per month to correct Linux networking lockups associated with NFS disk mounting.}

\section{Probability Distributions}

The data for this thesis separates into two sets: the ``online" data which was analyzed in real time at the site after the $1^\text{st}$ of April, 2002, and the ``offline" data which was analyzed on computers at the University of Maryland several months after its collection.  Though the online analysis was stable long before April $1^\text{st}$, 2002, this date marks when the analysis began to archive the total exposure for each time scale.  While unimportant for identifying TeV signals, the total exposure is needed to determine upper limits (see Chapter \ref{Discussion:chap}).

The raw probability of the background producing a candidate signal is given by Equation \ref{Prob:Eq}.  Because of the approximation made in deriving Equation \ref{Prob:Eq}, the slope of a probability histogram on a log-log plot is slightly shallower than 1 (see Section \ref{WATSensitivity}).  Instead of correcting for this effect, the raw probability histograms are presented to show the actual output of the GRB search. 

An additional complication is introduced by the variable search density used in the analysis.  The GRB search program performs a sparse search of nearly independent locations over the full sky, and only searches all locations near large candidate signals (see Section \ref{SignalIdent}).  The raw probability histograms contain only the nearly independent locations from the sparse search with a raw probability less than $10^{-2}$.  These histograms are written to disk at 6 a.m. universal time --- or whenever a significant change in the detector is observed --- and characterize the daily performance of the analysis.\footnote{Since the analysis rate is $\sim$7 times faster than the typical data rate on the computer used for the offline search, the offline search can include up to one week of data in a daily histogram set.}

\begin{figure}
\begin{center}
\includegraphics[width=5.75in]{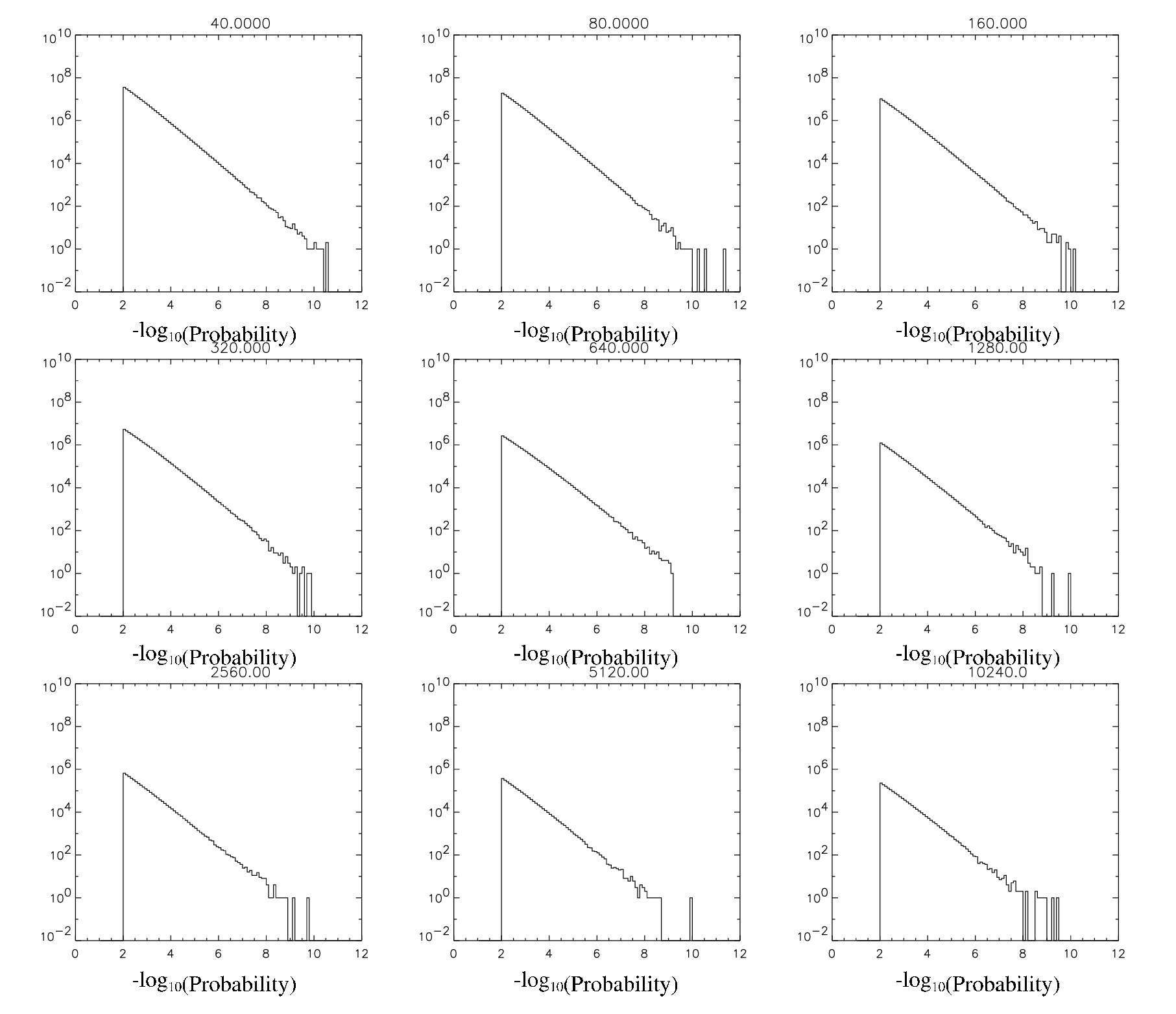}
\caption[Online Background Distributions]{Distribution of raw probabilities less than $10^{-2}$ observed in the online search data set for each of the nine time scales.  The duration of the search window in seconds is listed above each plot.}
\label{onlineResult}
\end{center}
\end{figure}

The online data covers the period from April $1^\text{st}$, 2002 to May $22^\text{nd}$, 2002.  The raw probability distributions are shown in Figure \ref{onlineResult} and the cumulative exposure in Table \ref{onlineExposure}.  All the data between April $1^\text{st}$, 2002 to May $22^\text{nd}$, 2002 is included, with no data cuts.  

\begin{table}
  \centering
\begin{tabular}{|c|c|}
\hline
   Search Duration (s) &  Total Exposure (days) \\ \hline
   40 s & 41.2  \\ \hline
   80 s & 41.2 \\ \hline
   160 s & 41.1 \\ \hline
   320 s & 41.1 \\ \hline
   640 s & 41.0 \\ \hline
   1280 s & 40.9 \\ \hline
   2560 s & 40.6 \\ \hline
   5124 s & 40.1 \\ \hline
   10240 s & 39.1 \\ \hline
\end{tabular}
  \caption[Total Online Exposure]{Total exposure in days for the online data set for each of the nine time scales.}
\label{onlineExposure}
\end{table}

The offline data covers the period from May $2^\text{nd}$, 2001 to March $1^\text{st}$, 2002.  The data from March of 2002 was not included because of an upgrade of the Milagro trigger which shifted the weight distributions.  The March data was analyzed online and no unusual events were identified,  however, this was before the archiving of cumulative exposure was implemented.  Consequently the March data was analyzed, but is not included in the online or offline data sets presented here.\footnote{The March data could be reanalyzed offline by carefully piecing together the proper weight distributions.  This re-analysis was not performed due to time constraints and the minimal effect of the extra exposure on the upper limit calculations.}

\begin{figure}
\begin{center}
\includegraphics[width=5.75in]{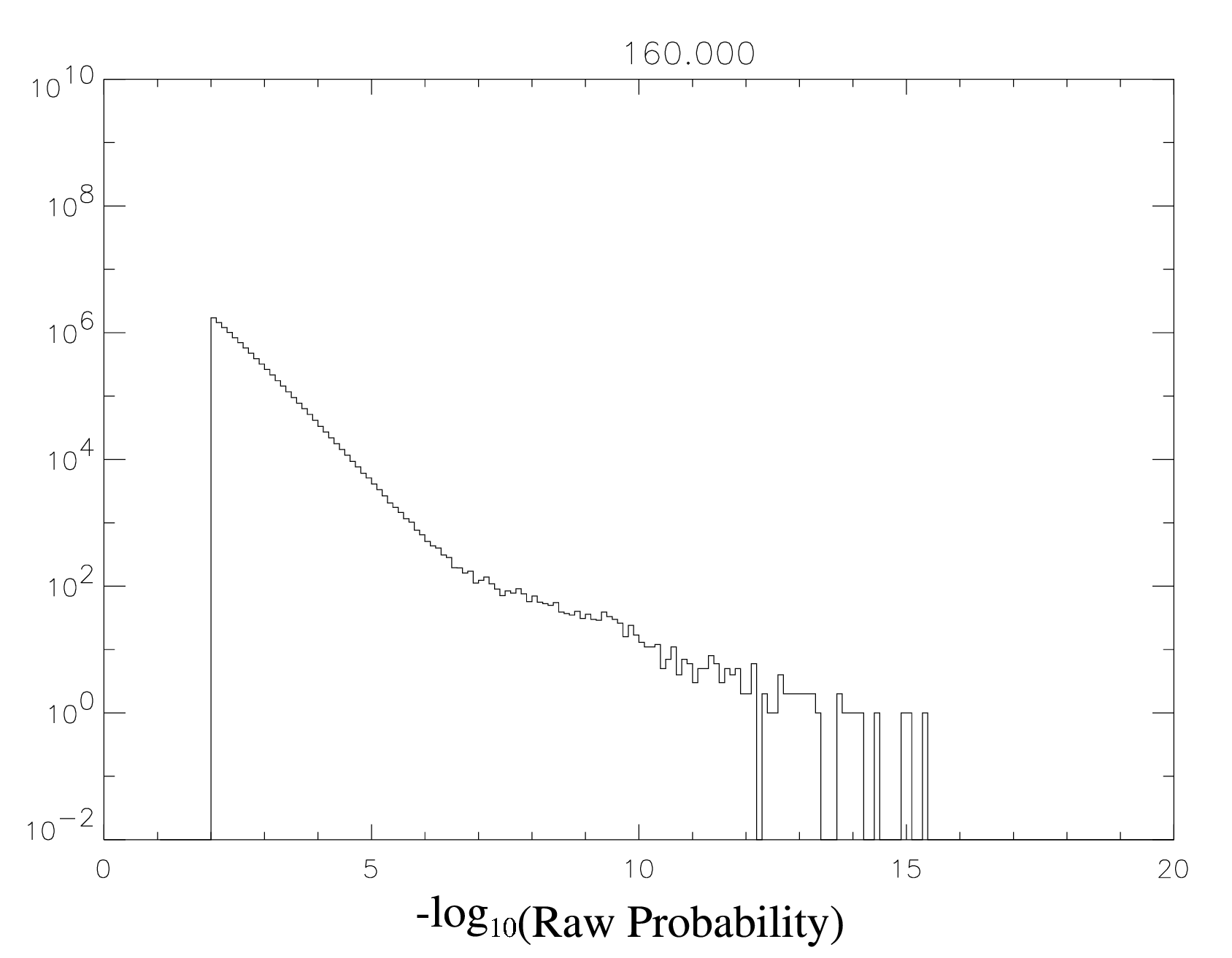}
\caption[Example of Poor Probability Distribution]{An example of a poor distribution of raw probabilities from the 160 s search in early October 2001 (part of the offline data set).  When distributions like this occur, many false alerts from different locations in the sky are generated.}
\label{badDistEx}
\end{center}
\end{figure}

Unlike the online data set, there are several offline probability distributions which were corrupted by unusual detector states missed by the on-the-fly data quality cuts (see Section \ref{EventSelection}).  These corrupted runs are easily identified by eye, and have a characteristic shallow tail of low probability events (see Figure \ref{badDistEx}).  After analyzing all of the offline distributions, nine data sets covering six time periods were deemed to be of ``poor" quality based on the raw probability distributions and removed.  Table \ref{ThrowDist} lists the time periods removed, with a brief description of the reason.

\begin{table}
  \centering
\begin{tabular}{|c|c|c|}
\hline
   $\sim$End Date & Duration & Reason\\ \hline
   5/30/2001 & 0.3 days &  2560 s and longer time scales had poor distributions \\ \hline
   6/13/2001 & 0.3 days &  640 s and longer time scales had poor distributions \\ \hline
   9/23/2001 & 3.6 days &  160 s to 2560 s scales had poor distributions \\ \hline
   10/09/2001 & 9.4 days &  all time scales had very poor distributions \\ \hline
   10/24/2001 & 0.6 days &  640 s to 2560 s scales had poor distributions \\ \hline
   12/15/2001 & 12.4 days &  4 long alerts from different locations \\ \hline
\end{tabular}
  \caption[Table of Removed Offline Sets]{This table lists the time periods which were deemed to have ``poor" probability distributions and removed, and a brief description of why.  Nine files covering six time periods were removed.}
\label{ThrowDist}
\end{table}

\begin{figure}
\begin{center}
\includegraphics[width=5.75in]{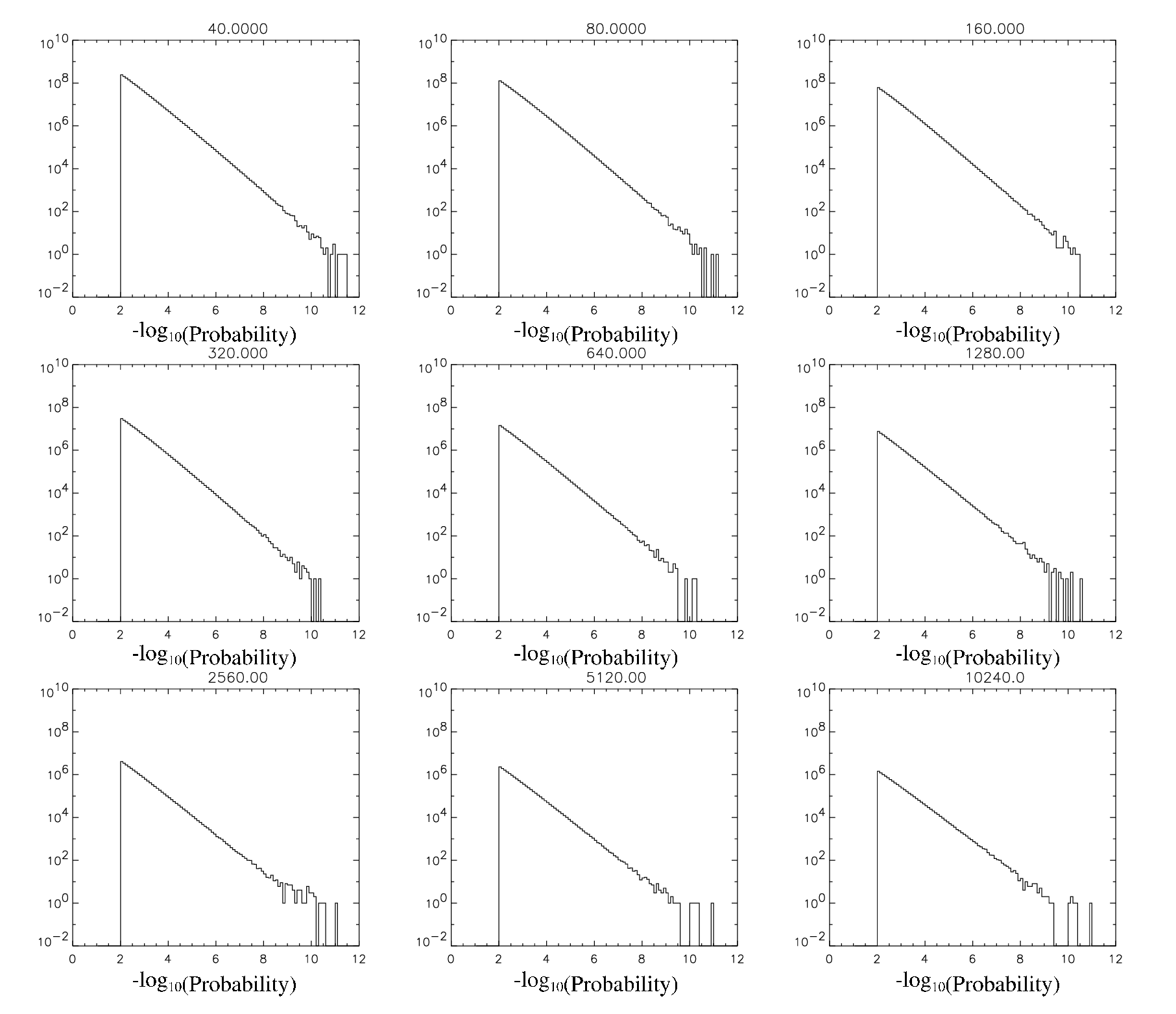}
\caption[Offline Background Distributions]{Distribution of raw probabilities below $10^{-2}$ for the offline May $2^\text{nd}$, 2001 to March $1^\text{st}$, 2002 data.  The duration of the search window in seconds is listed above each plot.}
\label{offlineResults}
\end{center}
\end{figure}

\begin{figure}
\begin{center}
\includegraphics[width=5.75in]{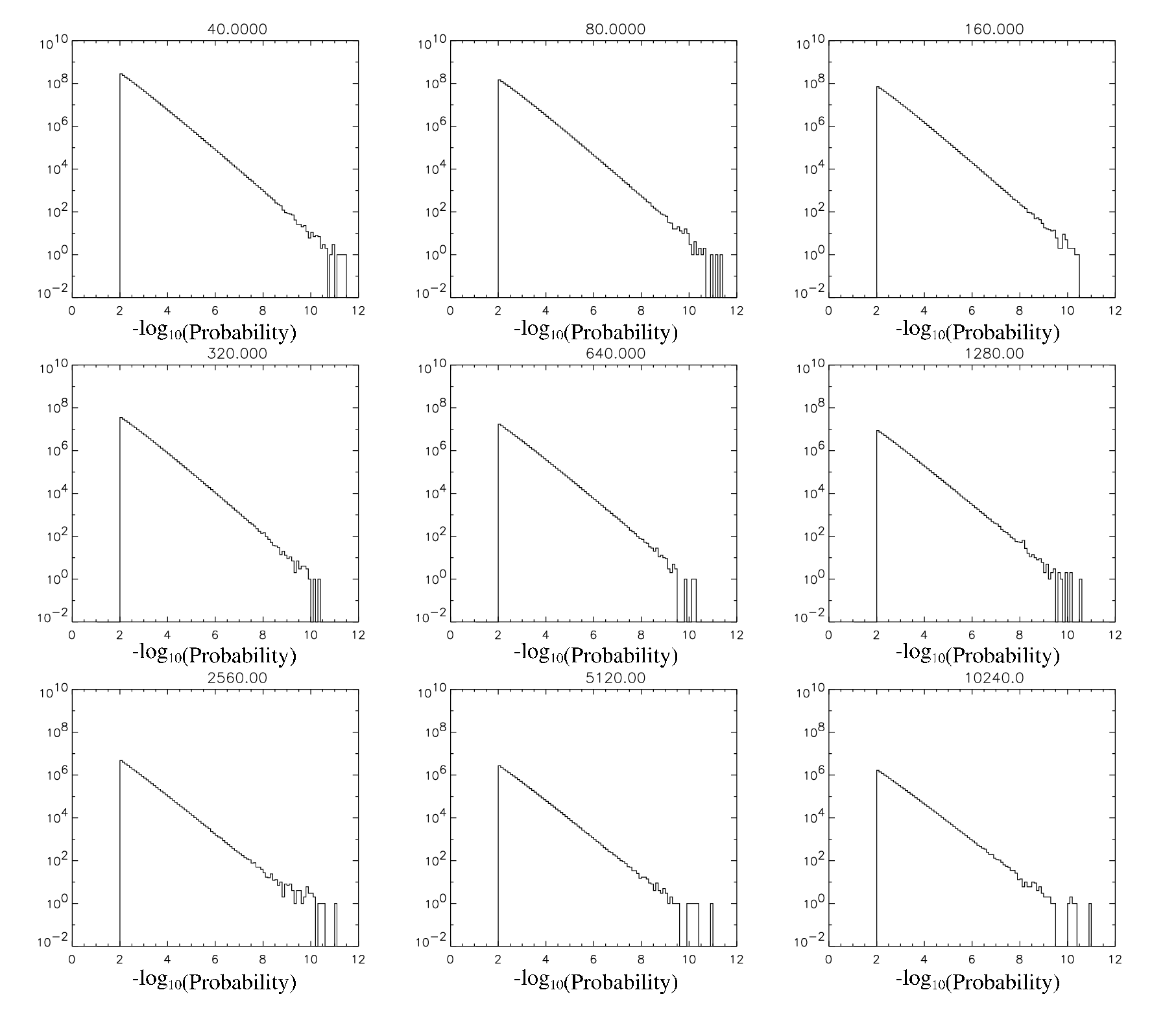}
\caption[Total Background Distributions]{Total distributions of raw probabilities below $10^{-2}$ for the combined online and offline data sets.  The duration of the search window in seconds is listed above each plot.}
\label{totalResult}
\end{center}
\end{figure}

The raw probability distributions for the offline data are shown in Figure \ref{offlineResults} with a list of the offline exposures in Table \ref{offlineExposure}.  The total exposure can be formed by combining the online and offline data sets. Figure \ref{totalResult} shows the combined online and offline raw probability distributions, with the total exposure for each time scale listed in Table \ref{totalExposure}.

\begin{table}
  \centering
\begin{tabular}{|c|c|}
\hline
   Search Duration (s) &  Total Exposure (days) \\ \hline
   40 s & 249.1  \\ \hline
   80 s & 249.0 \\ \hline
   160 s & 248.9 \\ \hline
   320 s & 248.6 \\ \hline
   640 s & 248.2 \\ \hline
   1280 s & 247.3 \\ \hline
   2560 s & 245.5 \\ \hline
   5124 s & 242.1 \\ \hline
   10240 s & 236.6 \\ \hline
\end{tabular}
  \caption[Total Offline Exposure]{Total exposure in days for the offline data set for each of the nine time scales.  The data sets listed in Table \ref{ThrowDist} are not included.}
\label{offlineExposure}
\end{table}

\begin{table}
  \centering
\begin{tabular}{|c|c|}
\hline
   Search Duration (s) &  Total Exposure (days) \\ \hline
   40 s & 290.2  \\ \hline
   80 s & 290.2 \\ \hline
   160 s & 290.0 \\ \hline
   320 s & 289.7 \\ \hline
   640 s & 289.2 \\ \hline
   1280 s & 288.1 \\ \hline
   2560 s & 286.1 \\ \hline
   5124 s & 282.2 \\ \hline
   10240 s & 275.7 \\ \hline
\end{tabular}
  \caption[Total Exposure Time]{Total exposure for the combined online and offline data sets.}
\label{totalExposure}
\end{table}

\section{Results}

No evidence for 40 s to 3 hour transient TeV emission was found in the Milagro data taken between  May $2^\text{nd}$ 2001 and May $22^\text{nd}$ 2002.  The raw probability distributions shown in Figure \ref{totalResult} cover $\sim$290 days of observation in the northern sky, with no signals below a probability of $10^{-12}$ being observed.  These observations are entirely consistent with no transient TeV emission of 40 s to 3 hours duration.

\chapter{Discussion}
\label{Discussion:chap}

\section{Introduction}

While one always hopes for a detection, the lack of a signal can be just as important scientifically.  Two sets of limits --- one for the observed radiation and one for the emitted radiation --- are calculated because of the attenuation of very high energy gamma rays by extragalactic background light.  The observed emission limits are model independent but difficult to compare with theoretical calculations, whereas the limits on emitted radiation can be directly compared to theory but depend on the predicted redshift distribution of GRBs and other model parameters. In this chapter I describe how to calculate upper limits using the weighted analysis technique, and then determine upper limits for both observed and emitted TeV emission from gamma-ray bursts. 

\section{Generating Limits with the Weighted Analysis Technique}
\label{WATLimits}

Although the weighted analysis technique should increase the sensitivity of the TeV transient search, it also complicates the calculation of upper limits.  This section describes how to determine the response of the Milagro detector using the weighted analysis technique, with the following sections convolving this detector response with various spectra to calculate specific upper limits.

In the weighted analysis technique, the response of the detector and analysis chain to an incident photon is represented by the photon probability density at the source location ( $p(\vec{k_s})$ where $\vec{k_s}$ is the position of the source, see Section \ref{WATSkymap}).  For Milagro, the detector response is a strong function of the zenith angle and energy of the initiating particle, but is azimuthally symmetric and constant in time.\footnote{There are small changes in the detector response with time, particularly with diurnal freezing and melting of surface water during the winter.  However, these have a minimal effect on the event reconstruction, and no diurnal changes in the $p(\vec{k})$ values have been observed.  Similarly, the very weak diurnal ``breathing" of Milagro's zenith angle response is unimportant in this analysis \cite{MilagroAsymmetry}.  Because the trials factors associated with the transient searches are so large, a fairly strong signal is required to make a detection.  Consequently the 1 part in 10,000 effects some analyses must correct for are completely irrelevant to the transient searches.}  Thus the Milagro detector can be characterized by determining the spectrum of detector responses as a function of energy and zenith angle.  Since the detector response is determined by $p(\vec{k_s})$, we need to determine the relative probability of an incident gamma-ray causing $p(\vec{k_s})$ of a certain value to be added to the sky map, depending on the energy and zenith angle of the incident gamma-ray.

The energy and angle dependent detector response was determined by using a Monte Carlo simulation of 22 million gamma-ray initiated EAS with zenith angles from 0 -- 45 degrees and energies from 100 GeV -- 21 TeV thrown over a 1 $\text{km}^2$ area.  This Monte Carlo simulation modeled both the EAS development and the detection of the shower by the Milagro detector.  The simulated detector signal was then propagated through the Milagro reconstruction code (Section \ref{MilagroReconstruction}) and the weighted analyses implementation (Section \ref{Skymap}) to determine the photon probability density at the source position.  The photon probability density was then added to a three dimensional histogram $g(p_{i},\theta_j, E_k)$, with 1001 photon probability density bins, seven zenith angle bins, and 101 logarithmically spaced energy bins.\footnote{Photon probability densities of zero are included in the histogram.  The probability density bins range from 0 -- 1 and the energy bins from 100 GeV -- 21 TeV.  The zenith angle bins are seven degrees wide, but because only events up to 45 degrees are analyzed the seventh bin only contains showers from 42 -- 45 degrees.} The histogram $g(p_{i},\theta_j, E_k)$ was then normalized for every combination of $\theta_j$ and $E_k$ so that $g(p_i,\theta_j, E_k)$ represents the probability of the photon density $p_i$ being added to the source location as a function of zenith angle and energy.

$g(p_i,\theta_j, E_k)$ is the characterization of the Milagro detector and analysis response. The response to an individual source (at zenith angle $\theta_s$ with spectrum $P_s(E_k)$ ) can be generated by determining the number of photons $N_s$ that would be incident on the 1 $\text{km}^2$ simulation area, and then randomly selecting photon probability densities $p_i$ according to the probability distribution $P_s(E_k)g(p_{i},\theta_s, E_k)$.  The photon probability densities $p_i$ represent the detector's response to the individual gamma-rays, with the total detector response to the simulated source given by 
\begin{equation}
\label{ }
w_s=\sum_{i=0}^{N_s} p_i.
\end{equation}

The response to an expected source $w_s$ can be used with slightly modified versions of the analysis to add fake signals to the sky maps.  If the resulting probability falls below a predetermined threshold  --- $10^{-12}$ for this study --- the simulated signal would have been identified by the analysis. Upper limits can be determined by adding many fake sources of various zenith angles, luminosities and spectra, and recording the percentage of the simulated sources which are detected.

\section{Observer Frame Limits}
\label{LocalLimits}

The first set of limits we wish to calculate is the observed TeV transient luminosities excluded by this study at the $90\%$ confidence level.  The question is:  what is the observed TeV transient luminosity --- as a function of zenith angle, duration, and spectrum --- at which Milagro detects at least $90\%$ of the events?  


The spectral dependence of the luminosity limit is problematic because of the incredible diversity of possible spectra.  Not only are there very few predictions of the emitted TeV spectrum, but the emitted spectrum must then be convolved with the absorption by extragalactic background light to determine the observed spectrum \cite{Jelley}.  In general, every source needs to be individually modeled.  For this section I ignore these effects, and choose two plausible spectra as observed local to the earth.  The first spectrum is $E^{-2.0}$ which has equal energy per logarithmic energy interval.  While somewhat harder than the mode of $E^{-2.25}$ observed by BATSE at MeV energies \cite{BATSEspectralcatalog}, this is well within the scatter of observed spectra and serves as a reasonable model for an inverse Compton spectral bump at TeV energies.  The second spectrum is somewhat softer at $E^{-2.4}$, and brackets the observed BATSE MeV spectrum.  These examples are only used to pick a few plausible points in the phase space of all possible spectra, and are not intended to be exhaustive. 

For a given spectrum and luminosity, the percentage of events that would be detected is calculated with a slightly modified version of the online GRB search program.  A set of real data with no candidate signals (we have lots to choose from) is run though the analysis described in Chapters \ref{ImplementingWAT:chap} and \ref{GRBSearchProcedure:chap}, but without searching for the most significant fluctuations.  Instead the photon probability density from a simulated signal ($w_s$) is added to the sky map,\footnote{We want to use many different locations on the sky map to accurately represent the fluctuations seen in the absence of any signal.  However, using a unique sky map for each simulated GRB becomes computationally expensive, especially for the long time scales.  As a compromise each location on the sky map is used for 20 simulated signals.} and the probability of the background producing the observed signal is calculated (see Sections \ref{WATSourceID} and \ref {WATLimits}). If the probability falls below the detection threshold, the online GRB search would have found the signal.  By repeatedly adding simulated signals of a given spectrum and luminosity, the percentage of signals identified by the GRB search can be determined.  A single power-law spectrum is given by
\begin{equation}
\label{diffPhoton}
\frac{dN}{dE} = J\bigg(\frac{E}{E_0}\bigg)^{-\alpha},
\end{equation}
where $J$, in $\frac{\text{photons}}{\text{s } \text{cm}^2 \text{ TeV}}$ units, is the normalization factor which determines the signal intensity.  An iterative fitting algorithm adjusts the signal intensity until $90\%$ of the simulated events are detected by the GRB search.

\begin{figure}
\begin{center}
\includegraphics[width=5.75in]{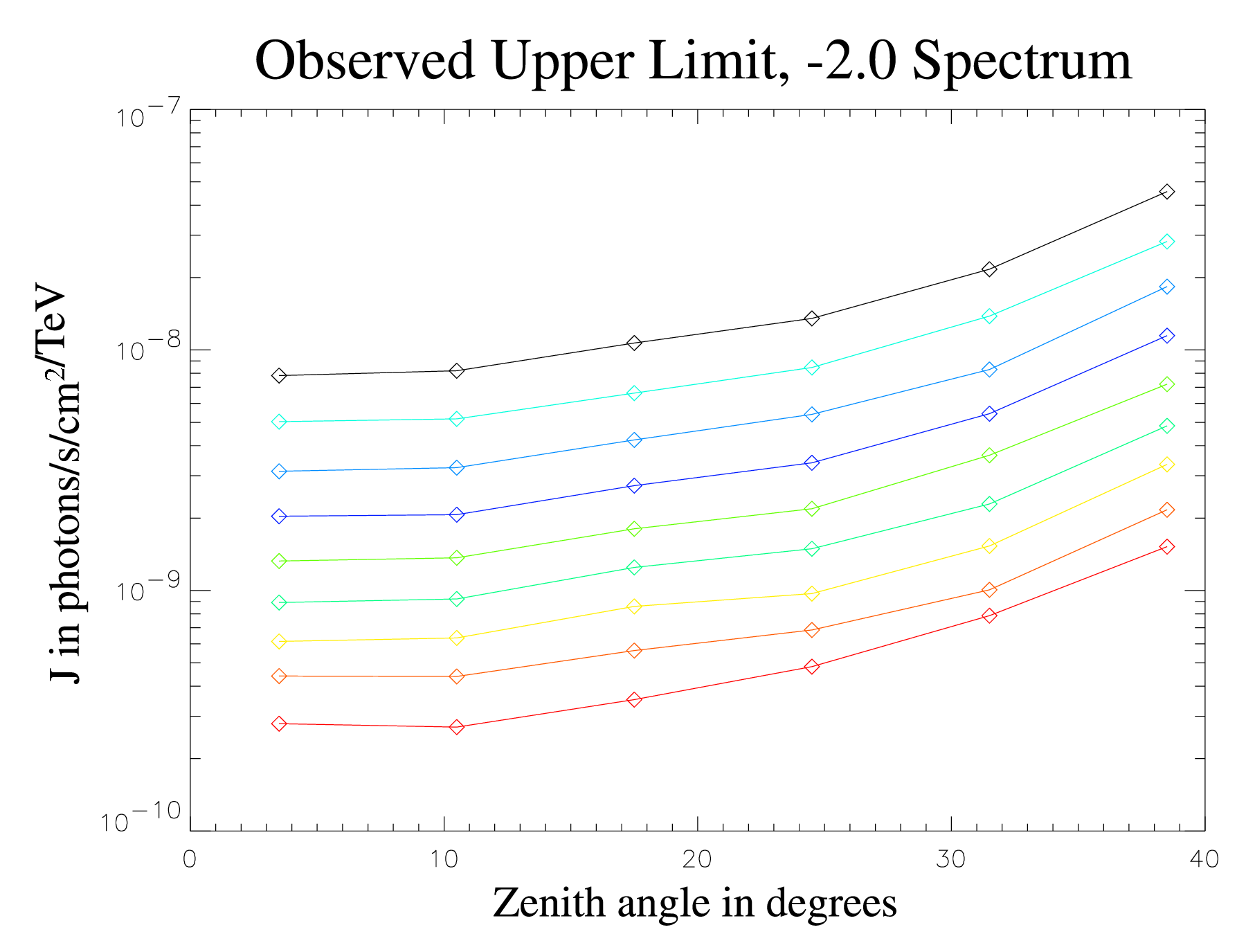}
\caption[Observer Fame Limits for $E^{-2.0}$ Spectrum]{The 90\% confidence upper limits for an $E^{-2.0}$ spectrum for all nine time scales as a function of zenith angle. The diamonds indicate the calculated limits on the normalization factor $J$ (see Equation \ref{diffPhoton}), with the lines of matching color providing visual interpolation between the points.  The time scales are from top to bottom: 40 s, 80 s, 160 s, 320 s, 640 s, 1280 s, 2560 s, 5120 s, and 10240 s.}
\label{LocalLimit2.0}
\end{center}
\end{figure}

\begin{figure}
\begin{center}
\includegraphics[width=5.75in]{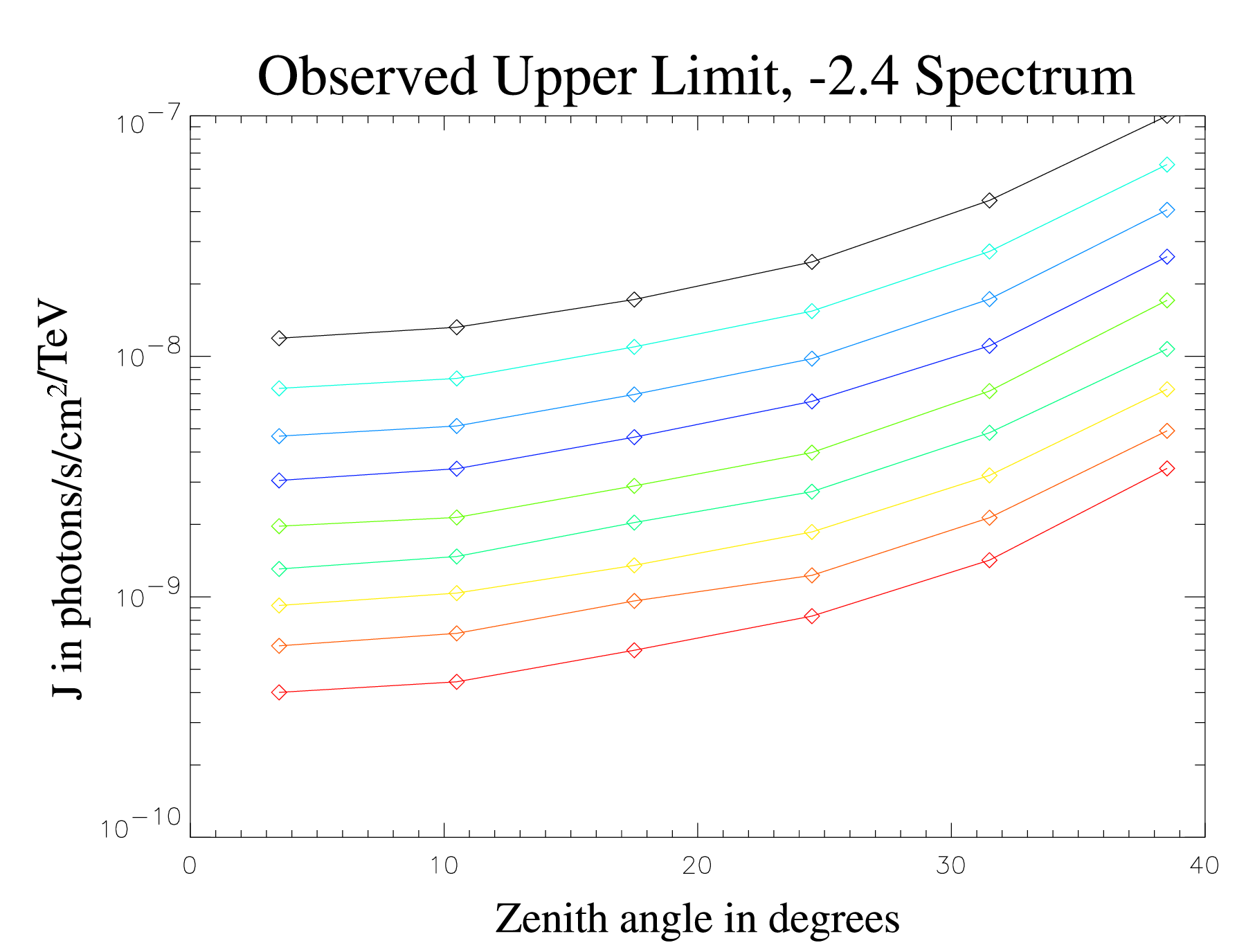}
\caption[Observer Fame Limits for $E^{-2.4}$ Spectrum] {The 90\% confidence upper limits for an $E^{-2.4}$ spectrum for all nine time scales as a function of zenith angle. The diamonds indicate the calculated limits on the normalization factor $J$ (see Equation \ref{diffPhoton}), with the lines of matching color providing visual interpolation between the points.  The time scales are from top to bottom: 40 s, 80 s, 160 s, 320 s, 640 s, 1280 s, 2560 s, 5120 s, and 10240 s.}
\label{LocalLimit2.4}
\end{center}
\end{figure}

The limits for observed spectra of $E^{-2.0}$ and $E^{-2.4}$ are presented in Figures \ref{LocalLimit2.0} and \ref{LocalLimit2.4} as a function of zenith angle and burst duration.
These upper limits may be directly compared to the limits obtained by McCullough \cite*{McCullough}, and are 4 -- 10 times more restrictive.\footnote{In McCullough \cite*{McCullough} there were several different detector configurations with separate upper limits for each, but the upper limits for all configurations were very similar.}  These represent the strongest observer-frame limits on TeV GRB emission obtained to date.

The errors for the upper limits are from the iterative fitting algorithm which required a step size $\le$5\% in $J$ for at least 300 simulated bursts in order to converge.  Since we are looking to identify at least 90\% of the bursts this leads to a Poisson error of 18\% on the 30 missed bursts in addition to the 5\% convergence window.  This leads to an error due to Monte Carlo statistics of 19\%.

Other systematic errors are harder to quantify and are due principally to uncertainty in the Monte Carlo simulation.  In many ways the Monte Carlo predicts the behavior of the detector very well.  However, Benbow \cite*{Benbow:thesis} has found evidence which suggests that the true PSF may be $\sim$30\% worse than predicted when using a hard compactness cut.  A small set of the upper limits were recalculated assuming that the true PSF in each region was 30\% worse than PSF used in the analysis, and the resulting limits were $\sim$30\% less restrictive.  It is possible that the effect is larger or smaller than this $\sim$30\% value depending on which events are affected\footnote{Benbow \cite*{Benbow:thesis} uses an optimal binned analysis so all events are treated equally.  It is uncertain whether all PSF regions are affected equally and thus the magnitude of the effect on this analysis is unclear (see Chapter \ref{Characterization:chap}).} and how dependent the effect is upon the compactness parameter.  Because Milagro is somewhat data limited at  40 s - 3 hour time scales, this analysis uses a much looser effective compactness ``cut" (see Section \ref{Pgammasection}) and may be less sensitive to the observed effect.  

Since indications are that errors in the Monte Carlo are probably detrimental, I will estimate the systematic uncertainty in the limits as +40\%/-20\%.  Combining with the error from Monte Carlo statistics, this gives a total estimated error of +44\%/-27\%.

\section{Intrinsic Limits}
\label{IntrinsicLimitsSection}

The observer frame limits are useful for comparing the sensitivity of this study to previous work, and can be used to set limits on local TeV transient sources.  However, it is difficult to use these limits to set direct constraints on GRB emission due to the absorption of TeV photons by extragalactic background light (EBL).  The observed spectrum becomes a convolution of the emitted spectrum, the EBL absorption (which depends on cosmology and the star formation history of the universe), and the intrinsic distance distribution of GRBs --- none of which have been well measured.  Upper limits on TeV radiation from GRBs are thus necessarily model dependent.  In an effort to place the current observations in context, I have chosen a set of reasonable\footnote{Different readers may disagree on how ``reasonable" these model parameters are, but an attempt has been made to choose parameters which are popular in current theoretical papers.} assumptions about the emitted spectrum, the EBL absorption, the cosmology, and the distance distribution of GRBs, and calculated limits within this framework.

\begin{figure}
\begin{center}
\includegraphics[width=5.75in]{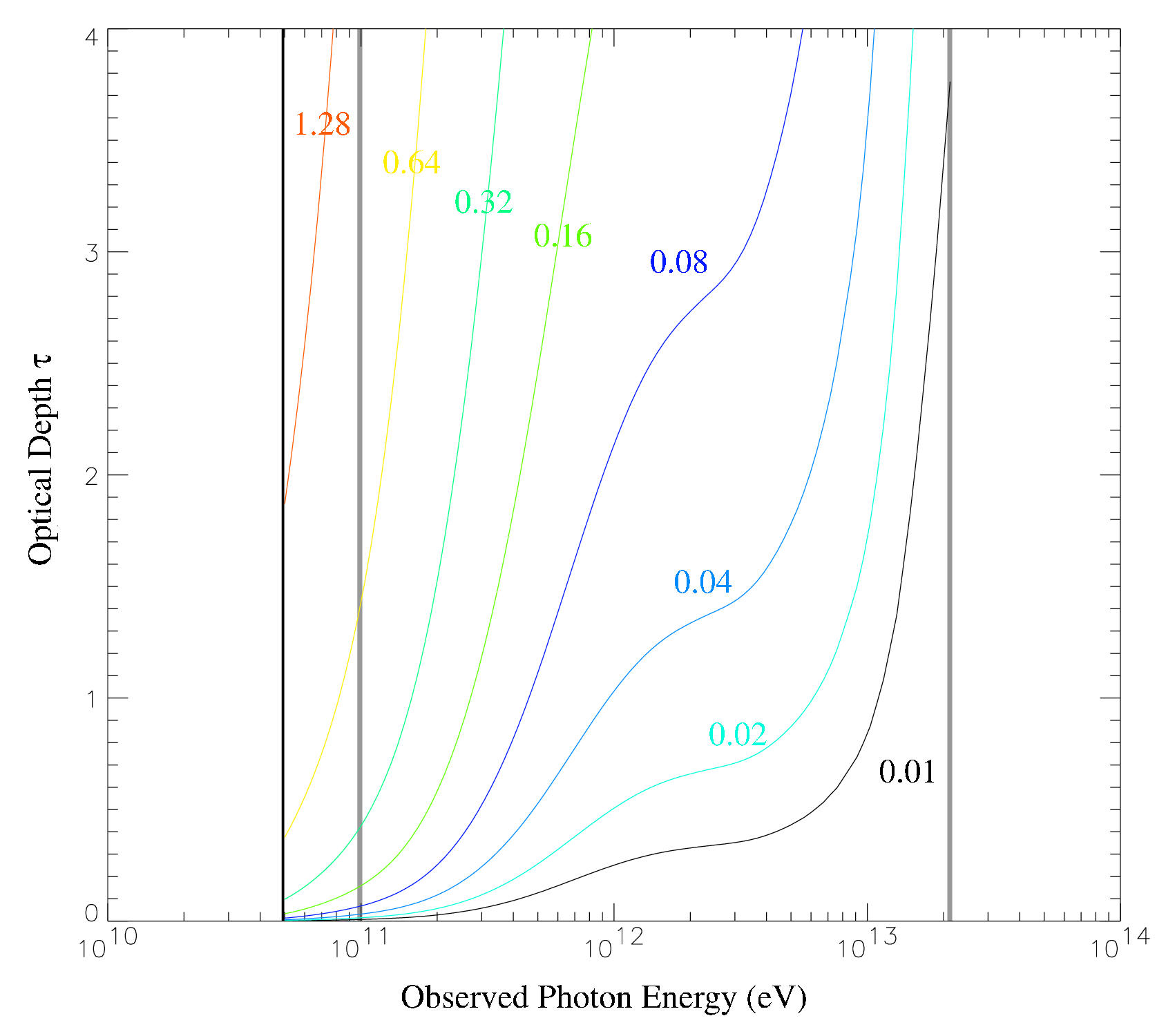}
\caption[Optical Depth Due to Extragalactic Background Light]{The optical depth $\tau$ as a function of observed photon energy and redshift.  The z value of the optical depth curve is indicated by the number with matching color.  The attenuation factor is given by $e^{-\tau}$.   The vertical black line indicates the lower limit of the optical depth calculation, while the vertical grey lines indicate the limits of the Monte Carlo simulation and correspond to the approximate observation band of the Milagro telescope for GRBs.  Data courtesy Bullock \cite*{James:EBL}.}
\label{attnplot}
\end{center}
\end{figure}

The absorption of TeV gamma-rays by extragalactic background light has been extensively modeled \cite{Primack:ProbingGalaxyFormationWGammaRays,Stecker:EBL}, but direct measurements of the EBL are not particularly restrictive.  For these upper limits, I have chosen results from the semi-analytic modeling of Bullock \cite*{James:EBL} shown in Figure \ref{attnplot}.  The calculation assumes a flat $\Lambda$CDM cosmology with $\Omega_M = 0.3$, collisional starburst effects and a Kennicut initial mass function.  While these calculations depend on the star formation history, the stellar initial mass function, the cosmology, and the star formation mechanism, all of the models agree fairly closely on the absorption of high energy gamma rays \cite{Primack:ProbingGalaxyFormationWGammaRays}.  The net effect of the EBL absorption is to constrain the redshift distribution of GRBs observable with Milagro to z $\lesssim$ 0.5.  

The most controversial element of the theoretical framework is the distance distribution of GRBs.  In the collapsar model very high mass stars are the progenitors of GRBs.  Because of the short lifespans of high mass stars, GRBs are closely associated with the star formation rate in this model \cite{Woosley:collapsars}.  Although there has been some evidence for an association for GRBs with star formation \cite{Holland:3HST,Bloom:GRBDist}, the correlation is still a source of strong debate within the community.  Even if the collapsar model is correct, formation of the massive spinning helium star that immediately precedes the GRB may be easier for low metallicity stars.  The dependence on metallicity could lead to a collapsar population which declines earlier than the star formation rate, with very few GRBs at low redshift.  Alternatively, there are progenitor models based on the mergers of binary systems --- containing various combinations of white dwarfs, neutron stars, and black holes --- which are long lived and whose distribution trails the star formation rate.  These models would predict nearby GRBs to be relatively common.  The excellent paper by Fryer et al.\ \cite*{Freyer:GRBformationRate} calculates the formation rate of GRB progenitors for most of the black hole based models --- including three types of collapsars and many binary merger models --- with a review of the theoretical uncertainties.

For this upper limit calculation I have chosen to use the star formation rate (SFR) to predict the distance distribution of GRBs.  This is currently a popular conjecture in the GRB community, and lies somewhere in the middle with respect to the number of low redshift GRBs. The best determination to date of the SFR is by Somerville et al.\ \cite*{Somerville:HighZGalaxies} based on semi-analytic modeling and collisional starbursts.  This theoretical work nicely fits the observable data, and provides a fairly robust determination of the SFR, particularly at low redshift.  A three part power-law fit to the SFR determined by Somerville et al.\ \cite*{Somerville:HighZGalaxies} is used to calculate the distance distribution of GRBs for these upper limit calculations.

The predicted TeV emission spectrum of GRBs is highly model dependent, with everything from very hard rising spectra to very soft spectra proposed.  For these upper limits a spectrum of $E^{-2.0}$ was chosen.  The sensitivity of Milagro to distant objects is dominated by the number of photons at a few hundred GeV where the absorption by the EBL is less severe (see Figure \ref{attnplot}).   Because the EBL absorption creates such a narrow window of observable gamma-ray energies, the sensitivity becomes dominated by the luminosity at a few hundred GeV and is less dependent on the emitted spectrum.  

The cosmology for these upper limit calculations uses the fashionable $\Lambda$CDM cosmology, with $\Omega_M = 0.3$, $\Omega_\Lambda= 0.7$, and $h=0.65$, which is commonly used in theoretical calculations including the EBL absorption and SFR calculations by Bullock \cite*{James:EBL} and Somerville et al.\ \cite*{Somerville:HighZGalaxies}.\footnote{The SFR calculation by Somerville \cite*{Somerville:HighZGalaxies} uses $h = 0.7$, but this does not affect the calculation which is stated in terms of $h$.}  It is important when performing the upper limit calculation to correct for cosmological effects.  In particular the comoving volume element determines the volume of space as a function of redshift, and the luminosity distance determines the apparent luminosity of a source. Following the development by Hogg \cite*{Hogg:Distance} the comoving volume element $dV_c$ for a flat cosmology is given by
\begin{equation}
\label{ }
dV_c = \frac{D_H^3}{E(z)}\Biggl[\int_0^z \frac{dz'}{E(z')}\Biggr]^2 d\Omega dz,
\end{equation}
where $E(z) = \sqrt{\Omega_M(1+z)^3 + \Omega_\Lambda}$ and $D_H = 3000h^{-1}$ Mpc. The comoving volume element is used in conjunction with the SFR to determine the redshift distribution of GRBs.  The luminosity distance $D_L$ is given by
\begin{equation}
\label{ }
D_L = (1+z)D_H\int_0^z \frac{dz'}{E(z')},
\end{equation}
and is used to determine the photon flux.  Given an isotropic luminosity at TeV energies $L_{TeV}$, the photon flux depends on $L_{TeV}$ and $D_L$ according to
\begin{equation}
\label{ }
\text{Photon Flux} \propto \frac{L_{TeV}}{4\pi D_L^2}.
\end{equation}
Given an emitted spectrum, the size of the Monte Carlo throw area, and the duration of the GRB, the number of photons that would be incident on the upper atmosphere in the absence of absorption can be directly calculated.

Given the theoretical model parameters, the upper limits are calculated by constructing a set of fake GRBs.  Since we have an ensemble of events which were not observed, we want to determine the number of TeV GRBs needed for Milagro to have a 90\% chance of observing at least one event. Using the Poisson probability this is equivalent to determining the number of TeV GRBs needed for Milagro to observe on average 2.3 bursts.  The only model parameter we have not yet determined is the isotropic gamma ray luminosity at TeV energies $L_{TeV}$, measured in ergs/s.  The result of the limit calculation can then be presented as the frequency of TeV GRBs at a given isotropic TeV luminosity which can be ruled out with current observations.

To determine the limits, a $L_{TeV}$ value was chosen and a large sample of fake GRBs generated.\footnote{Typically 5000 bursts were generated, though 1000 bursts were used for some of the longer time scales to reduce the computational demands.}  The distance of each GRB was generated according to the product of the SFR times the comoving volume element, with a maximum redshift of 1.28.  The number of photons that would impact the 1 $\text{km}^2$ throw area of the Monte Carlo simulation in the absence of EBL absorption can then be determined from the luminosity distance of the burst.  This gives the raw number of photons that need to be generated for each GRB.  The energy of each photon is then chosen according to the emitted $E^{-2.0}$ spectrum, and then the probability of the photon being absorbed by the EBL is calculated using the data by Bullock \cite*{James:EBL}.\footnote{If the redshift falls between the calculated redshift values, a linear interpolation is used to estimate the appropriate optical depth.}  For the photons which are not absorbed, the Monte Carlo based detector response is then used determine whether the GRB would have been detected by Milagro (see Section  \ref{WATLimits}).

\begin{figure}
\begin{center}
\includegraphics[width=5.75in]{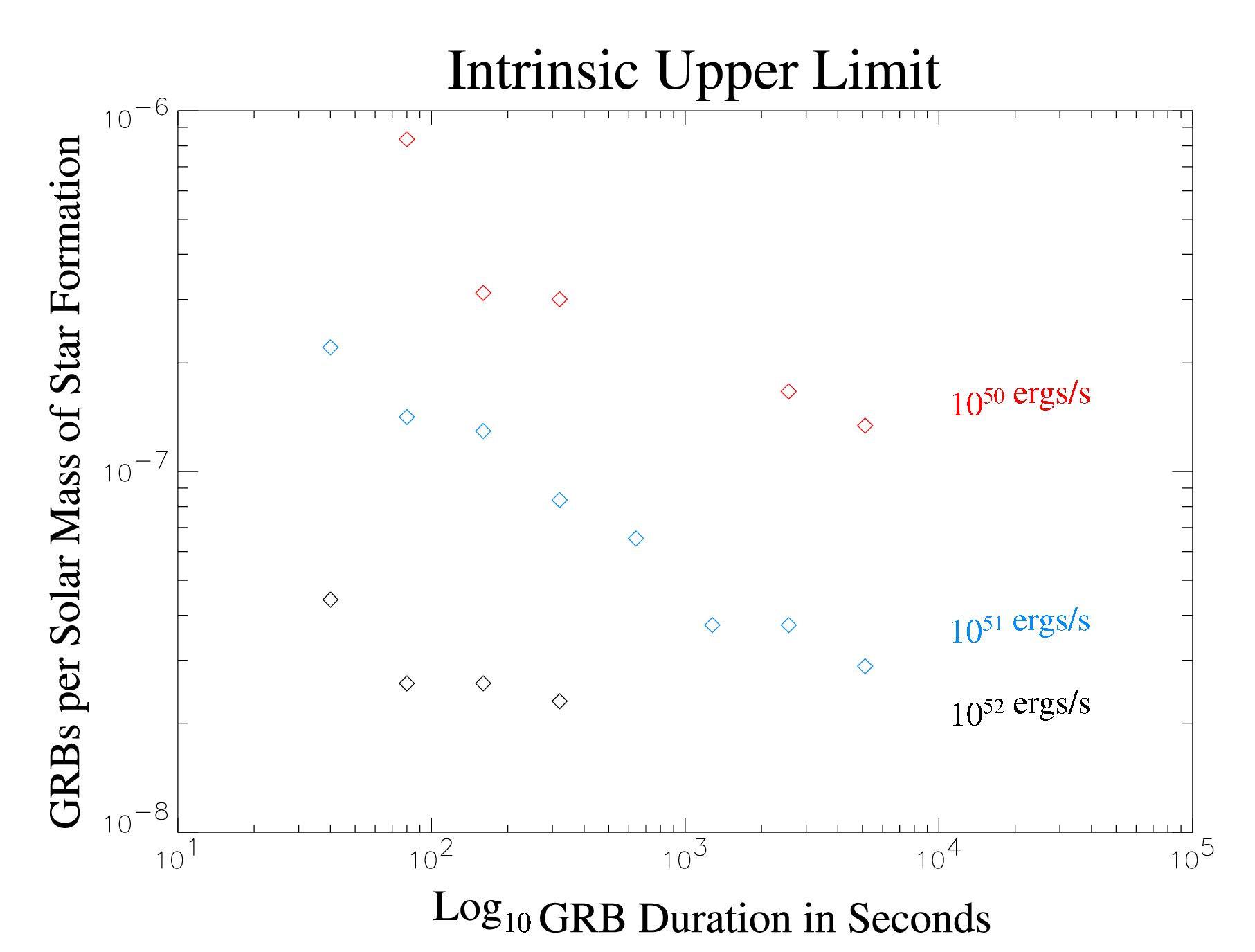}
\caption[Intrinsic Limits, TeV GRBs per Solar Mass of Star Formation]{Upper limits on the number of GRBs per solar mass of star formation versus burst duration in seconds.  Three intrinsic isotropic TeV luminosities from 100 GeV -- 21 TeV are shown: $10^{52}$ ergs/s in black, $10^{51}$ ergs/s in blue, and $10^{50}$ ergs/s in red.}
\label{IntrinsicLimitTSSFR}
\end{center}
\end{figure}

\begin{figure}
\begin{center}
\includegraphics[width=5.75in]{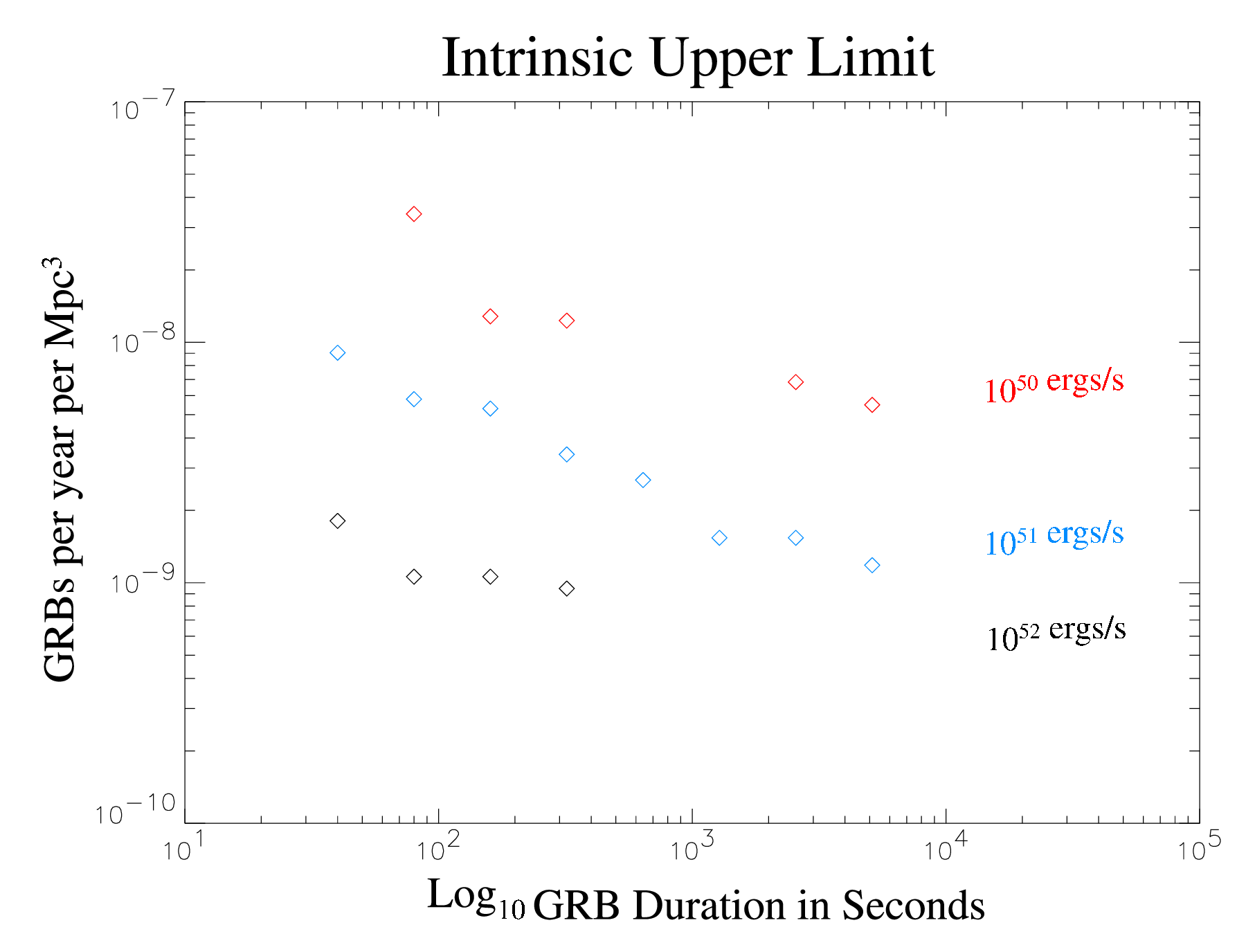}
\caption[Intrinsic Limits, TeV GRBs/year/Mpc$^3$]{Upper limits on the number of GRBs/year/Mpc$^3$.  This is the same data presented in Figure \ref{IntrinsicLimitTSSFR} in different units.  Three intrinsic isotropic TeV luminosities are shown: $10^{52}$ ergs/s in black, $10^{51}$ ergs/s in blue, and $10^{50}$ ergs/s in red.}
\label{IntrinsicLimitTSMpc}
\end{center}
\end{figure}

\begin{figure}
\begin{center}
\includegraphics[width=5.75in]{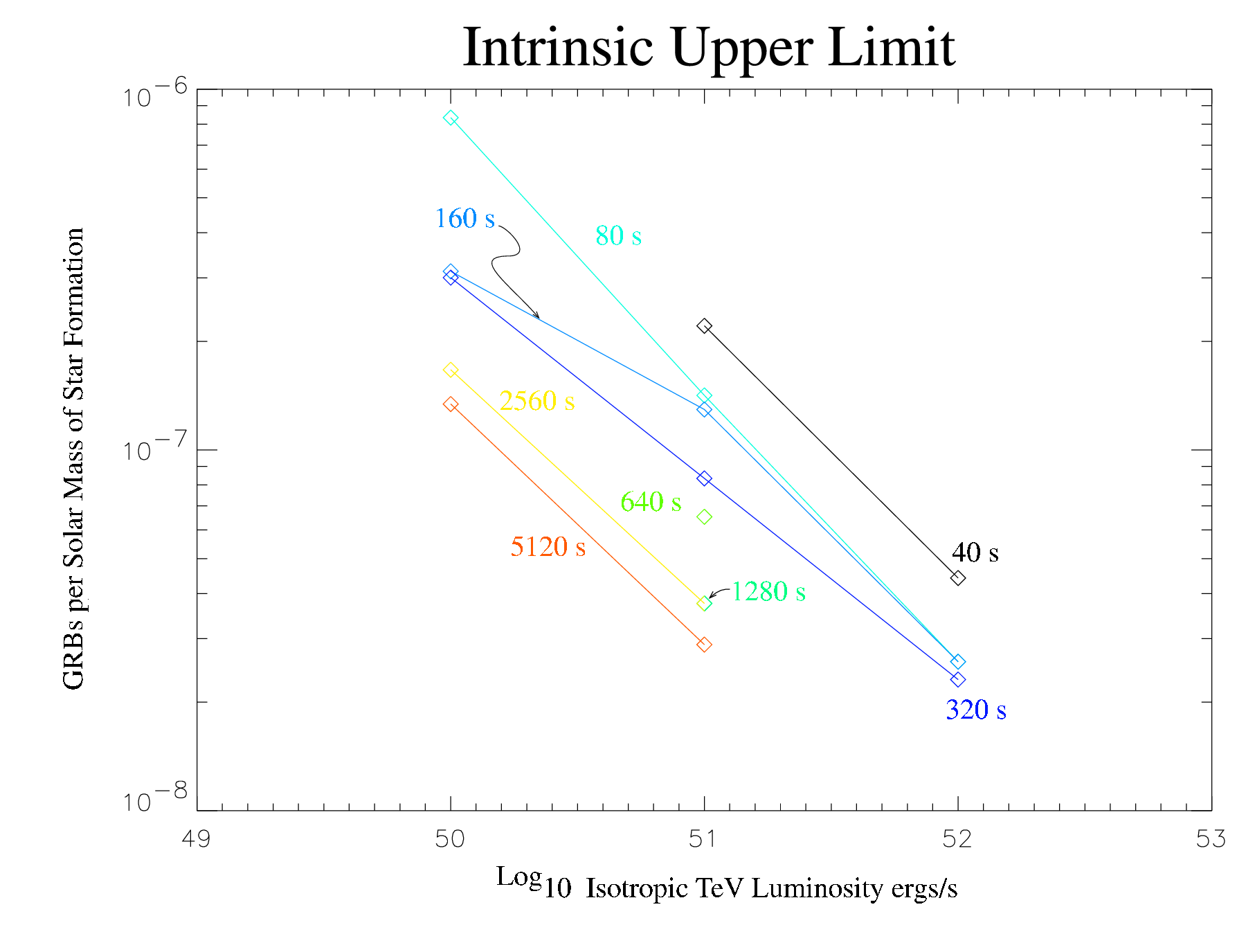}
\caption[Intrinsic Limits, TeV GRBs per Solar Mass View 2]{Upper limits on the number of GRBs per solar mass of star formation versus isotropic TeV luminosity for all nine time scales.  This is the same data and units as Figure \ref{IntrinsicLimitTSSFR}, plotted versus a different horizontal axis to show the trend to lower limits with increased luminosity.  The time scale of each curve is indicated by the label with matching color.}
\label{IntrinsicLimitLumSFR}
\end{center}
\end{figure}

The upper limits for all nine time scales are shown in Figures \ref{IntrinsicLimitTSSFR}, \ref{IntrinsicLimitTSMpc}, and \ref{IntrinsicLimitLumSFR} for GRBs which follow the SFR in a flat $\Lambda$CDM cosmology\footnote{$\Omega_M = 0.3$, $\Omega_\Lambda= 0.7$, and $h=0.65$.} with an emitted TeV spectrum of $E^{-2.0}$ and an EBL absorption based on collisional starbursts and a Kennicut initial mass function.
 The three figures contain the same data points with different units and axis to ease comparison with theoretical models.

The systematic errors for these upper limits are identical to the observer frame limits generated in Section \ref{LocalLimits}.  The errors due to Monte Carlo statistics are somewhat higher due to the small number of simulated GRBs which were detected, with values of $\sim$20\% for the $10^{52}$ ergs/s points to a maximum of 50\% for one of the $10^{50}$ ergs/s points.\footnote{If the error exceeded 50\% no limit was set.}$^,$\footnote{A missing geometric correction in the calculation of the intrinsic limits was discovered too late for correction in this thesis, and is believed to increase the intrinsic limits by $\lesssim$5\%.}  In addition there is significant uncertainty in some of the model parameters.  Of particular note is the uncertainty in the SFR at low redshift which some measurements indicate may be as much as an order of magnitude higher than in the model used here.  These upper limits are inherently dependent on the theoretical framework in which they were calculated and the specifics of the model parameters.  Because of the dependence on the theory, these limits apply only for the model in which they were calculated.

\section{Conclusion}

This search for 40 s -- 3 hour TeV transient emission observed by Milagro between May $2^\text{nd}$, 2001 and May $22^\text{nd}$, 2002 represents the most sensitive search for moderate duration unidentified TeV transients yet performed.  Because of the absorption of very high energy gamma rays by extragalactic background light and the diverse predictions of the distance distribution of GRB progenitors and their TeV emission, no model independent limits on the TeV emission of GRBs can be set.  The limits in Section \ref{LocalLimits} detail the sensitivity of the Milagro search and the limits in Section \ref{IntrinsicLimitsSection} detail the constraints on the frequency of moderate duration TeV emission for one set of model parameters.  Interested parties are encouraged to contact the author to convert the observations presented in this thesis into upper limits for particular theoretical models of very high energy GRB emission.


\appendix
\chapter{Probability of the Background Producing a Signal-Like Event}
\label{ProbAppendix}

The probability calculation represented by Equation \ref{CorrectProbEq} is an approximation of the true probability.  The question of how to determine the correct probability when there are two independent variables ($w_{obs}$ and $N_{obs}$) can be answered using an argument analogous to the one used by Feldman and Cousins \cite*{FeldmanCousins} in their excellent paper on confidence limits.

\begin{figure}
\begin{center}
\includegraphics[width=5.75in]{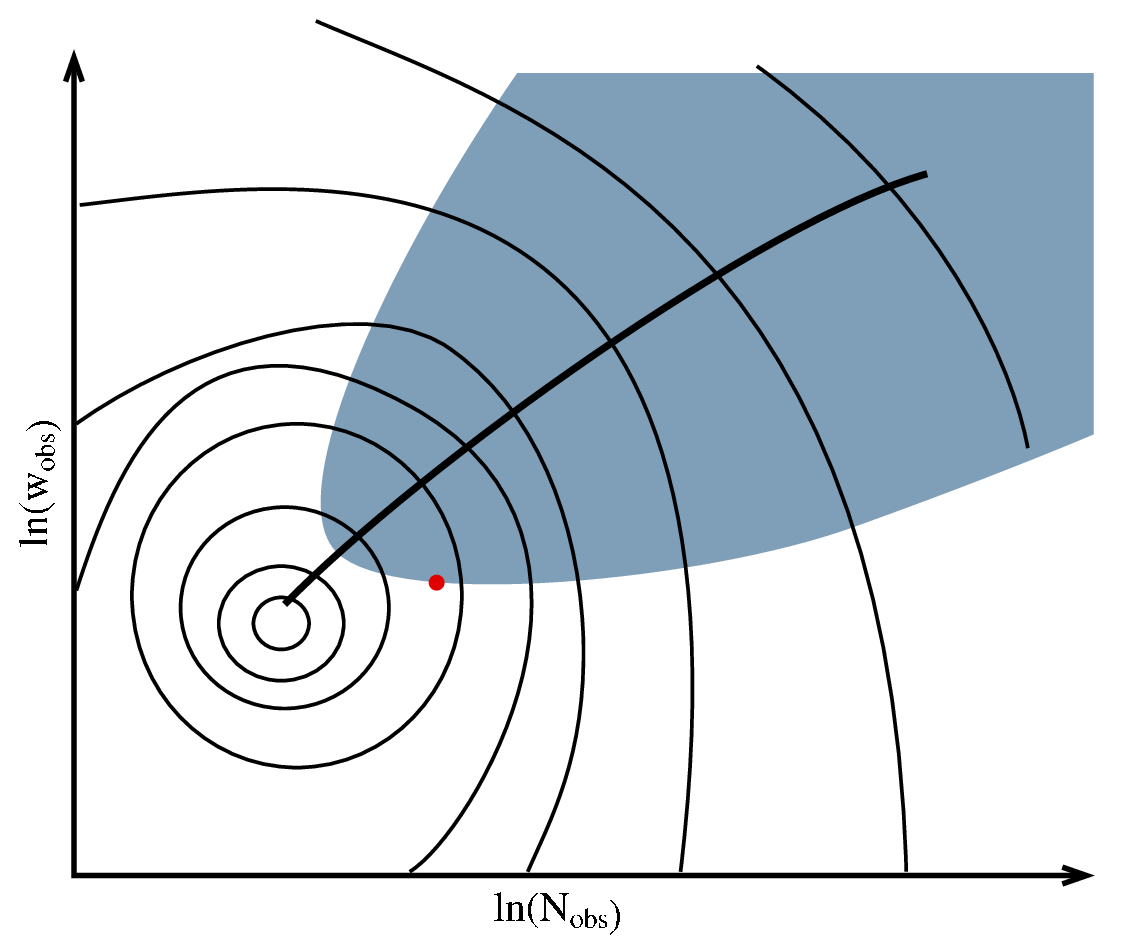}
\caption[Determining the True Probability]{A cartoon of how to determine the probability of the background producing an event which is more signal-like than an observation.  The contour lines are the probability density of the background producing a particular combination of $w_{obs}$ and $N_{obs}$;  the heavy line indicates the trend where signal events are likely to be.  The red dot is a particular observation, with the blue region including all of the points where the likelihood ratio is equal to or more signal-like than the current observation.  Please see the text for a full description.}
\label{ProbabilityDiagram1}
\end{center}
\end{figure}

Since we have two independent variables, the probability density distributions will be two dimensional.  The probability density distribution for the background is represented by the contour lines in Figure \ref{ProbabilityDiagram1}.  Given a theoretical model, a similar probability density distribution can be constructed for the expected signal.\footnote{In the case of Milagro the model would need to supply both an energy spectrum and a luminosity spectrum.  The probability density of the expected signal as a function of $w_{obs}$ and $N_{obs}$ could then be determined using a Monte Carlo simulation.}  The ratio of the background probability density to the signal probability density can then be determined for every point in the field.  This likelihood ratio is analogous to Equation 4.1 in Feldman and Cousins \cite*{FeldmanCousins}. Feldman and Cousins used the likelihood ratio to determine the order in which terms should be summed, then defined the confidence interval by adding terms until a predetermined probability was reached ({\it ie.}\ 90\%). The problem under consideration here is very similar.  Instead of using the sum to determine the location of a confidence limit, we have an observation and want to determine the sum.  Since the likelihood ratio of the observed point is known, we can determine the probability of the background producing an event which is more signal-like by integrating the probability density of the background at all points where the likelihood ratio of background to signal is equal to or lower (more signal-like) than the likelihood ratio at the observed location.

This process is shown graphically in Figure \ref{ProbabilityDiagram1}.  
In this example, signal events tend to have both higher $w_{obs}$ and $N_{obs}$, as indicated by the heavy line.  An observation is indicated by the red dot.  The blue region includes all of the positions where the likelihood ratio is as signal-like or more than the observed position.  The probability of the background producing an event which is more signal-like is then given by integrating the background probability density (shown by contour lines) in the blue region.  This defines the probability of the background producing an observation when there are two or more independent variables.

\begin{figure}
\begin{center}
\includegraphics[width=5.75in] {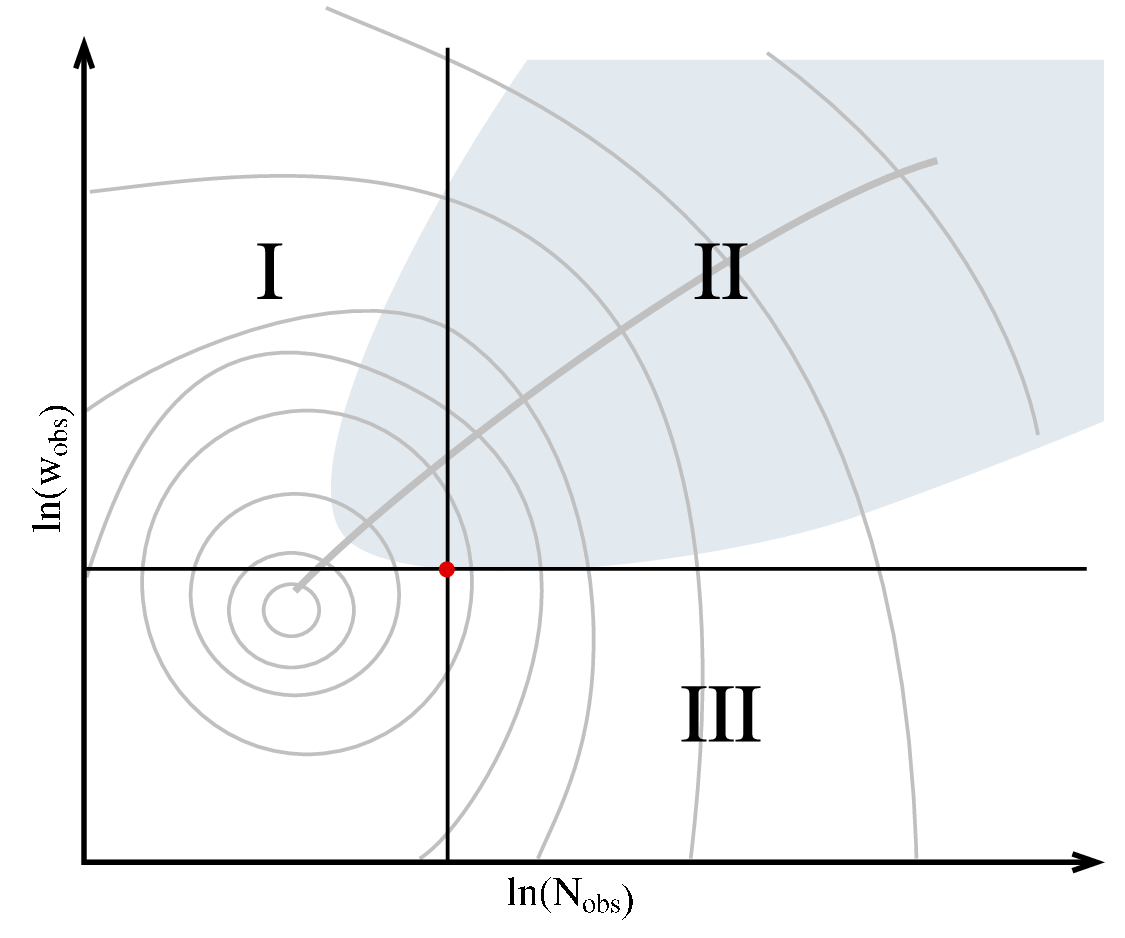}
\caption[Approximation to the True Probability]{This is the same diagram as Figure \ref{ProbabilityDiagram1}, with lines indicating the $w_{obs}$ and $N_{obs}$ values.  Equation \ref{CorrectProbEq} approximates the true probability by integrating the background probability density in region II.}
\label{ProbabilityDiagram2}
\end{center}
\end{figure}

In the case of Milagro, the problem is complicated by the lack of a good model, making any determination of the likelihood ratio uncertain.  Without a model, the question becomes how can the appropriate integration region be approximated?  The reason we have two independent measures of the probability density is the truncation of the PSFs used in this implementation of the weighted analysis technique (see Table \ref{Cutoff:table}).  Because of the truncation, we expect a signal to increase both the $w_{obs}$ and $N_{obs}$ values.  Equation \ref{CorrectProbEq} provides a first order approximation of the probability of the background creating a more signal-like event by integrating the background probability density for all $w$ and $N$ greater than the observed value, as depicted by region II in Figure \ref{ProbabilityDiagram2}.  This is not the ideal integration region, but gives a rough indication of the relevant probability.

The effect of the approximation in Equation \ref{CorrectProbEq} is to slightly degrade the sensitivity of the analysis through two separate mechanisms.  First, for a true signal event, the probability of the background producing an event that signal-like or more is only approximated.  A second more subtle effect is caused by the value of the probability integral not having a one-to-one mapping to the integration region.  Using the approximation, two events may have the same total probability without integrating the same region.  Because the integral should include all locations as signal-like or more than the current observation --- and thus all events with equal or lower probability --- not including the other event location in the integral systematically underestimates the probability.  This flattens the probability distribution as seen in Figure \ref{backdist2} and reduces the sensitivity of the analysis to weak signals.

It is interesting to note that both a binned analysis and a ``pure" weighted analysis can be obtained by integrating different regions of  Figure \ref{ProbabilityDiagram2}.  An integral of all $N$ greater than the observation (regions II and III) is equivalent to a binned analysis and recreates the familiar integral Poisson distribution.  Similarly, integrating all $w$ greater than the observation (regions I and II) recreates a ``pure" weighted technique with no dependence on the number of observed events.

A better implementation of the weighted analysis technique could be achieved by modeling an expected signal; determining the likelihood-based integration regions; and forming a table in $w_{obs}$ and $N_{obs}$ of the resulting background probability density integrals.  The table could then be used online to quickly look up the correct probability, and is an interesting avenue for future work.

\chapter{Detailed Plots}
\label{PlotAppendix}

In this appendix are the detailed plots showing the functional fits and comparisons for all the regions, not just the example regions used for figures in Chapter \ref{Characterization:chap}.

\begin{figure}
\begin{center}
\includegraphics[width=5.1in]{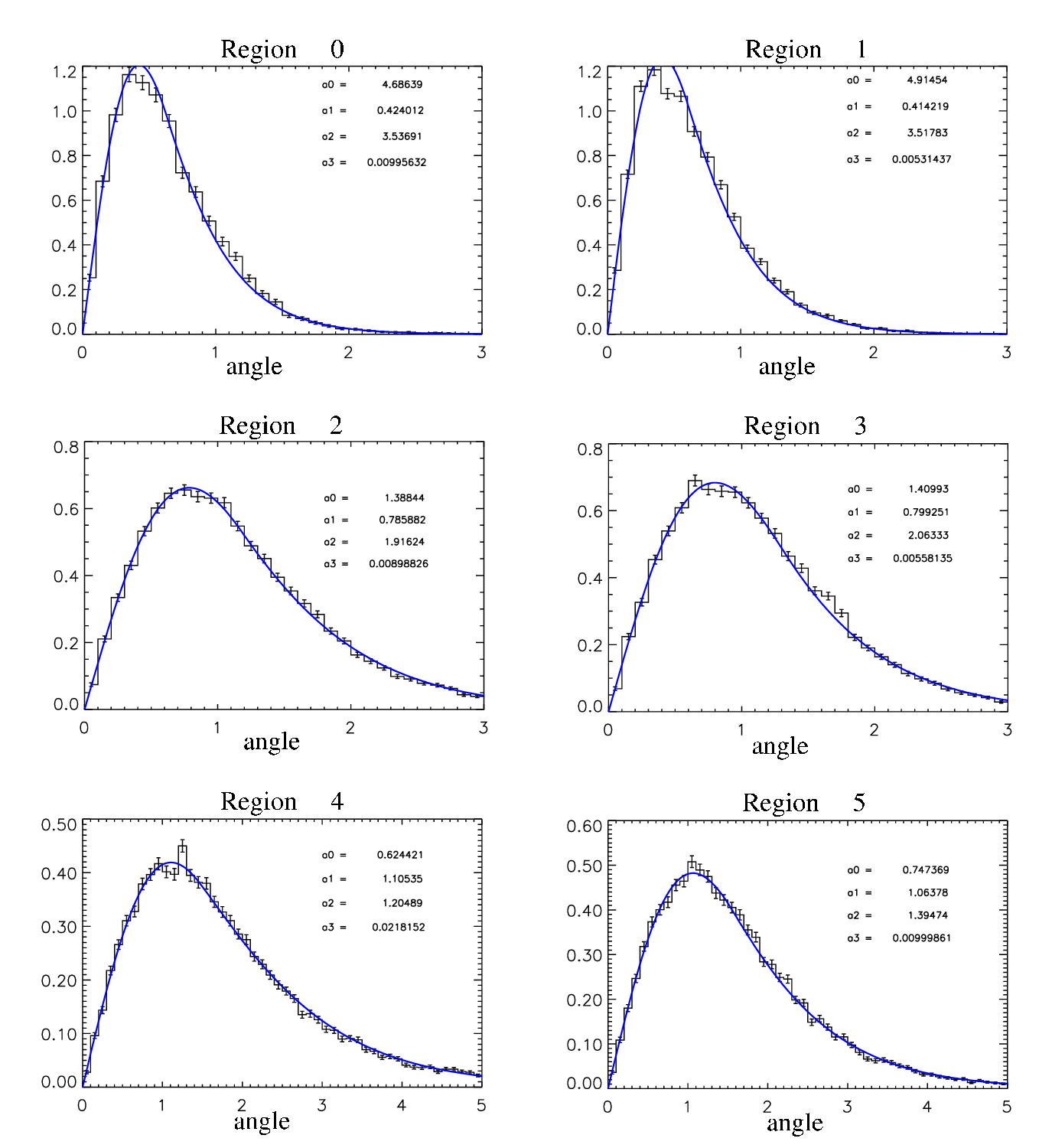}
\caption[PSF Fits To Deleo/2 Distributions Part 1]{PSF fits to deleo/2 for regions 0 -- 5. The fit parameters are displayed in the corner of each graph.  The fits are not constrained to be normalized, so that they accurately represent the PSF at each distance.}
\label{v3DataRegionFits6}
\end{center}
\end{figure}

\begin{figure}
\begin{center}
\includegraphics[width=5.1in]{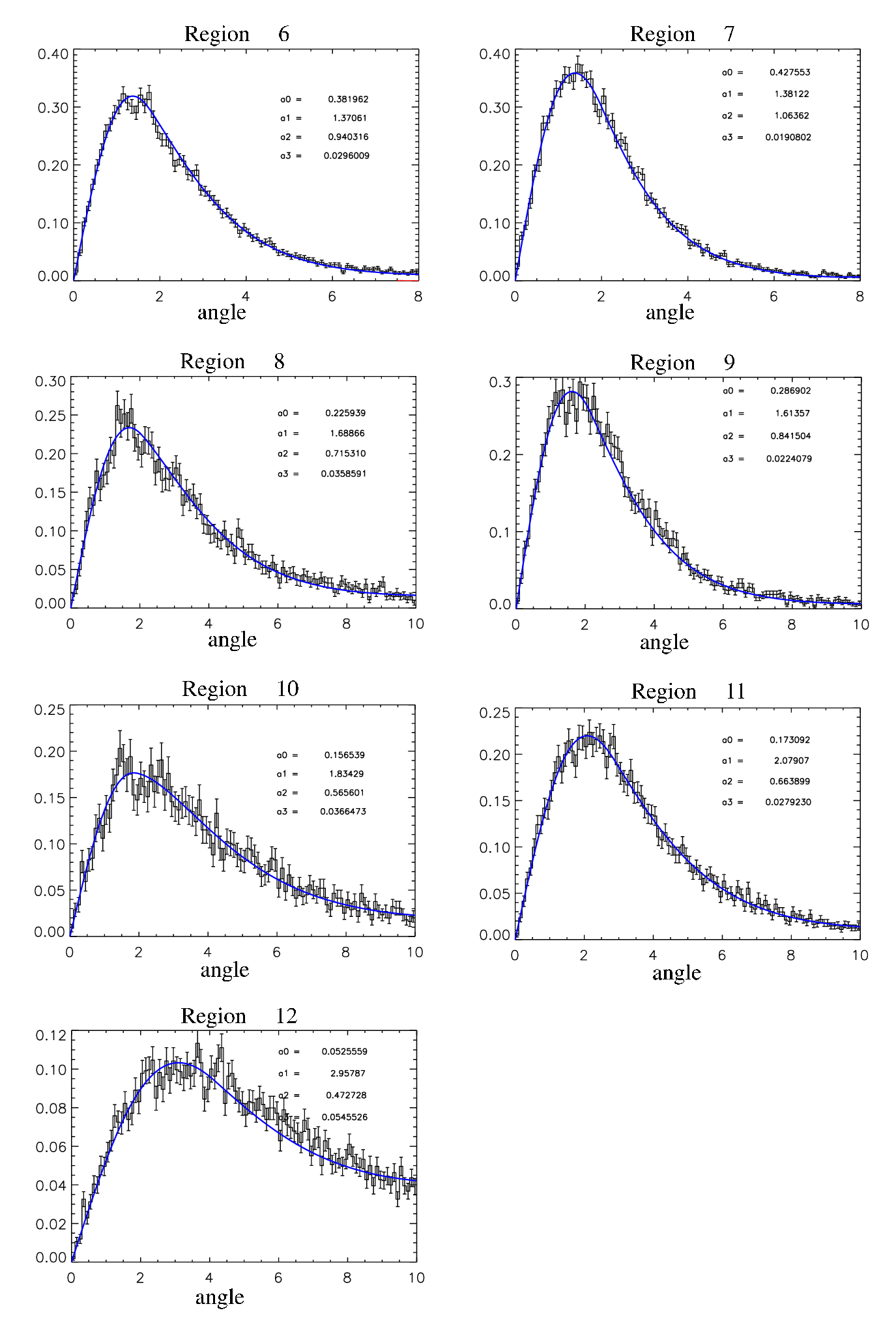}
\caption[PSF Fits To Deleo/2 Distributions Part 2]{Same as \ref{v3DataRegionFits6} for regions 6 -- 12.}
\label{v3DataRegionFits6B}
\end{center}
\end{figure}

\begin{figure}
\begin{center}
\includegraphics[width=5.1in]{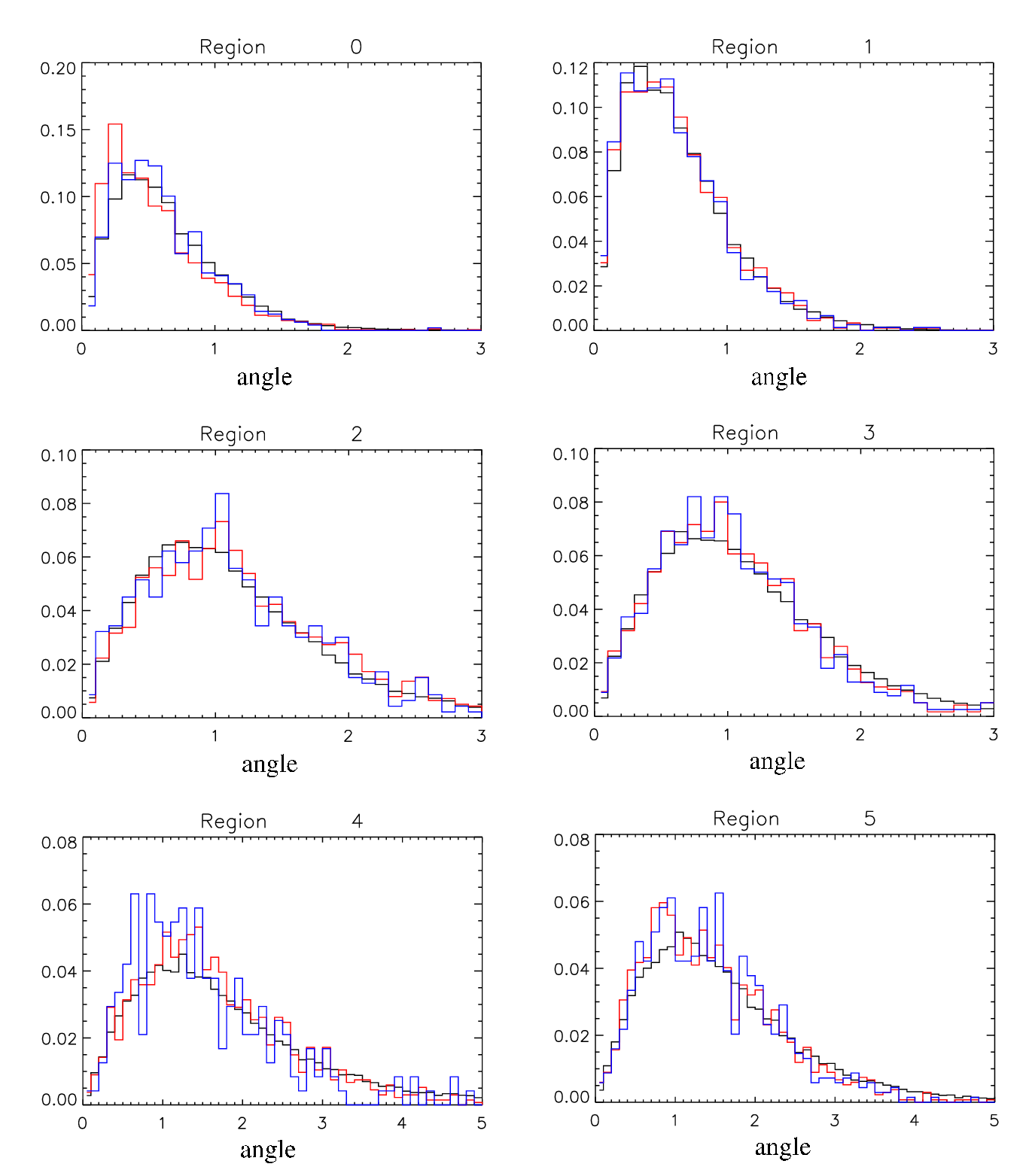}
\caption[Deleo/2 Distributions for Data and Monte Carlo Part 1]{The deleo/2 distributions for regions 0 -- 5. The data distribution is in black, the Monte Carlo gamma initiated showers are in red, and the same Monte Carlo showers with a very hard compactness cut of 3.0 are in blue.}
\label{v3dat53mcgmcdhdeleo}
\end{center}
\end{figure}

\begin{figure}
\begin{center}
\includegraphics[width=5.1in]{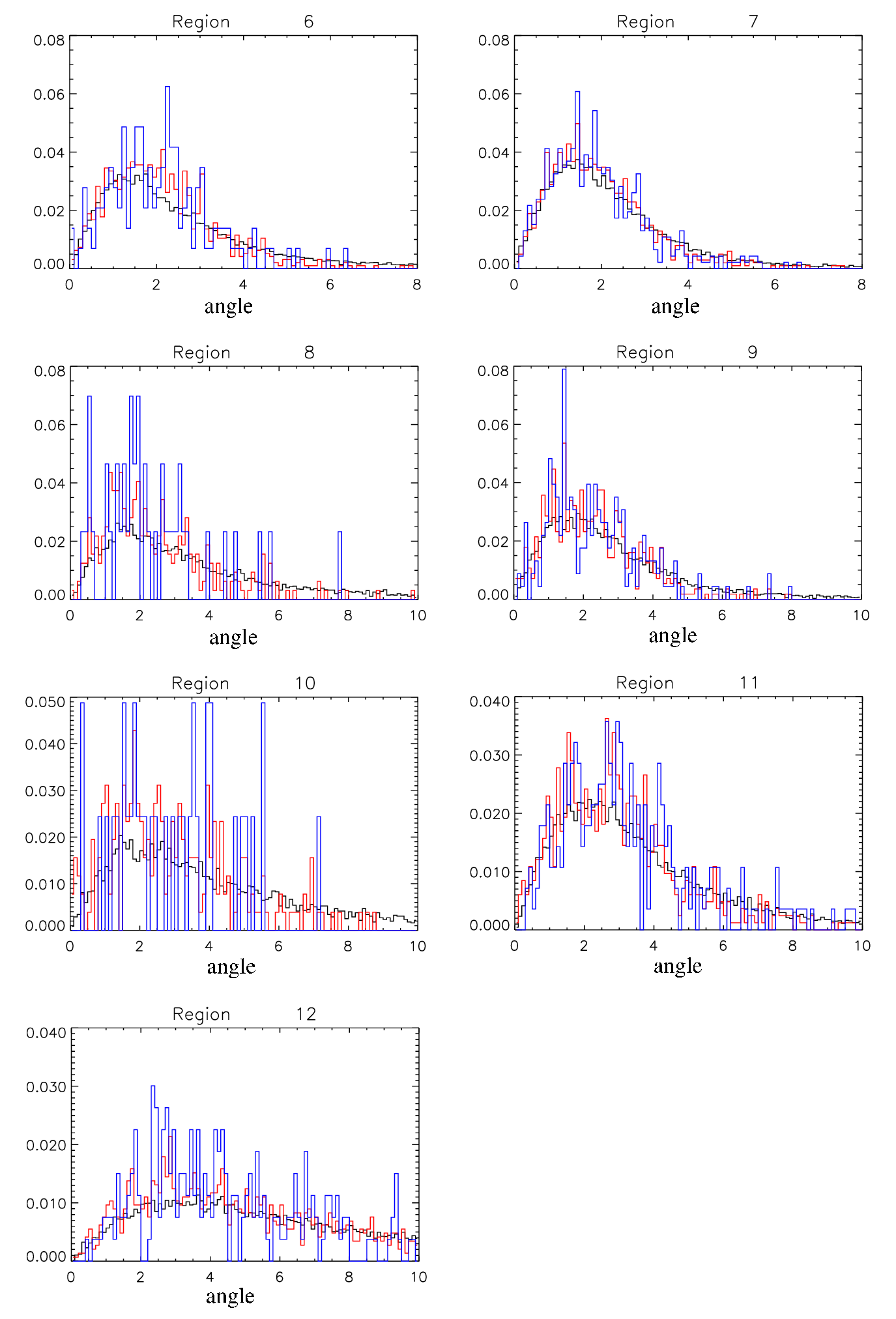}
\caption[Deleo/2 Distributions for Data and Monte Carlo Part 2]{Same as \ref{v3dat53mcgmcdhdeleo} for regions 6 -- 12.}
\label{v3dat53mcgmcdhdeleoB}
\end{center}
\end{figure}

\begin{figure}
\begin{center}
\includegraphics [width=5.1in]{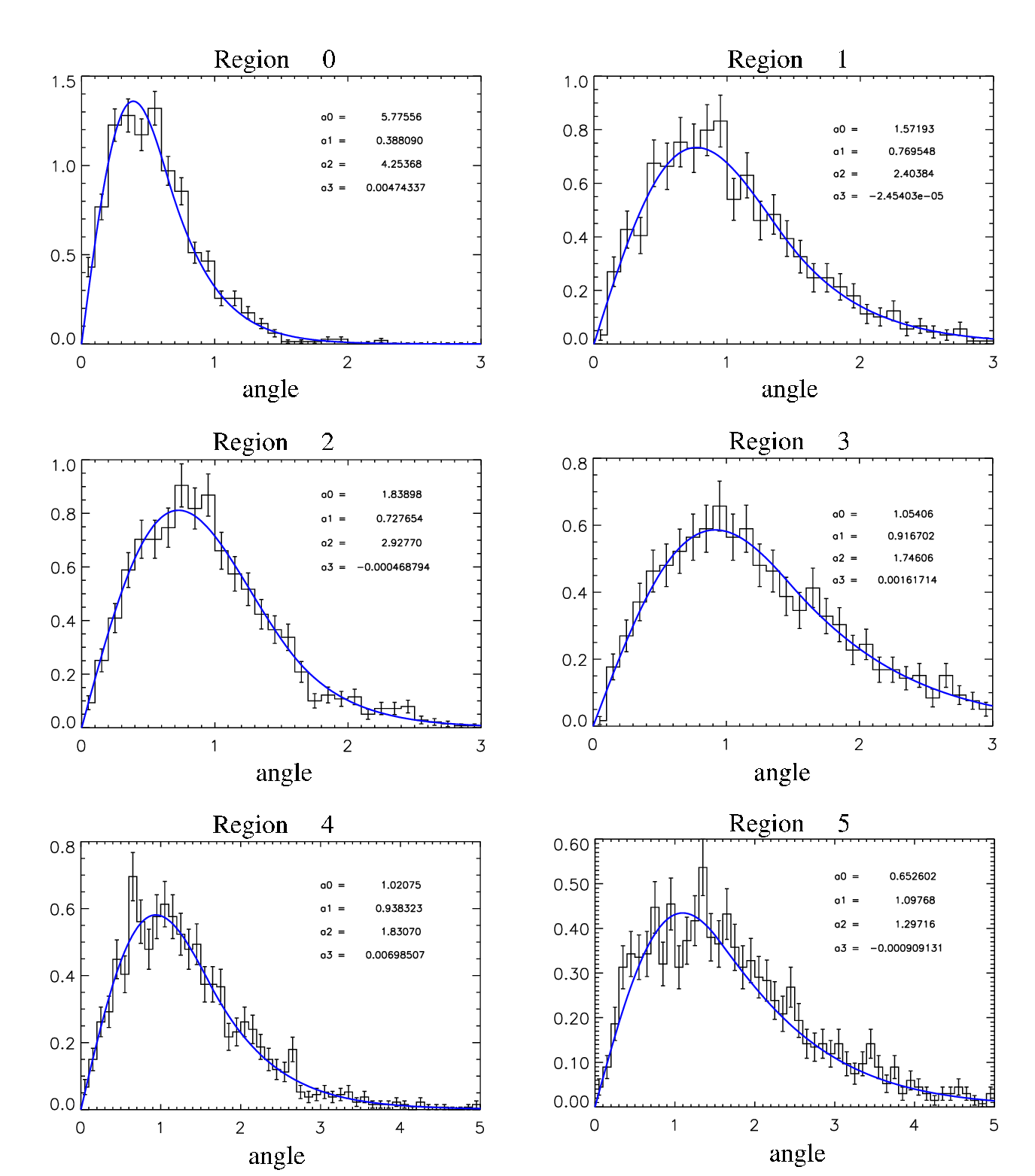}
\caption[PSF Fits to Monte Carlo Distributions Part 1] {PSF fits to the gamma-ray Monte Carlo angle difference distributions for regions 0 -- 5. The fit parameters are displayed in the corner of each graph.  The fits are not constrained to be normalized, so that they accurately represent the PSF at each distance.}
\label{v3mcgRegionFits6}
\end{center}
\end{figure}

\begin{figure}
\begin{center}
\includegraphics [width=5.1in]{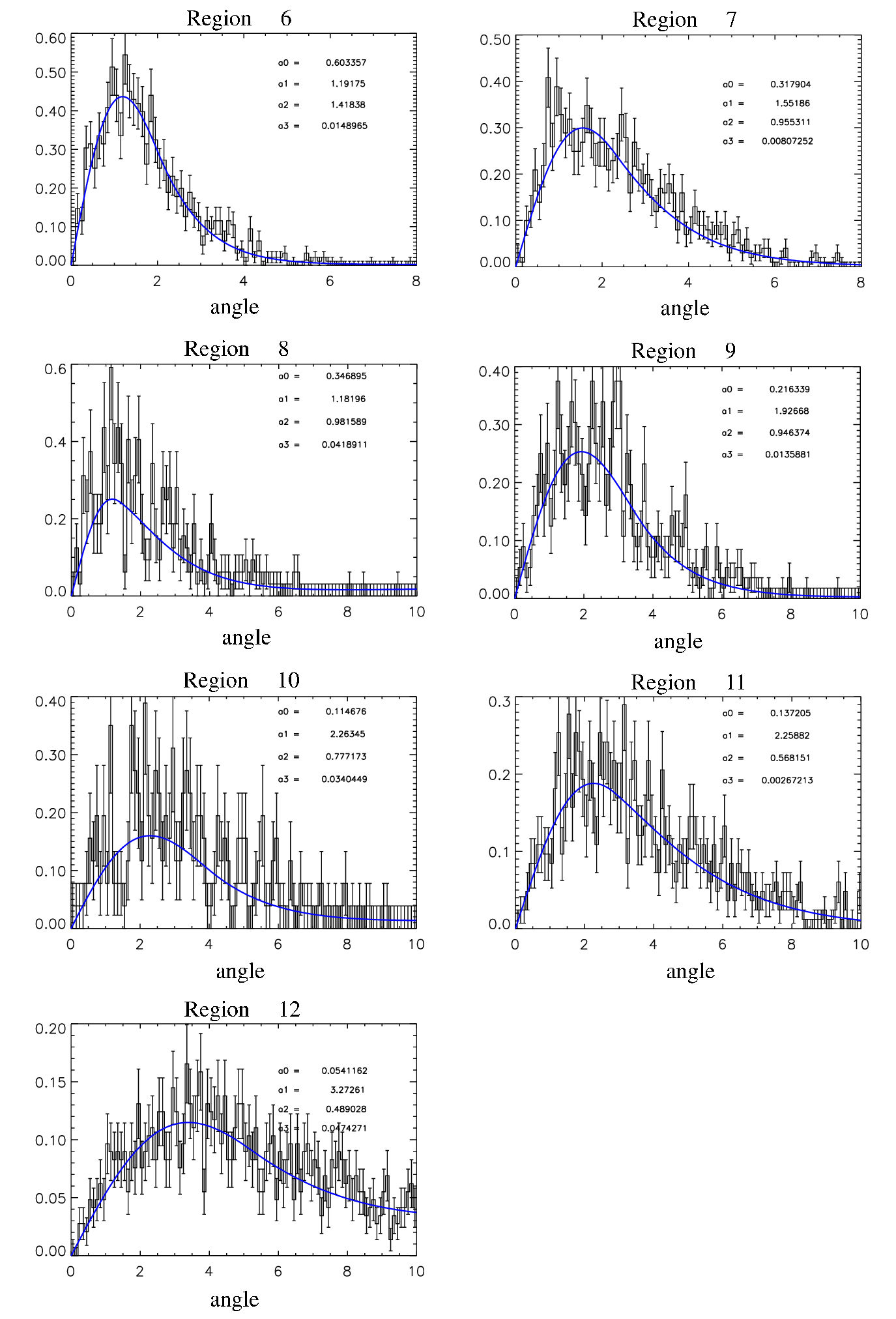}
\caption[PSF Fits to Monte Carlo Distributions Part 2] {Same as \ref{v3mcgRegionFits6} for regions 6 -- 12.}
\label{v3mcgRegionFits6B}
\end{center}
\end{figure}

\begin{figure}
\begin{center}
\includegraphics [width=5.1in]{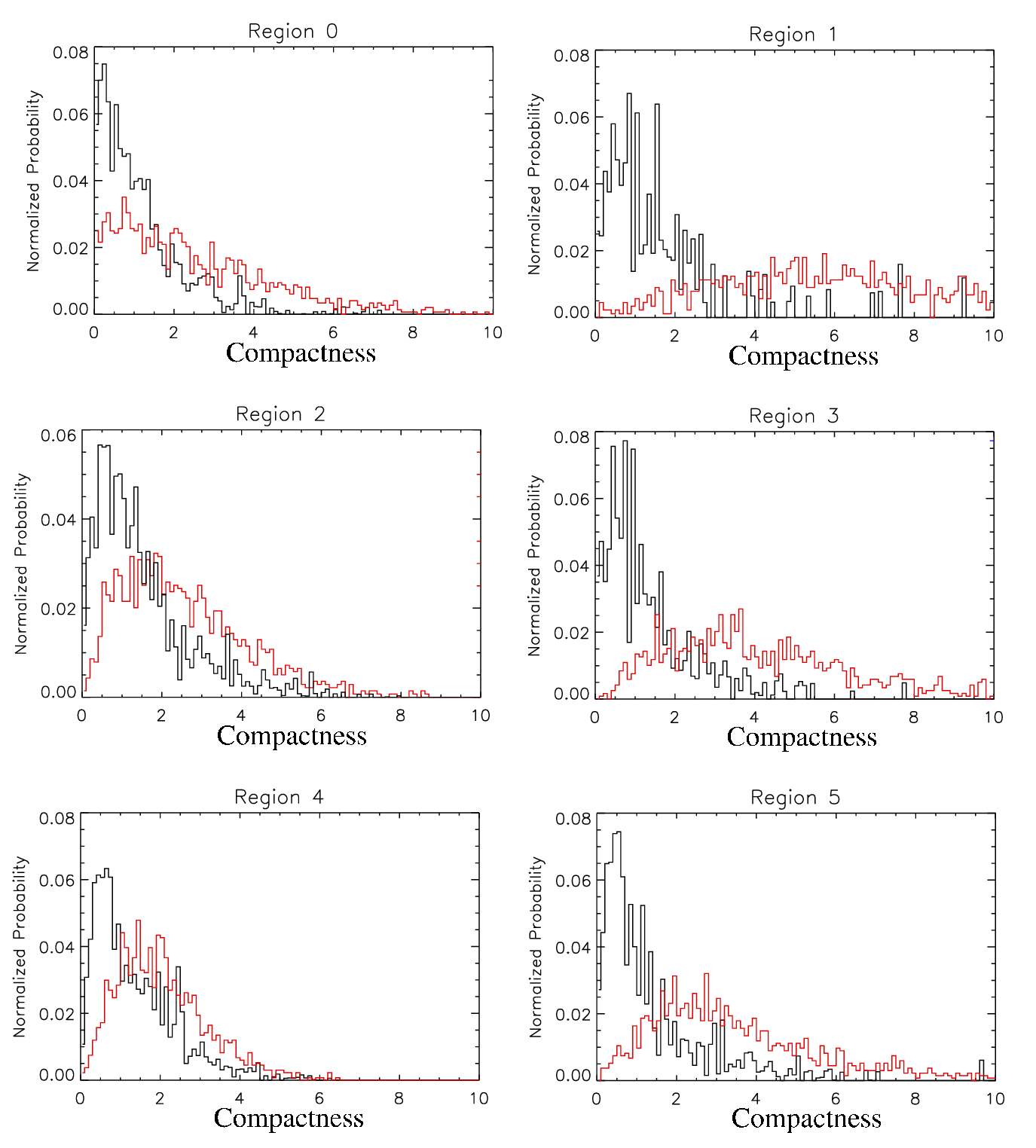}
\caption[Compactness Distributions by Region] {The probability distributions for regions 0 -- 5 plotted as a function of the compactness parameter for Monte Carlo protons (black) and Monte Carlo gamma rays (red).}
\label{X2_RegionSimplified}
\end{center}
\end{figure}

\begin{figure}
\begin{center}
\includegraphics [width=5.1in]{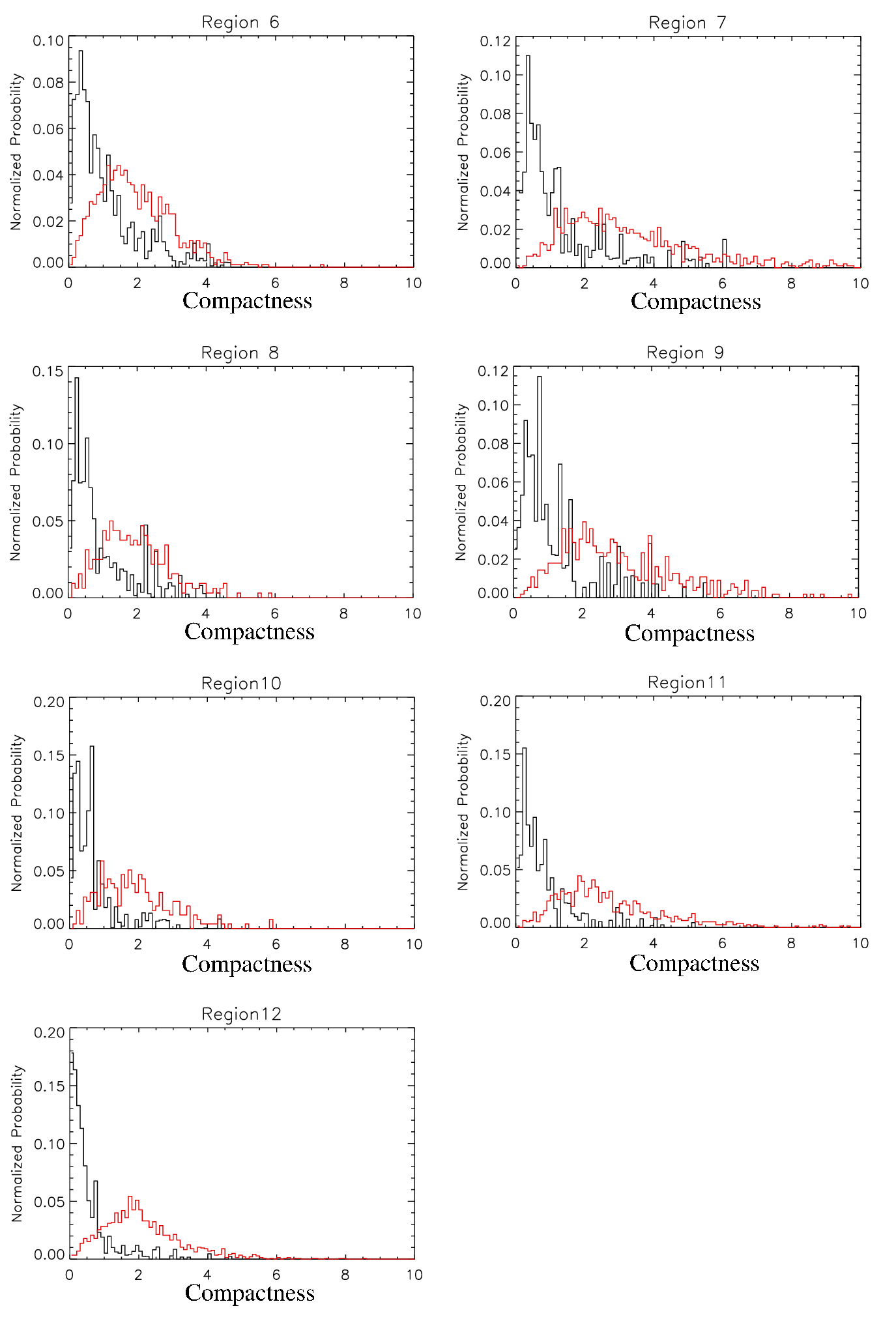}
\caption[Compactness Distributions by Region] {Same as \ref{X2_RegionSimplified} for regions 6 -- 12.}
\label{X2_RegionSimplifiedB}
\end{center}
\end{figure}

\begin{figure}
\begin{center}
\includegraphics[width=5.1in]{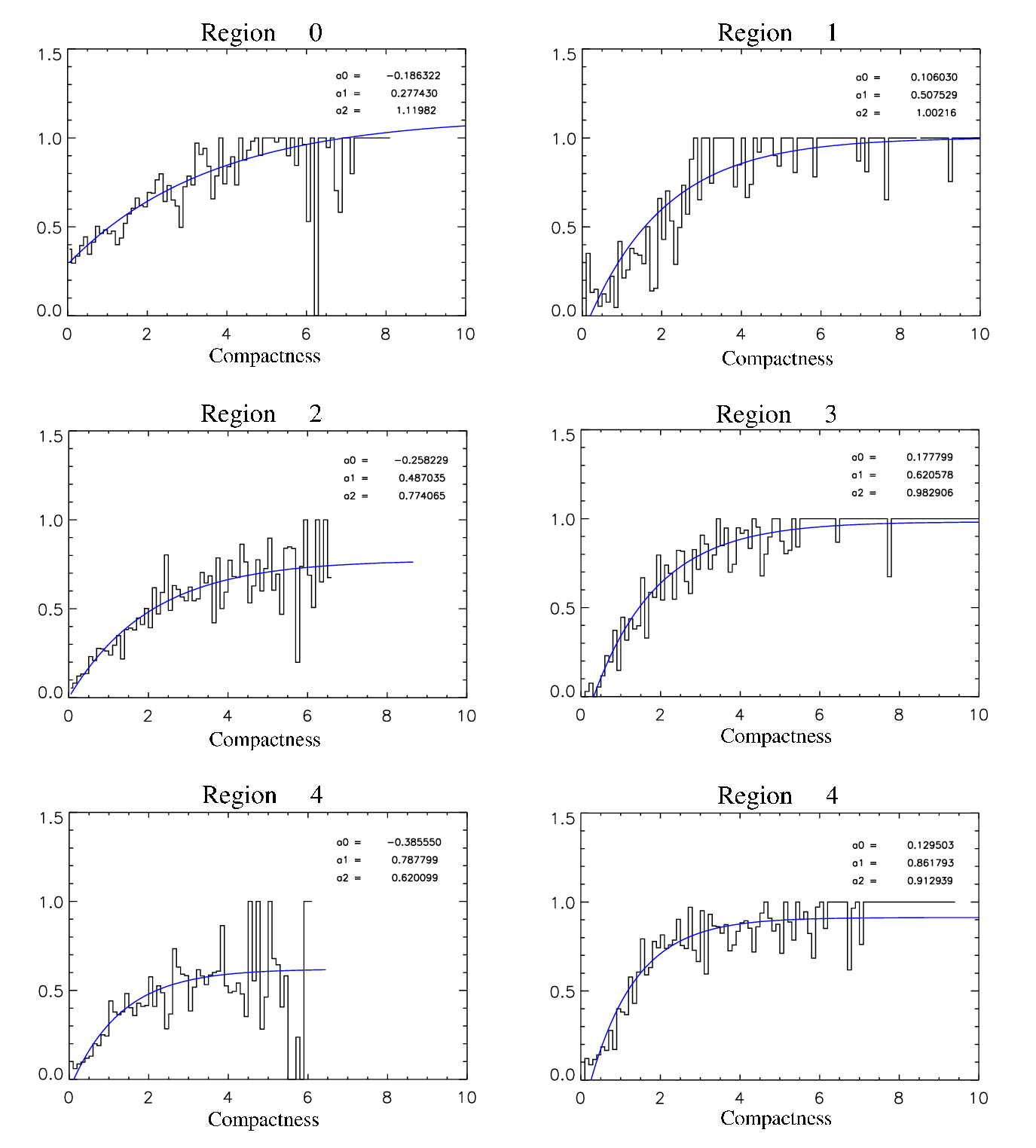}
\caption[$P_\gamma$ vs. Compactness Fits by Region]{The $P_\gamma$ distributions for regions 0 -- 5 with the associated fits in blue. The plots are $P_\gamma$ vs. compactness, with the fit parameters listed on each plot.  For clarity the error bars on the individual data points are not shown, but they were used in the fit.}
\label{X2_RegionFits}
\end{center}
\end{figure}

\begin{figure}
\begin{center}
\includegraphics[width=5.1in]{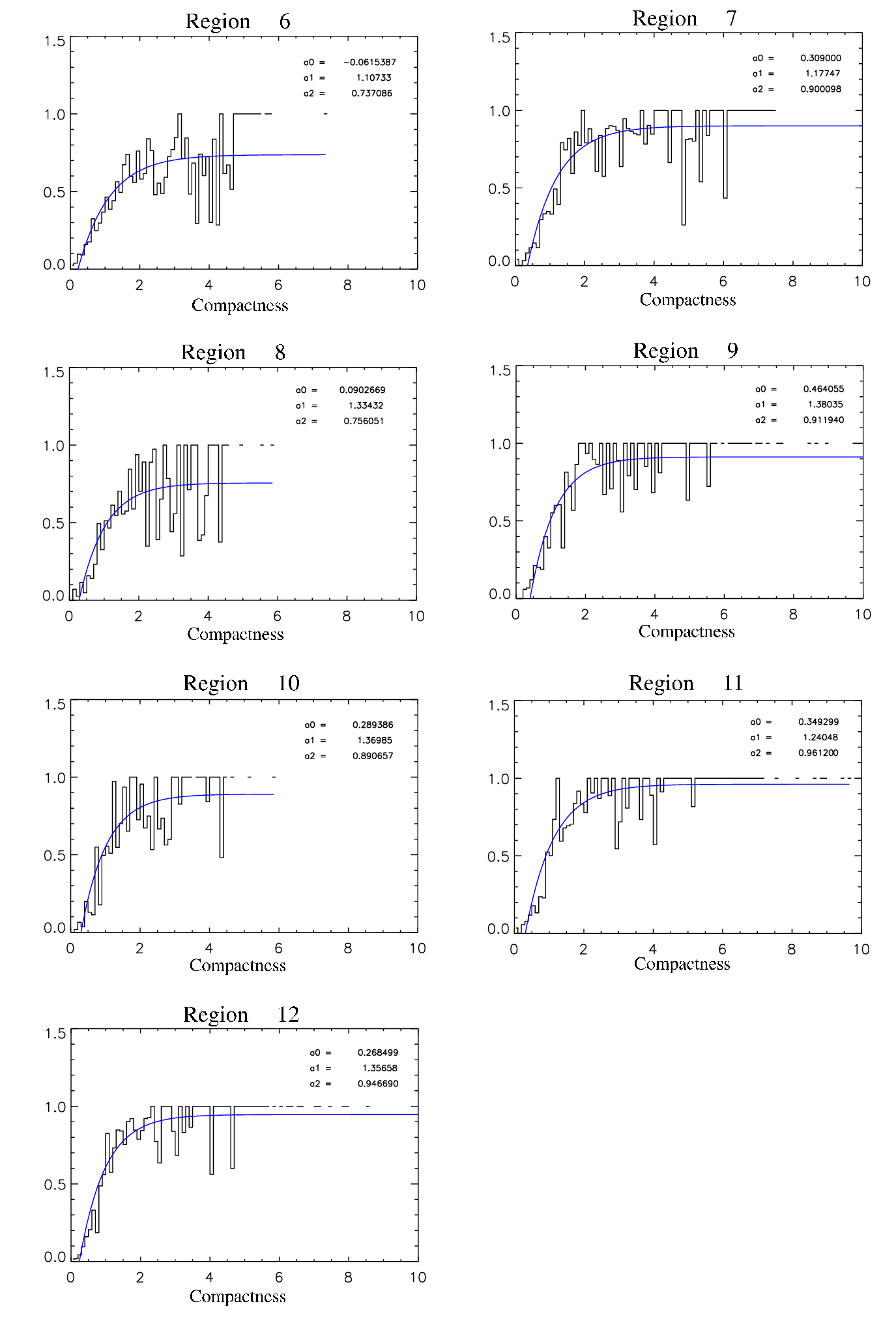}
\caption[$P_\gamma$ vs. Compactness Fits by Region] {Same as \ref{X2_RegionFits} for regions 6 -- 12.}
\label{X2_RegionFitsB}
\end{center}
\end{figure}


\nocite{*}
\bibliographystyle{styles/apjuc}

\end{document}